\documentclass[a4paper,11pt]{article}
\usepackage{jheppub} 
\usepackage[T1]{fontenc} 
\usepackage{amsmath,amsthm}
\usepackage{amssymb, amsfonts}
\usepackage{bbm}
\usepackage{enumitem}
\usepackage{dcolumn}
\usepackage{bm}

\usepackage[T1]{fontenc} 
\usepackage{amsmath,amsthm}
\usepackage{amssymb, amsfonts}
\usepackage{bbm}
\usepackage{graphicx}
\usepackage{dsfont}
\usepackage{relsize}
\usepackage{stmaryrd}
\usepackage{lmodern}
\usepackage{slantsc}
\usepackage{scalefnt}
\usepackage{wasysym}
\usepackage{xspace}
\usepackage{float}
\usepackage{boxedminipage}
\usepackage{ifpdf}
\usepackage{multirow, hhline}
\usepackage{slashed}
\usepackage{xifthen}
\usepackage{tensor}
\usepackage{array, nicematrix}
\usepackage{tabu}
\usepackage{tabularx}
\usepackage{pbox}
\usepackage{multirow}
\usepackage{tikz}
\usepackage{overpic}
\usepackage{color}
\usepackage{diagbox}
\usepackage{makecell}
\usetikzlibrary{arrows, matrix, calc, scopes, decorations.markings}
\usepackage[mathcal]{euscript}
\usepackage{booktabs}
\usepackage{autobreak, subfig, diagbox, mathrsfs}
\usepackage{mathbbol}
\usepackage[skins,breakable]{tcolorbox}
\usepackage{hyperref}
\usepackage{adjustbox}
\allowdisplaybreaks[4]

\newcommand{\NN}{\mathbb{N}}
\newcommand{\Z}{\mathbb{Z}}
\newcommand{\A}{\mathcal{A}}
\newcommand{\B}{\mathcal{B}}
\newcommand{\M}{\mathcal{M}}
\newcommand{\Hil}{\mathcal{H}}
\newcommand{\G}{\mathcal{G}}
\newcommand{\Cent}{\mathcal{Z}}
\newcommand{\T}{\mathcal{T}}
\newcommand{\Bimod}{{\mathsf{Bimod}}}

\newcommand{\C}{\mathbb{C}}
\newcommand{\V}{\mathcal{V}}
\newcommand{\D}{\mathcal{D}}

\newcommand{\Fus}{\mathscr{F}}
\newcommand{\F}{\mathcal{F}}
\newcommand{\RR}{\mathbb{R}}
\newcommand{\CC}{\mathbb{C}}
\newcommand{\dash}{{\text{-}}}
\newcommand{\idm}{{\mathds{1}}}
\newcommand{\nulm}{{\mathbb{O}}}
\newcommand{\Out}{{{\tt Out}}}
\renewcommand{\mod}{{\hspace{2pt}{\tt mod}\hspace{2pt}}}

\newcommand{\df}{{\mathrm{d}}}

\newcommand{\dk}[2][1]{{\ifthenelse{\equal{#1}{1}}{\frac{\df{#2}}{2\pi}}{\frac{\df^{#1}{#2}}{(2\pi)^{#1}}}}}

\renewcommand{\Vec}{{\tt Vec}}
\newcommand{\Rep}{{\tt Rep}}

\newcommand{\Cat}{{\mathcal{C}}}

\newcommand{\MA}{A^F_5}

\newcommand{\eqs}[1]{{\texorpdfstring{${#1}$}{Lg}}}

\newcommand{\ket}[1]{{\left\vert{#1}\right\rangle}}

\newcommand{\mat}[1]{{\begin{pmatrix}#1\end{pmatrix}}}

\newcommand{\eq}[1]{\begin{align*}#1\end{align*}}
\newcommand{\eqn}[2][0]{\ifthenelse{\equal{#1}{0}}{\begin{equation}\begin{aligned}#2\end{aligned}\end{equation}}{\begin{equation}\begin{aligned}#2\end{aligned}\label{#1}\end{equation}}}

\usetikzlibrary{shapes.geometric, snakes}
\tikzset{>=latex}
\usetikzlibrary{arrows, matrix, calc, scopes, decorations.markings}
\usetikzlibrary{decorations.pathmorphing, positioning}
\tikzset{snake it/.style={decorate, decoration={snake,amplitude=0.2mm,segment length=1mm}}}
\tikzset{->-/.style={decoration={
			 markings,
			 mark=at position .5*\pgfdecoratedpathlength+2pt with {\arrow{>}}},postaction={decorate}}}

\tikzset{-<-/.style={decoration={
			 markings,
			 mark=at position .5*\pgfdecoratedpathlength+2pt with {\arrow{<}}},postaction={decorate}}}

\newenvironment{example}[2]{\begin{tcolorbox}[title={{\textbf{\textit{Example (#1): Step #2}}}},colback=white, colframe=teal!80!darkgray!100!, breakable=true, width=\linewidth, enhanced]}{\end{tcolorbox}}

\newenvironment{exampleb}{\begin{tcolorbox}[title={{\textbf{\textit{Example}}}},colback=white, colframe=teal!80!darkgray!100!, breakable=true, width=\linewidth, enhanced]}{\end{tcolorbox}}

\newtabulinestyle{dash=off 2pt}

{\null\,\vcenter\bgroup\scriptsize
\arraycolsep=.13885em
\hbox\bgroup$\array{@{}#1@{}}}
{\endarray$\egroup\egroup\,\null}
\newcommand{\Lattice}{
\begin{tikzpicture}[baseline={([yshift=-.8ex]current bounding box.center)},scale=1]
\draw [decorate,decoration={snake,segment length=.12cm, amplitude=.04cm}] (1.7,6.8) -- (2.3,6.8);
\draw [] (1.7,6.4) -- (1.7,6.8);
\draw [] (1.7,6.8) -- (1.7,7.2);
\draw [] (1.7,7.2) -- (2.4,7.6);
\draw [] (2.4,7.6) -- (3.1,7.2);
\draw [] (2.4,6.) -- (1.7,6.4);
\draw [] (3.1,6.4) -- (2.4,6.);
\draw [] (2.4,5.2) -- (2.4,5.6);
\draw [] (2.4,5.6) -- (2.4,6.);
\draw [decorate,decoration={snake,segment length=.12cm, amplitude=.04cm}] (2.4,5.6) -- (3.,5.6);
\draw [decorate,decoration={snake,segment length=.12cm, amplitude=.04cm}] (3.1,6.8) -- (3.7,6.8);
\draw [] (3.1,6.4) -- (3.1,6.8);
\draw [] (3.1,6.8) -- (3.1,7.2);
\draw [] (3.1,7.2) -- (3.8,7.6);
\draw [] (3.8,7.6) -- (4.5,7.2);
\draw [] (3.8,6.) -- (3.1,6.4);
\draw [] (4.5,6.4) -- (3.8,6.);
\draw [] (3.8,5.2) -- (3.8,5.6);
\draw [] (3.8,5.6) -- (3.8,6.);
\draw [decorate,decoration={snake,segment length=.12cm, amplitude=.04cm}] (3.8,5.6) -- (4.4,5.6);
\draw [decorate,decoration={snake,segment length=.12cm, amplitude=.04cm}] (4.5,6.8) -- (5.1,6.8);
\draw [] (4.5,6.4) -- (4.5,6.8);
\draw [] (4.5,6.8) -- (4.5,7.2);
\draw [] (4.5,7.2) -- (5.2,7.6);
\draw [] (5.2,7.6) -- (5.9,7.2);
\draw [] (5.2,6.) -- (4.5,6.4);
\draw [] (5.9,6.4) -- (5.2,6.);
\draw [] (5.2,5.2) -- (5.2,5.6);
\draw [] (5.2,5.6) -- (5.2,6.);
\draw [decorate,decoration={snake,segment length=.12cm, amplitude=.04cm}] (5.2,5.6) -- (5.8,5.6);
\draw [decorate,decoration={snake,segment length=.12cm, amplitude=.04cm}] (5.9,6.8) -- (6.5,6.8);
\draw [] (5.9,6.4) -- (5.9,6.8);
\draw [] (5.9,6.8) -- (5.9,7.2);
\draw [] (5.9,7.2) -- (6.6,7.6);
\draw [] (6.6,7.6) -- (7.3,7.2);
\draw [] (6.6,6.) -- (5.9,6.4);
\draw [] (7.3,6.4) -- (6.6,6.);
\draw [] (6.6,5.2) -- (6.6,5.6);
\draw [] (6.6,5.6) -- (6.6,6.);
\draw [decorate,decoration={snake,segment length=.12cm, amplitude=.04cm}] (6.6,5.6) -- (7.2,5.6);
\draw [decorate,decoration={snake,segment length=.12cm, amplitude=.04cm}] (1.7,4.4) -- (2.3,4.4);
\draw [] (1.7,4.) -- (1.7,4.4);
\draw [] (1.7,4.4) -- (1.7,4.8);
\draw [] (1.7,4.8) -- (2.4,5.2);
\draw [] (2.4,5.2) -- (3.1,4.8);
\draw [] (2.4,3.6) -- (1.7,4.);
\draw [] (3.1,4.) -- (2.4,3.6);
\draw [] (2.4,3.2) -- (2.4,3.6);
\draw [decorate,decoration={snake,segment length=.12cm, amplitude=.04cm}] (3.1,4.4) -- (3.7,4.4);
\draw [] (3.1,4.) -- (3.1,4.4);
\draw [] (3.1,4.4) -- (3.1,4.8);
\draw [] (3.1,4.8) -- (3.8,5.2);
\draw [] (3.8,5.2) -- (4.5,4.8);
\draw [] (3.8,3.6) -- (3.1,4.);
\draw [] (4.5,4.) -- (3.8,3.6);
\draw [] (3.8,3.2) -- (3.8,3.6);
\draw [decorate,decoration={snake,segment length=.12cm, amplitude=.04cm}] (4.5,4.4) -- (5.1,4.4);
\draw [] (4.5,4.) -- (4.5,4.4);
\draw [] (4.5,4.4) -- (4.5,4.8);
\draw [] (4.5,4.8) -- (5.2,5.2);
\draw [] (5.2,5.2) -- (5.9,4.8);
\draw [] (5.2,3.6) -- (4.5,4.);
\draw [] (5.9,4.) -- (5.2,3.6);
\draw [] (5.2,3.2) -- (5.2,3.6);
\draw [decorate,decoration={snake,segment length=.12cm, amplitude=.04cm}] (5.9,4.4) -- (6.5,4.4);
\draw [] (5.9,4.) -- (5.9,4.4);
\draw [] (5.9,4.4) -- (5.9,4.8);
\draw [] (5.9,4.8) -- (6.6,5.2);
\draw [] (6.6,5.2) -- (7.3,4.8);
\draw [] (6.6,3.6) -- (5.9,4.);
\draw [] (7.3,4.) -- (6.6,3.6);
\draw [] (6.6,3.2) -- (6.6,3.6);
\draw [decorate,decoration={snake,segment length=.12cm, amplitude=.04cm}] (7.3,6.8) -- (7.9,6.8);
\draw [] (7.3,6.4) -- (7.3,6.8);
\draw [] (7.3,6.8) -- (7.3,7.2);
\draw [] (7.3,7.2) -- (7.6,7.4);
\draw [] (8.,6.) -- (7.3,6.4);
\draw [] (8.3,6.2) -- (8.,6.);
\draw [] (8.,5.2) -- (8.,5.6);
\draw [] (8.,5.6) -- (8.,6.);
\draw [decorate,decoration={snake,segment length=.12cm, amplitude=.04cm}] (8.,5.6) -- (8.6,5.6);
\draw [decorate,decoration={snake,segment length=.12cm, amplitude=.04cm}] (7.3,4.4) -- (7.9,4.4);
\draw [] (7.3,4.) -- (7.3,4.4);
\draw [] (7.3,4.4) -- (7.3,4.8);
\draw [] (7.3,4.8) -- (8.,5.2);
\draw [] (8.,5.2) -- (8.3,5.);
\draw [] (8.,3.6) -- (7.3,4.);
\draw [] (8.,3.2) -- (8.,3.6);
\draw [] (6.6,7.6) -- (6.6,8.);
\draw [] (5.2,7.6) -- (5.2,8.);
\draw [] (3.8,7.6) -- (3.8,8.);
\draw [] (2.4,7.6) -- (2.4,8.);
\draw [] (8.3,3.8) -- (8.,3.6);
\draw [] (1.7,4.) -- (1.3,3.8);
\draw [] (1.7,4.8) -- (1.3,5.);
\draw [] (1.7,6.4) -- (1.3,6.2);
\draw [] (1.7,7.2) -- (1.3,7.4);
\end{tikzpicture}
}

\newcommand{\Vertex}[3]{
\begin{tikzpicture}[baseline={([yshift=-.8ex]current bounding box.center)},scale=.7]
\draw [] (1.,0.) -- (1.,1.);
\draw [] (1.,1.) -- (0.1,1.5);
\draw [] (1.,1.) -- (1.9,1.5);

\node [centered, rotate=0.] at (1.2,.5) {\scriptsize ${#1}$};
\node [centered, rotate=0.] at (.5,1.5) {\scriptsize ${#2}$};
\node [centered, rotate=0.] at (1.5,1.5) {\scriptsize ${#3}$};
\node [centered, rotate=0.] at (1.,1.25) {\scriptsize $v$};
\end{tikzpicture}
}

\newcommand{\FatLattice}{
\begin{tikzpicture}[baseline={([yshift=-.8ex]current bounding box.center)},scale=.85]
\draw [] (0.1,0.057) -- (0.1,0.9);
\draw [] (-0.1,0.057) -- (-0.1,1.943);
\draw [] (0.1,1.1) -- (0.1,1.943);
\draw [] (0.1,1.943) -- (1.714,2.886);
\draw [] (0.1,0.057) -- (1.714,-0.886);
\draw [] (1.714,2.886) -- (3.328,1.943);
\draw [] (1.714,-0.886) -- (3.328,0.057);
\draw [] (3.328,1.943) -- (3.328,0.057);
\draw [] (-0.1,0.057) -- (-0.907,-0.414);
\draw [] (-0.1,1.943) -- (-0.907,2.414);
\draw [] (0.,-0.114) -- (-0.807,-0.586);
\draw [] (0.,2.114) -- (-0.807,2.586);
\draw [] (0.,2.114) -- (1.614,3.057);
\draw [] (1.614,3.057) -- (1.614,4.057);
\draw [] (1.814,3.057) -- (1.814,4.057);
\draw [] (1.814,3.057) -- (3.428,2.114);
\draw [] (3.428,2.114) -- (4.235,2.586);
\draw [] (3.528,1.943) -- (4.335,2.414);
\draw [] (0.,-0.114) -- (1.614,-1.057);
\draw [] (1.614,-1.057) -- (1.614,-2.057);
\draw [] (1.814,-1.057) -- (1.814,-2.057);
\draw [] (1.814,-1.057) -- (3.428,-0.114);
\draw [] (3.428,-0.114) -- (4.235,-0.586);
\draw [] (3.528,0.057) -- (4.335,-0.414);
\draw [] (3.528,0.057) -- (3.528,0.9);
\draw [] (3.528,1.943) -- (3.528,1.1);
\draw [decorate,decoration={snake,segment length=.12cm, amplitude=.04cm}] (0.1,0.9) -- (1.7,0.9);
\draw [decorate,decoration={snake,segment length=.12cm, amplitude=.04cm}] (0.1,1.1) -- (1.7,1.1);
\draw [decorate,decoration={snake,segment length=.12cm, amplitude=.04cm}] (3.528,0.9) -- (4.2,0.9);
\draw [decorate,decoration={snake,segment length=.12cm, amplitude=.04cm}] (3.528,1.1) -- (4.2,1.1);

\node [centered, rotate=0.] at (0.,0.5) {\scriptsize $x$};
\node [centered, rotate=0.] at (0.,1.5) {\scriptsize $y$};
\node [centered, rotate=0.] at (-0.3,0.5) {\scriptsize $g_i$};
\node [centered, rotate=0.] at (0.3,0.5) {\scriptsize $g_j$};
\node [centered, rotate=0.] at (-0.3,1.5) {\scriptsize $g_i$};
\node [centered, rotate=0.] at (0.3,1.5) {\scriptsize $g_j$};
\node [centered, rotate=0.] at (1.,1.) {\scriptsize $z$};
\node [centered, rotate=0.] at (1.,1.3) {\scriptsize $g_j$};
\node [centered, rotate=0.] at (1.,0.7) {\scriptsize $g_j$};
\node [centered, rotate=0.] at (0.857,2.5) {\scriptsize $u$};
\node [centered, rotate=0.] at (0.715,2.7) {\scriptsize $g_k$};
\node [centered, rotate=0.] at (1.045,2.3) {\scriptsize $g_j$};
\node [centered, rotate=0.] at (2.571,2.5) {\scriptsize $v$};
\node [centered, rotate=0.] at (2.421,2.3) {\scriptsize $g_j$};
\node [centered, rotate=0.] at (2.631,2.75) {\scriptsize $g_l$};
\end{tikzpicture}
}

\newcommand{\PlaquetteSrc}{
\begin{tikzpicture}[baseline={([yshift=-.8ex]current bounding box.center)},scale=.5]
\draw [->-] (3.2,5.6) -- (3.2,6.8);
\draw [->-] (3.2,7.2) -- (4.6,8.);
\draw [->-] (4.6,8.) -- (6.,7.2);
\draw [->-] (6.,5.6) -- (4.6,4.8);
\draw [->-] (4.6,4.8) -- (3.2,5.6);
\draw [->-] (6.,7.2) -- (6.,5.6);
\draw [decorate,decoration={snake,segment length=.12cm, amplitude=.04cm}, ->] (3.2,6.4) -- (4.4,6.4);
\draw [->-] (2.6,7.6) -- (3.2,7.2);
\draw [->-] (4.6,8.8) -- (4.6,8.);
\draw [->-] (6.6,7.6) -- (6.,7.2);
\draw [->-] (6.6,5.2) -- (6.,5.6);
\draw [->-] (4.6,4.) -- (4.6,4.8);
\draw [->-] (2.6,5.2) -- (3.2,5.6);
\draw [->-] (3.2,6.6) -- (3.2,7.2);
\node [centered, rotate=0.] at (2.4,7.8) {\scriptsize $e_1$};
\node [centered, rotate=0.] at (4.95,8.8) {\scriptsize $e_2$};
\node [centered, rotate=0.] at (6.9,7.8) {\scriptsize $e_3$};
\node [centered, rotate=0.] at (6.9,5.) {\scriptsize $e_4$};
\node [centered, rotate=0.] at (5.,4.) {\scriptsize $e_5$};
\node [centered, rotate=0.] at (2.3,5.) {\scriptsize $e_6$};
\node [centered, rotate=0.] at (2.85,6.) {\scriptsize $i_7$};
\node [centered, rotate=0.] at (2.85,6.9) {\scriptsize $i_1$};
\node [centered, rotate=0.] at (3.6,7.9) {\scriptsize $i_2$};
\node [centered, rotate=0.] at (5.6,7.9) {\scriptsize $i_3$};
\node [centered, rotate=0.] at (6.4,6.4) {\scriptsize $i_4$};
\node [centered, rotate=0.] at (5.6,4.9) {\scriptsize $i_5$};
\node [centered, rotate=0.] at (3.7,4.9) {\scriptsize $i_6$};
\node [centered, rotate=0.] at (4.7,6.4) {\scriptsize $p$};
\end{tikzpicture}
}

\newcommand{\PlaquetteTar}{
\begin{tikzpicture}[baseline={([yshift=-.8ex]current bounding box.center)},scale=.5]
\draw [->-] (3.2,5.6) -- (3.2,6.8);
\draw [->-] (3.2,7.2) -- (4.6,8.);
\draw [->-] (4.6,8.) -- (6.,7.2);
\draw [->-] (6.,5.6) -- (4.6,4.8);
\draw [->-] (4.6,4.8) -- (3.2,5.6);
\draw [->-] (6.,7.2) -- (6.,5.6);
\draw [decorate,decoration={snake,segment length=.12cm, amplitude=.04cm}] (3.2,6.4) -- (4.4,6.4);
\draw [->-] (2.6,7.6) -- (3.2,7.2);
\draw [->-] (4.6,8.8) -- (4.6,8.);
\draw [->-] (6.6,7.6) -- (6.,7.2);
\draw [->-] (6.6,5.2) -- (6.,5.6);
\draw [->-] (4.6,4.) -- (4.6,4.8);
\draw [->-] (2.6,5.2) -- (3.2,5.6);
\draw [->-] (3.2,6.6) -- (3.2,7.2);
\node [centered, rotate=0.] at (2.4,7.8) {\scriptsize $e_1$};
\node [centered, rotate=0.] at (4.95,8.8) {\scriptsize $e_2$};
\node [centered, rotate=0.] at (6.9,7.8) {\scriptsize $e_3$};
\node [centered, rotate=0.] at (6.9,5.) {\scriptsize $e_4$};
\node [centered, rotate=0.] at (5.,4.) {\scriptsize $e_5$};
\node [centered, rotate=0.] at (2.3,5.) {\scriptsize $e_6$};
\node [centered, rotate=0.] at (2.85,6.) {\scriptsize $j_7$};
\node [centered, rotate=0.] at (2.85,6.9) {\scriptsize $j_1$};
\node [centered, rotate=0.] at (3.6,7.9) {\scriptsize $j_2$};
\node [centered, rotate=0.] at (5.6,7.9) {\scriptsize $j_3$};
\node [centered, rotate=0.] at (6.4,6.4) {\scriptsize $j_4$};
\node [centered, rotate=0.] at (5.6,4.9) {\scriptsize $j_5$};
\node [centered, rotate=0.] at (3.7,4.9) {\scriptsize $j_6$};
\node [centered, rotate=0.] at (4.7,6.4) {\scriptsize $1$};
\end{tikzpicture}
}

\newcommand{\PachnerOne}{\begin{tikzpicture}[baseline={([yshift=-.8ex]current bounding box.center)},scale=.6]
\draw [->-] (0.,0.2) -- (.5,1.);
\draw [->-] (0.,1.8) -- (.5,1.);
\draw [->-] (2.,1.) -- (.5,1.);
\draw [->-] (2.5,1.8) -- (2.,1.);
\draw [->-] (2.5,0.2) -- (2.,1.);
\node at (1.25, 1.3) {\scriptsize $m$};
\node at (-.15, 2.) {\scriptsize $a$};
\node at (-.15, 0.) {\scriptsize $b$};
\node at (2.65, 2.) {\scriptsize $d$};
\node at (2.65, 0.) {\scriptsize $c$};
\end{tikzpicture}\qquad = \qquad \sum_{n\in L_{\Fus}}\ \sqrt{d_md_n}\ G^{abm}_{cdn}\quad \begin{tikzpicture}[baseline={([yshift=-.8ex]current bounding box.center)},scale=.6]
\draw [->-] (0.,0.) -- (.8,.5);
\draw [->-] (1.6,0.) -- (.8,.5);
\draw [->-] (.8,.5) -- (.8,2.);
\draw [->-] (0.,2.5) -- (.8,2.);
\draw [->-] (1.6,2.5) -- (.8,2.);
\node at (1.05, 1.3) {\scriptsize $n$};
\node at (-.15, 2.6) {\scriptsize $a$};
\node at (-.15, -.1) {\scriptsize $b$};
\node at (1.75, 2.6) {\scriptsize $d$};
\node at (1.75, -0.1) {\scriptsize $c$};
\end{tikzpicture}}

\newcommand{\PachnerTwo}{\begin{tikzpicture}[baseline={([yshift=-.8ex]current bounding box.center)},scale=.6]
\draw [->-] (1, 0) -- (1, 1);
\draw [->-] (1, 1) arc (270:90:.5);
\draw [->-] (1, 1) arc (-90:90:.5);
\draw [->-] (1, 2) -- (1, 3);
\node at (1.25, .3) {\scriptsize $a$};
\node at (1., 1.5) {\scriptsize \color{red}$\times$};
\node at (.2, 1.5) {\scriptsize $x$};
\node at (1.8, 1.5) {\scriptsize $y$};
\node at (1.25, 2.7) {\scriptsize $b$};
\end{tikzpicture}\qquad = \qquad \sqrt{\frac{d_xd_y}{d_a}}\ \delta_{ab}\ \delta_{xya^\ast}\quad \begin{tikzpicture}[baseline={([yshift=-.8ex]current bounding box.center)},scale=.5]
\draw [->-] (1, 0) -- (1, 3);
\node at (1.25, 1.5) {\scriptsize $a$};
\end{tikzpicture}}

\newcommand{\PachnerThree}{\begin{tikzpicture}[baseline={([yshift=-.8ex]current bounding box.center)},scale=.5]
\draw [->-] (1, 0) -- (1, 3);
\node at (1.25, 1.5) {\scriptsize $a$};
\end{tikzpicture}\quad = \quad \frac{1}{D}\sum_{xy\in L_{\Fus}} \sqrt{\frac{d_xd_y}{d_a}}\ \delta_{xya^\ast} \quad \begin{tikzpicture}[baseline={([yshift=-.8ex]current bounding box.center)},scale=.6]
\draw [->-] (1, 0) -- (1, 1);
\draw [->-] (1, 1) arc (270:90:.5);
\draw [->-] (1, 1) arc (-90:90:.5);
\draw [->-] (1, 2) -- (1, 3);
\node at (1.25, .3) {\scriptsize $a$};
\node at (.2, 1.5) {\scriptsize $x$};
\node at (1.8, 1.5) {\scriptsize $y$};
\node at (1.25, 2.7) {\scriptsize $a$};
\end{tikzpicture}}

\newcommand{\PachnerFour}{
\begin{tikzpicture}[baseline={([yshift=-.8ex]current bounding box.center)},scale=.6]
\draw [->-] (3.2,5.6) -- (3.2,6.8);
\draw [->-] (3.2,7.2) -- (4.6,8.);
\draw [->-] (4.6,8.) -- (6.,7.2);
\draw [->-] (6.,5.6) -- (4.6,4.8);
\draw [->-] (4.6,4.8) -- (3.2,5.6);
\draw [->-] (6.,7.2) -- (6.,5.6);
\draw [decorate,decoration={snake,segment length=.12cm, amplitude=.04cm}, ->] (3.2,6.6) -- (4.4,6.6);
\draw [->-] (2.6,7.6) -- (3.2,7.2);
\draw [->-] (4.6,8.8) -- (4.6,8.);
\draw [->-] (6.6,7.6) -- (6.,7.2);
\draw [->-] (6.6,5.2) -- (6.,5.6);
\draw [->-] (4.6,4.) -- (4.6,4.8);
\draw [->-] (2.6,5.2) -- (3.2,5.6);
\draw [->-] (3.2,6.6) -- (3.2,7.2);
\node [centered, rotate=0.] at (2.5,7.8) {\scriptsize $e_1$};
\node [centered, rotate=0.] at (2.9,6.) {\scriptsize $i_0$};
\node [centered, rotate=0.] at (2.9,6.9) {\scriptsize $i_1$};
\node [centered, rotate=0.] at (3.6,7.8) {\scriptsize $i_2$};
\node [centered, rotate=0.] at (4.7,6.6) {\scriptsize $p$};
\end{tikzpicture}\ =\quad\sum_{j\in L_\Fus}\sqrt{d_{i_1}d_j}\ G^{i_2^\ast e_1i_1}_{i_0 p^\ast j}\ 
\begin{tikzpicture}[baseline={([yshift=-.8ex]current bounding box.center)},scale=.6]
\draw [->-] (3.2,5.6) -- (3.2,7.2);
\draw [->-] (3.2,7.2) -- (3.55,7.4);
\draw [->-] (3.55,7.4) -- (4.6,8.);
\draw [->-] (4.6,8.) -- (6.,7.2);
\draw [->-] (6.,5.6) -- (4.6,4.8);
\draw [->-] (4.6,4.8) -- (3.2,5.6);
\draw [->-] (6.,7.2) -- (6.,5.6);
\draw [->-] (2.6,7.6) -- (3.2,7.2);
\draw [->-] (4.6,8.8) -- (4.6,8.);
\draw [->-] (6.6,7.6) -- (6.,7.2);
\draw [->-] (6.6,5.2) -- (6.,5.6);
\draw [->-] (4.6,4.) -- (4.6,4.8);
\draw [->-] (2.6,5.2) -- (3.2,5.6);
\draw [decorate,decoration={snake,segment length=.12cm, amplitude=.04cm}, ->] (3.59,7.42) -- (4.1,6.5);
\node [centered, rotate=0.] at (2.5,5.) {\scriptsize $e_n$};
\node [centered, rotate=0.] at (2.5,7.8) {\scriptsize $e_1$};
\node [centered, rotate=0.] at (2.9,6.5) {\scriptsize $i_0$};
\node [centered, rotate=0.] at (3.3,7.6) {\scriptsize $j$};
\node [centered, rotate=0.] at (3.9,8.) {\scriptsize $i_2$};
\node [centered, rotate=0.] at (4.3,6.3) {\scriptsize $p$};
\node [centered, rotate=0.] at (3.7,5.1) {\scriptsize $i_n$};
\end{tikzpicture}
}

\newcommand{\PachnerFive}{
\begin{tikzpicture}[baseline={([yshift=-.8ex]current bounding box.center)},scale=.6]
\draw [->-] (3.2,5.6) -- (3.2,6.1);
\draw [->-] (3.2,6.1) -- (3.2,6.7);
\draw [->-] (3.2,6.7) -- (3.2,7.2);
\draw [->-] (3.2,7.2) -- (4.6,8.);
\draw [->-] (4.6,8.) -- (6.,7.2);
\draw [->-] (6.,5.6) -- (4.6,4.8);
\draw [->-] (4.6,4.8) -- (3.2,5.6);
\draw [->-] (6.,7.2) -- (6.,5.6);
\draw [decorate,decoration={snake,segment length=.12cm, amplitude=.04cm}, ->] (3.2,6.1) -- (4.4,6.1);
\draw [decorate,decoration={snake,segment length=.12cm, amplitude=.04cm}, ->] (3.2,6.7) -- (4.4,6.7);
\draw [->-] (2.6,7.6) -- (3.2,7.2);
\draw [->-] (4.6,8.8) -- (4.6,8.);
\draw [->-] (6.6,7.6) -- (6.,7.2);
\draw [->-] (6.6,5.2) -- (6.,5.6);
\draw [->-] (4.6,4.) -- (4.6,4.8);
\draw [->-] (2.6,5.2) -- (3.2,5.6);
\node [centered, rotate=0.] at (2.9,5.8) {\scriptsize $i_0$};
\node [centered, rotate=0.] at (2.9,7.1) {\scriptsize $i_1$};
\node [centered, rotate=0.] at (4.7,6.7) {\scriptsize $r$};
\node [centered, rotate=0.] at (4.7,6.1) {\scriptsize $s$};
\node [centered, rotate=0.] at (2.9,6.4) {\scriptsize $j$};
\end{tikzpicture}\quad = \quad \sum_{u\in L_\Fus}\sqrt{d_jd_p}\ G^{r^\ast i_1^\ast j}_{i_0s^\ast p}\quad 
\begin{tikzpicture}[baseline={([yshift=-.8ex]current bounding box.center)},scale=.6]
\draw [->-] (3.2,5.6) -- (3.2,6.8);
\draw [->-] (3.2,7.2) -- (4.6,8.);
\draw [->-] (4.6,8.) -- (6.,7.2);
\draw [->-] (6.,5.6) -- (4.6,4.8);
\draw [->-] (4.6,4.8) -- (3.2,5.6);
\draw [->-] (6.,7.2) -- (6.,5.6);
\draw [decorate,decoration={snake,segment length=.12cm, amplitude=.04cm}, ->] (3.2,6.6) -- (4.4,6.6);
\draw [->-] (2.6,7.6) -- (3.2,7.2);
\draw [->-] (4.6,8.8) -- (4.6,8.);
\draw [->-] (6.6,7.6) -- (6.,7.2);
\draw [->-] (6.6,5.2) -- (6.,5.6);
\draw [->-] (4.6,4.) -- (4.6,4.8);
\draw [->-] (2.6,5.2) -- (3.2,5.6);
\draw [->-] (3.2,6.6) -- (3.2,7.2);
\node [centered, rotate=0.] at (2.9,6.) {\scriptsize $i_0$};
\node [centered, rotate=0.] at (2.9,6.9) {\scriptsize $i_1$};
\node [centered, rotate=0.] at (4.7,6.6) {\scriptsize $p$};
\end{tikzpicture}
}

\newcommand{\PlaquetteMsr}{
\begin{tikzpicture}[baseline={([yshift=-.8ex]current bounding box.center)},scale=.55]
\draw [->-] (3.2,5.6) -- (3.2,6.8);
\draw [->-] (3.2,7.2) -- (4.6,8.);
\draw [->-] (4.6,8.) -- (6.,7.2);
\draw [->-] (6.,5.6) -- (4.6,4.8);
\draw [->-] (4.6,4.8) -- (3.2,5.6);
\draw [->-] (6.,7.2) -- (6.,5.6);
\draw [->-] (2.6,7.6) -- (3.2,7.2);
\draw [->-] (4.6,8.8) -- (4.6,8.);
\draw [->-] (6.6,7.6) -- (6.,7.2);
\draw [->-] (6.6,5.2) -- (6.,5.6);
\draw [->-] (4.6,4.) -- (4.6,4.8);
\draw [->-] (2.6,5.2) -- (3.2,5.6);
\draw [->-] (3.2,6.6) -- (3.2,7.2);
\node [centered, rotate=0.] at (2.4,7.8) {\scriptsize $e_1$};
\node [centered, rotate=0.] at (4.95,8.8) {\scriptsize $e_2$};
\node [centered, rotate=0.] at (6.9,7.8) {\scriptsize $e_3$};
\node [centered, rotate=0.] at (6.9,5.) {\scriptsize $e_4$};
\node [centered, rotate=0.] at (5.,4.) {\scriptsize $e_5$};
\node [centered, rotate=0.] at (2.3,5.) {\scriptsize $e_6$};
\node [centered, rotate=0.] at (2.85,6.) {\scriptsize $i_7$};
\node [centered, rotate=0.] at (2.85,6.9) {\scriptsize $i_1$};
\node [centered, rotate=0.] at (3.6,7.9) {\scriptsize $i_2$};
\node [centered, rotate=0.] at (5.6,7.9) {\scriptsize $i_3$};
\node [centered, rotate=0.] at (6.4,6.4) {\scriptsize $i_4$};
\node [centered, rotate=0.] at (5.6,4.9) {\scriptsize $i_5$};
\node [centered, rotate=0.] at (3.7,4.9) {\scriptsize $i_6$};

\draw [decorate,decoration={snake,segment length=.12cm, amplitude=.04cm}, ->] (3.2,6.4) -- (5.,6.4);
\draw [->] (3.8, 6.4) arc (180:0:.8);
\draw [] (4.4, 6.4) arc (-180:0:0.5);
\node [centered, rotate=0.] at (4.8,6.7) {\scriptsize $p$};
\node [centered, rotate=0.] at (4.1,6.1) {\scriptsize $t$};
\node [centered, rotate=0.] at (3.5,6.1) {\scriptsize $p$};
\node [centered, rotate=0.] at (5.65,6.4) {\scriptsize $s$};
\node [centered, rotate=0.] at (4.6,5.4) {\color{red} $\times$};
\end{tikzpicture}
}

\newcommand{\ExcitedA}{
\begin{tikzpicture}[baseline={([yshift=-.8ex]current bounding box.center)},scale=.7]
\draw [->-] (3.2,5.6) -- (3.2,7.2);
\node [centered, rotate=0.] at (2.9,6.4) {\scriptsize $j$};
\end{tikzpicture}
}

\newcommand{\ExcitedB}{
\begin{tikzpicture}[baseline={([yshift=-.8ex]current bounding box.center)},scale=.8]
\draw [->-] (3.2,5.6) -- (3.2,6.1);
\draw [->-] (3.2,6.1) -- (3.2,6.7);
\draw [->-] (3.2,6.7) -- (3.2,7.2);
\draw [decorate,decoration={snake,segment length=.12cm, amplitude=.04cm}, ->] (3.2,6.7) -- (4.,6.7);
\draw [decorate,decoration={snake,segment length=.12cm, amplitude=.04cm}, ->] (3.2,6.1) -- (2.4,6.1);
\node [centered, rotate=0.] at (3.,5.8) {\scriptsize $j$};
\node [centered, rotate=0.] at (3.,6.4) {\scriptsize $k$};
\node [centered, rotate=0.] at (3.4,7.) {\scriptsize $j$};
\node [centered, rotate=0.] at (4.2,6.7) {\scriptsize $q$};
\node [centered, rotate=0.] at (2.2,6.1) {\scriptsize $p^\ast$};
\end{tikzpicture}
}

\newcommand{\ExcitedC}{
\begin{tikzpicture}[baseline={([yshift=-.8ex]current bounding box.center)},scale=1.2]
\draw [->-] (3.2,5.6) -- (3.2,7.2);
\node [centered, rotate=0.] at (2.8,6.4) {\scriptsize $x_{g_ig_j}$};
\end{tikzpicture}
}

\newcommand{\ExcitedD}{
\begin{tikzpicture}[baseline={([yshift=-.8ex]current bounding box.center)},scale=1.2]
\draw [->-] (3.2,5.6) -- (3.2,6.1);
\draw [->-] (3.2,6.1) -- (3.2,6.7);
\draw [->-] (3.2,6.7) -- (3.2,7.2);
\draw [decorate,decoration={snake,segment length=.12cm, amplitude=.04cm}, ->] (3.2,6.7) -- (4.,6.7);
\draw [decorate,decoration={snake,segment length=.12cm, amplitude=.04cm}, ->] (3.2,6.1) -- (2.4,6.1);
\node [centered, rotate=0.] at (2.9,5.8) {\scriptsize $x_{g_ig_j}$};
\node [centered, rotate=0.] at (2.9,6.4) {\scriptsize $x_{g_ig_j}$};
\node [centered, rotate=0.] at (2.9,7.) {\scriptsize $x_{g_ig_j}$};
\node [centered, rotate=0.] at (4.5,6.7) {\scriptsize $(\idm_{g_jg_j})_b$};
\node [centered, rotate=0.] at (1.8,6.1) {\scriptsize $(\idm_{g_ig_i})_a$};
\end{tikzpicture}
}

\newcommand{\FrobeniusA}{\begin{tikzpicture}[baseline={([yshift=-.8ex]current bounding box.center)},scale=.8]
\draw [red, ->-] (.3, 0.5) -- (1, 1);
\draw [red, ->-] (1.7, 0.5) -- (1, 1);
\draw [red, ->-] (1., 2.) -- (1, 1);
\fill [red] (1., 1.) circle (.15);
\node at (1.3, 1.7) {\scriptsize $a_\alpha$};
\node at (.2, .8) {\scriptsize $b_\beta$};
\node at (1.75, .8) {\scriptsize $c_\gamma$};
\end{tikzpicture}\qquad := \qquad f_{a_\alpha b_\beta c_\gamma}\quad \begin{tikzpicture}[baseline={([yshift=-.8ex]current bounding box.center)},scale=.8]
\draw [->-] (.3, 0.5) -- (1, 1);
\draw [->-] (1.7, 0.5) -- (1, 1);
\draw [->-] (1., 2.) -- (1, 1);
\node at (1.25, 1.6) {\scriptsize $a_\alpha$};
\node at (.3, .8) {\scriptsize $b_\beta$};
\node at (1.6, .8) {\scriptsize $c_\gamma$};
\end{tikzpicture}}

\newcommand{\FrobeniusB}{\T\ \sum_{\gamma = 0}^{n_c}\begin{tikzpicture}[baseline={([yshift=-.8ex]current bounding box.center)},scale=.7]
\draw [red, ->-] (0.,0.2) -- (.5,1.);
\draw [red, ->-] (0.,1.8) -- (.5,1.);
\draw [red, ->-] (2.,1.) -- (.5,1.);
\draw [red, ->-] (2.5,1.8) -- (2.,1.);
\draw [red, ->-] (2.5,0.2) -- (2.,1.);
\fill [red] (.5, 1.) circle (.15);
\fill [red] (2., 1.) circle (.15);
\node at (1.25, 1.3) {\scriptsize $c_\gamma$};
\node at (-.1, 2.) {\scriptsize $s_\sigma$};
\node at (-.1, 0.) {\scriptsize $a_\alpha$};
\node at (2.6, 2.) {\scriptsize $r_\rho$};
\node at (2.8, 0.) {\scriptsize $b_\beta$};
\end{tikzpicture}\qquad =\qquad \sum_{t_\tau\in L_{\mathcal{A}}}\quad \begin{tikzpicture}[baseline={([yshift=-.8ex]current bounding box.center)},scale=.7]
\draw [red, ->-] (0.,0.) -- (.8,.5);
\draw [red, ->-] (1.6,0.) -- (.8,.5);
\draw [red, ->-] (.8,.5) -- (.8,2.);
\draw [red, ->-] (0.,2.5) -- (.8,2.);
\draw [red, ->-] (1.6,2.5) -- (.8,2.);
\fill [red] (.8, .5) circle (.15);
\fill [red] (.8, 2.) circle (.15);
\node at (1.2, 1.3) {\scriptsize $t_\tau$};
\node at (-.2, 2.7) {\scriptsize $s_\sigma$};
\node at (-.2, -.2) {\scriptsize $a_\alpha$};
\node at (1.9, 2.7) {\scriptsize $r_\rho$};
\node at (1.9, -0.2) {\scriptsize $b_\beta$};
\end{tikzpicture}\qquad =: \qquad \begin{tikzpicture}[baseline={([yshift=-.8ex]current bounding box.center)},scale=.7]
\draw [red, ->-] (0.,0.) -- (.8,.5);
\draw [red, ->-] (1.6,0.) -- (.8,.5);
\draw [red, thick, densely dashed] (.8,.5) -- (.8,2.);
\draw [red, ->-] (0.,2.5) -- (.8,2.);
\draw [red, ->-] (1.6,2.5) -- (.8,2.);
\fill [red] (.8, .5) circle (.15);
\fill [red] (.8, 2.) circle (.15);
\node at (-.2, 2.7) {\scriptsize $s_\sigma$};
\node at (-.2, -.2) {\scriptsize $a_\alpha$};
\node at (1.9, 2.7) {\scriptsize $r_\rho$};
\node at (1.9, -0.2) {\scriptsize $b_\beta$};
\end{tikzpicture}}

\newcommand{\FrobeniusC}{\T\quad \begin{tikzpicture}[baseline={([yshift=-.8ex]current bounding box.center)},scale=.7]
\draw [red, ->-] (1, 0) -- (1, 1);
\draw [red, thick, densely dashed] (1, 1) arc (-90:270:.5);
\draw [red, ->-] (1, 2) -- (1, 3);
\fill [red] (1., 1.) circle (.1);
\fill [red] (1., 2.) circle (.1);
\node at (1.35, .3) {\scriptsize $a_\alpha$};
\node at (1.35, 2.7) {\scriptsize $b_\beta$};
\node [red] at (1., 1.5) {$\times$};
\end{tikzpicture}\qquad = \qquad d_{\mathcal{A}}\ \delta_{ab}\ \delta_{\alpha\beta}\quad \begin{tikzpicture}[baseline={([yshift=-.8ex]current bounding box.center)},scale=.7]
\draw [red, ->-] (1, 0) -- (1, 3);
\node at (1.35, 1.5) {\scriptsize $a_\alpha$};
\end{tikzpicture}}

\newcommand{\BimoduleA}{\begin{tikzpicture}[baseline={([yshift=-.8ex]current bounding box.center)},scale=1.]
\draw [blue, ->-] (1, 0) -- (1, .6);
\draw [blue, snake it, ultra thick] (1, .6) -- (1, 1.3);
\draw [blue, ->-] (1, 1.3) -- (1, 2.);
\draw [red, -<-] (1, .6) -- (.3, .6);
\draw [orange, -<-] (1, 1.3) -- (1.7, 1.3);
\node at (0.1, .6) {\scriptsize $a_\alpha$};
\node at (1.9, 1.3) {\scriptsize $b_\beta$};
\node at (1.3, .2) {\scriptsize $x_\chi$};
\node at (1.25,.9) {\scriptsize $M$};
\node at (1.25,1.8) {\scriptsize $z_\zeta$};
\end{tikzpicture}\qquad := \qquad\sum_{y\in L_\Fus}\ [P_M]^{a_\alpha b_\beta}_{x_\chi y z_\zeta}\quad\begin{tikzpicture}[baseline={([yshift=-.8ex]current bounding box.center)},scale=1.]
\draw [->-] (1, 0) -- (1, .6);
\draw [->-] (1, .6) -- (1, 1.3);
\draw [->-] (1, 1.3) -- (1, 2.);
\draw [-<-] (1, .6) -- (.3, .6);
\draw [-<-] (1, 1.3) -- (1.7, 1.3);
\node at (0.1, .6) {\scriptsize $a_\alpha$};
\node at (1.9, 1.3) {\scriptsize $b_\beta$};
\node at (1.25, .2) {\scriptsize $x_\chi$};
\node at (1.15,.9) {\scriptsize $y$};
\node at (1.25,1.8) {\scriptsize $z_\zeta$};
\end{tikzpicture}
}

\newcommand{\BimoduleB}{\T\ \sum_{y_\upsilon\in L_M}
\begin{tikzpicture}[baseline={([yshift=-.8ex]current bounding box.center)},scale=.7]
\draw [blue, ->-] (1, 0) -- (1, 1.);
\draw [blue, snake it, ultra thick] (1, 1.) -- (1, 1.8);
\draw [blue, ->-] (1, 1.8) -- (1, 2.6);
\draw [blue, snake it, ultra thick] (1, 2.6) -- (1, 3.4);
\draw [blue, ->-] (1, 3.4) -- (1, 4.4);

\draw [red, -<-] (1, 1.) -- (0, 1.);
\draw [orange, -<-] (1, 1.8) -- (2, 1.8);
\draw [red, -<-] (1, 2.6) -- (0, 2.6);
\draw [orange, -<-] (1, 3.4) -- (2, 3.4);

\node at (1.3, .5) {\scriptsize $x_\chi$};
\node at (-.3, 1.) {\scriptsize $a_\alpha$};
\node at (1.3, 1.4) {\scriptsize $M$};
\node at (2.3, 1.8) {\scriptsize $r_\rho$};
\node at (1.3, 2.2) {\scriptsize $y_\upsilon$};
\node at (-.3, 2.6) {\scriptsize $b_\beta$};
\node at (1.3, 3.) {\scriptsize $M$};
\node at (2.3, 3.4) {\scriptsize $s_\sigma$};
\node at (1.3, 3.9) {\scriptsize $z_\zeta$};
\end{tikzpicture}\qquad = \qquad
\begin{tikzpicture}[baseline={([yshift=-.8ex]current bounding box.center)},scale=.7]
\draw [blue, ->-] (1, 0) -- (1, 1.5);
\draw [blue, snake it, ultra thick] (1, 1.5) -- (1, 2.9);
\draw [blue, ->-] (1, 2.9) -- (1, 4.4);
\draw [red, -<-] (-.5, 1.5) -- (-1.3, 2.5);
\draw [red, -<-] (-.5, 1.5) -- (-1.3, .5);
\draw [red, densely dashed, very thick] (1, 1.5) -- (-.5, 1.5);
\fill [red] (-.5, 1.5) circle (0.15);
\draw [orange, densely dashed, very thick] (1, 2.9) -- (2.5, 2.9);
\draw [orange, -<-] (2.5, 2.9) -- (3.3, 3.9);
\draw [orange, -<-] (2.5, 2.9) -- (3.3, 1.9);
\fill [orange] (2.5, 2.9) circle (0.15);
\node at (1.35, .7) {\scriptsize $x_\chi$};
\node at (1.3, 2.2) {\scriptsize $M$};
\node at (1.35, 3.7) {\scriptsize $z_\zeta$};
\node at (-1.5, .3) {\scriptsize $a_\alpha$};
\node at (-1.5, 2.7) {\scriptsize $b_\beta$};
\node at (3.5, 1.7) {\scriptsize $r_\rho$};
\node at (3.5, 4.1) {\scriptsize $s_\sigma$};
\end{tikzpicture}}

\newcommand{\BimoduleC}{\T\quad
\begin{tikzpicture}[baseline={([yshift=-.8ex]current bounding box.center)},scale=1.]
\draw [blue, ->-] (2.,3.2) -- (2.,2.4);
\draw [blue, ultra thick, snake it] (2.,2.4) -- (2.,1.6);
\draw [blue, ->-] (2.,1.6) -- (2.,1.);
\draw [blue, ->-] (.2,-.2) -- (.95,.3);
\draw [blue, ultra thick, snake it] (.95,.3) -- (1.55,.7);
\draw [blue, ->-] (1.55,.7) -- (2.,1.);
\draw [blue, ->-] (3.8,-.2) -- (3.05,.3);
\draw [blue, ultra thick, snake it] (3.05,0.3) -- (2.45,0.7);
\draw [blue, ->-] (2.45,0.7) -- (2.,1.);
\draw [red, thick, densely dashed] (2., 2.4) ..controls (3., 2.4) and (2.9, 1.15) .. (2.45, 0.7);
\draw [purple, thick, densely dashed] (2., 1.6) ..controls (1., 1.6) and (.4, .8) .. (.95, 0.3);
\draw [orange, thick, densely dashed] (1.55, .7) ..controls (2., .1) and (2.6, -.3) .. (3.05, 0.3);
\node at (2.25, 2.9) {\scriptsize $x_\chi$};
\node at (2.25, 1.9) {\scriptsize $M_1$};
\node at (2.25, 1.3) {\scriptsize $r_\rho$};
\node at (.4, 0.2) {\scriptsize $y_\upsilon$};
\node at (1.1, 0.7) {\scriptsize $M_2$};
\node at (1.7, 1.) {\scriptsize $s_\sigma$};
\node at (2.15, .7) {\scriptsize $t_\tau$};
\node at (2.9, .65) {\scriptsize $M_3$};
\node at (3.5, .2) {\scriptsize $z_\zeta$};
\node [orange] at (2.4, .35) {$\times$};
\node [red] at (2.4, 1.6) {$\times$};
\node [purple] at (1.3, 1.1) {$\times$};
\end{tikzpicture}\qquad = \qquad d_{\mathcal{A}_1}d_{\mathcal{A}_2}d_{\mathcal{A}_3}\ [\Delta_{M_1M_2M_3}]^{x_\chi y_\upsilon z_\zeta}_{r_\rho s_\sigma t _\tau}\begin{tikzpicture}[baseline={([yshift=-.8ex]current bounding box.center)},scale=1.]
\draw [->-] (2.,3.2) -- (2.,1.);
\draw [->-] (.2,-.2) -- (2.,1.);
\draw [->-] (3.8,-.2) -- (2.,1.);
\node at (2.25, 2.) {\scriptsize $x_\chi$};
\node at (.8, .5) {\scriptsize $y_\upsilon$};
\node at (3.2, .5) {\scriptsize $z_\zeta$};
\end{tikzpicture}}

\newcommand{\DW}{
\begin{tikzpicture}[baseline={([yshift=-.8ex]current bounding box.center)},scale=.5]
\draw [->-, blue, ultra thick] (0.,0.)--(0.,4.);

\node [centered, rotate=0.] at (0.,-0.3) {$x_{g_ig_j}$};
\node [centered, rotate=0.] at (-2.,2.) {\Large $g_i$};
\node [centered, rotate=0.] at (2.,2.) {\Large $g_j$};
\end{tikzpicture}
\qquad=\qquad
\begin{tikzpicture}[baseline={([yshift=-.8ex]current bounding box.center)},scale=.5]
\draw [->-, blue, ultra thick] (0.,4.)--(0.,0.);

\node [centered, rotate=0.] at (0.,-0.3) {$x_{g_jg_i}$};
\node [centered, rotate=0.] at (-2.,2.) {\Large $g_i$};
\node [centered, rotate=0.] at (2.,2.) {\Large $g_j$};
\end{tikzpicture}
}

\newcommand{\BimoduleD}{\begin{tikzpicture}[baseline={([yshift=-.8ex]current bounding box.center)},scale=1.]
\draw [blue, thick, ->-] (.3, 0.5) -- (1, 1);
\draw [blue, thick, ->-] (1.7, 0.5) -- (1, 1);
\draw [blue, thick, ->-] (1., 2.) -- (1, 1);
\fill [blue] (1., 1.) circle (.15);
\node at (1.3, 1.7) {\scriptsize $M_1$};
\node at (.2, .8) {\scriptsize $M_2$};
\node at (1.75, .8) {\scriptsize $M_3$};
\end{tikzpicture}\qquad := \sum_{x_\chi\in L_{M_1}}\sum_{y_\upsilon\in L_{M_2}}\sum_{z_\zeta\in L_{M_3}}\mathcal{V}_{M_1M_2M_3}^{x_\chi y_\upsilon z_\zeta}\quad \begin{tikzpicture}[baseline={([yshift=-.8ex]current bounding box.center)},scale=.8]
\draw [->-] (.3, 0.5) -- (1, 1);
\draw [->-] (1.7, 0.5) -- (1, 1);
\draw [->-] (1., 2.) -- (1, 1);
\node at (1.3, 1.6) {\scriptsize $x_\chi$};
\node at (.3, .8) {\scriptsize $y_\upsilon$};
\node at (1.6, .8) {\scriptsize $z_\zeta$};
\end{tikzpicture}}

\newcommand{\BimoduleE}{G^{M_1M_2M}_{M_3M_4M'} = \frac{1}{d_{\A_1}^{N_{\A_1}-\frac{1}{2}}d_{\A_2}^{N_{\A_2}-\frac{1}{2}}\sqrt{d_{M_1}d_{M_2}d_{M_3}d_{M_4}d_{M}d_{M'}}}\ \ \mathcal{T}\ \ \begin{tikzpicture}[baseline={([yshift=-.8ex]current bounding box.center)},scale=.8]
\draw [blue, thick, -<-] (0, 1) -- (2, 1);
\draw [blue, thick, ->-] (1, 2) arc (90:180:1);
\draw [blue, thick, ->-] (1, 0) arc (270:180:1);
\draw [blue, thick, -<-] (2, 1) arc (0:90:1);
\draw [blue, thick, -<-] (2, 1) arc (0:-90:1);
\draw [blue, thick, -<-] (2, 1) arc (0:-90:1);
\draw [blue, thick] (1, 2) arc [start angle = 180, end angle = 90, x radius = .7, y radius = .5];
\draw [blue, thick] (1., 0) arc [start angle = 180, end angle = 270, x radius = .7, y radius = .5];
\draw [blue, thick, ->-] (1.7, 2.5) arc [start angle = 90, end angle = -90, x radius = 1.4, y radius = 1.5];
\node at (-.1, 1.7) {\scriptsize $M_1$};
\node at (-.1, .3) {\scriptsize $M_2$};
\node at (2.1, .3) {\scriptsize $M_3$};
\node at (2.1, 1.7) {\scriptsize $M_4$};
\node at (1., 1.3) {\scriptsize $M$};
\node at (3.4, 1.) {\scriptsize $M'$};
\node [red] at (1., 1.6) {$\times$};
\node [red] at (1., .6) {$\times$};
\node [red] at (2.6, 1.) {$\times$};
\fill [blue] (0, 1) circle (0.15);
\fill [blue] (2, 1) circle (0.15);
\fill [blue] (1, 2) circle (0.15);
\fill [blue] (1, 0) circle (0.15);
\end{tikzpicture}}

\newcommand{\BimoduleF}{\sum_{\alpha = 1}^{n^M_x}\sum_{\beta = 1}^{n^N_x}\begin{tikzpicture}[baseline={([yshift=-.8ex]current bounding box.center)},scale=1.]
\draw [thick, blue, ->- ] (0, 0) arc(-90:100:1);
\draw [thick, blue, ->- ] (0, 0) arc(270:80:1);
\fill [blue] (0, 0) circle (0.15);
\fill [blue] (0, 2) circle (0.15);
\draw [blue, ->-] (0, 0) -- (0, 1.);
\draw [blue, ->-] (0, 1.) -- (0, 2);
\node [] at (-1.4, 1.) {$M_1$};
\node [] at (1.4, 1.) {$M_2$};
\node [] at (.3, .5) {$x^M_\alpha$};
\node [] at (.3, 1.5) {$y^N_\beta$};
\node [red] at (.5, 1.) {$\times$};
\node [red] at (-.5, 1.) {$\times$};
\end{tikzpicture}\quad = \quad d_\mathcal{A}n^M_x\delta_{xy}\delta_{M_1M_2M_3}\delta_{MN}\sqrt{d_{M_1}d_{M_2}d_x}}

\newcommand{\BimoduleX}{
\begin{tikzpicture}[baseline={([yshift=-.8ex]current bounding box.center)},scale=.8]
\draw [->-] (1, 0) -- (1, .6);
\draw [->-] (1, .6) -- (1, 1.3);
\draw [->-] (1, 1.3) -- (1, 2.);
\draw [-<-] (1, .6) -- (.3, .6);
\draw [-<-] (1, 1.3) -- (1.7, 1.3);
\node at (0.1, .6) {\scriptsize $a_\alpha$};
\node at (1.9, 1.3) {\scriptsize $b_\beta$};
\node at (1.25, .2) {\scriptsize $x_\chi$};
\node at (1.15,.9) {\scriptsize $y$};
\node at (1.25,1.8) {\scriptsize $z_\zeta$};
\end{tikzpicture}
}

\newcommand{\TransB}{\mathcal{D}_v\quad
\begin{tikzpicture}[baseline={([yshift=-.8ex]current bounding box.center)},scale=.8]
\draw [->-] (0, 0) -- (1, .6);
\draw [->-] (2, 0) -- (1, .6);
\draw [-<-] (1, .6) -- (1, 1.8);
\node [] at (1.5, 1.4) {$M_1$};
\node [] at (0., .5) {$M_2$};
\node [] at (1.9, .5) {$M_3$};
\end{tikzpicture}\ \ := \ \ 
\begin{tikzpicture}[baseline={([yshift=-.8ex]current bounding box.center)},scale=.8]
\draw [->-, thick, blue] (0, 0) -- (1, .6);
\draw [->-, thick, blue] (2, 0) -- (1, .6);
\draw [-<-, thick, blue] (1, .6) -- (1, 1.8);
\fill [blue] (1, .6) circle (.15);
\node [] at (1.5, 1.4) {\scriptsize $M_1$};
\node [] at (0., .5) {\scriptsize $M_2$};
\node [] at (1.9, .5) {\scriptsize $M_3$};
\end{tikzpicture} \ \ = \ \ \sum_{x_\chi\in L_{M_1}} \sum_{y_\upsilon\in L_{M_2}} \sum_{z_\zeta\in L_{M_3}} \mathcal{V}_{M_1M_2M_3}^{x_\chi y_\upsilon z_\zeta}
\begin{tikzpicture}[baseline={([yshift=-.8ex]current bounding box.center)},scale=.8]
\draw [->-] (0, 0) -- (1, .6);
\draw [->-] (2, 0) -- (1, .6);
\draw [-<-] (1, .6) -- (1, 1.8);
\node [] at (1.4, 1.4) {\scriptsize $x_\chi$};
\node [] at (0.3, .5) {\scriptsize $y_\upsilon$};
\node [] at (1.7, .5) {\scriptsize $z_\zeta$};
\end{tikzpicture}
}

\newcommand{\TransE}{\varphi_\A\quad
\begin{tikzpicture}[baseline={([yshift=-.8ex]current bounding box.center)},scale=1.]
\draw (0, 0) -- (0, 2);
\node at (.2, 1.) {\scriptsize $a$};
\end{tikzpicture}\qquad := \qquad
\begin{tikzpicture}[baseline={([yshift=-.8ex]current bounding box.center)},scale=1.]
\draw [->-](1, 0) -- (1, 2.);
\node [] at (1.3, 1.) {\scriptsize $M_a$};
\end{tikzpicture}
}

\newcommand{\TransF}{E_e\quad
\begin{tikzpicture}[baseline={([yshift=-.8ex]current bounding box.center)},scale=1.]
\draw [blue, ->-] (0, 0) -- (.7, 0);
\draw [blue, ->-] (.7, 0) -- (1.4, 0);
\draw [blue, ->-] (0, 0) -- (-.3, -.5);
\draw [blue, ->-] (0, 0) -- (-.3, .5);
\draw [blue, ->-] (1.4, 0) -- (1.7, -.5);
\draw [blue, ->-] (1.4, 0) -- (1.7, .5);
\node [] at (-.3, 0.) {\scriptsize $v_1$};
\node [] at (1.7, 0.) {\scriptsize $v_2$};
\node [] at (.3, 0.3) {\scriptsize $x^{M_1}_\alpha$};
\node [] at (1.1, 0.3) {\scriptsize $x^{M_2}_\beta$};
\end{tikzpicture}\qquad := \qquad \sum_{i}(A^{x, M_1}_{\alpha, i})^\ast A^{x, M_2}_{\beta, i}\quad
\begin{tikzpicture}[baseline={([yshift=-.8ex]current bounding box.center)},scale=1.]
\draw [->-] (0, 0) -- (1., 0);
\draw [->-] (0, 0) -- (-.3, .5);
\draw [->-] (0, 0) -- (-.3, -.5);
\draw [->-] (1, 0) -- (1.3, .5);
\draw [->-] (1, 0) -- (1.3, -.5);
\node [] at (.5, 0.2) {\scriptsize $x$};
\end{tikzpicture}
}

\newcommand{\BimoduleG}{
\begin{tikzpicture}[baseline={([yshift=-.8ex]current bounding box.center)},scale=1.]
\draw [blue, ->-] (0, 0) -- (0, 0.5);
\draw [blue, snake it, ultra thick] (0, 0.5) -- (0, 1.);
\draw [blue, ->-] (0, 1) -- (0, 1.5);
\draw [red, -<-] (0, .5) -- (-.5, .5);
\draw [red, -<-] (0, 1.) -- (.5, 1.);

\node [centered, rotate=0.] at (.2,.15) {\scriptsize $x_\chi$};
\node [centered, rotate=0.] at (.2,.75) {\scriptsize $M$};
\node [centered, rotate=0.] at (.2,1.35) {\scriptsize $z_\zeta$};
\node [centered, rotate=0.] at (-.7,.5) {\scriptsize $a_\alpha$};
\node [centered, rotate=0.] at (.7,1.) {\scriptsize $b_\beta$};
\end{tikzpicture}
}

\newcommand{\BimoduleH}{
\begin{tikzpicture}[baseline={([yshift=-.8ex]current bounding box.center)},scale=.6]
\draw [->-] (3.2,5.6) -- (3.2,6.5);
\draw [->-] (3.2,6.5) -- (3.2,7.2);
\draw [->-] (3.2,7.2) -- (4.6,8.);
\draw [->-] (4.6,8.) -- (6.,7.2);
\draw [->-] (6.,7.2) -- (6.,5.6);
\draw [->-] (6.,5.6) -- (4.6,4.8);
\draw [->-] (4.6,4.8) -- (3.2,5.6);
\draw [decorate,decoration={snake,segment length=.12cm, amplitude=.04cm}, ->] (3.2,6.4) -- (4.2,6.4);
\node [centered, rotate=0.] at (2.7,6.) {\scriptsize $M_6$};
\node [centered, rotate=0.] at (2.7,6.9) {\scriptsize $M_0$};
\node [centered, rotate=0.] at (3.6,7.9) {\scriptsize $M_1$};
\node [centered, rotate=0.] at (5.6,7.9) {\scriptsize $M_2$};
\node [centered, rotate=0.] at (6.4,6.4) {\scriptsize $M_3$};
\node [centered, rotate=0.] at (5.6,4.9) {\scriptsize $M_4$};
\node [centered, rotate=0.] at (3.6,4.9) {\scriptsize $M_5$};
\node [centered, rotate=0.] at (4.5,6.4) {\scriptsize $M$};
\draw [->-] (2.6,7.6) -- (3.2,7.2);
\draw [->-] (4.6,8.8) -- (4.6,8.);
\draw [->-] (6.6,7.6) -- (6.,7.2);
\draw [->-] (6.6,5.2) -- (6.,5.6);
\draw [->-] (4.6,4.) -- (4.6,4.8);
\draw [->-] (2.6,5.2) -- (3.2,5.6);
\node [centered, rotate=0.] at (2.3,7.8) {\scriptsize $N_1$};
\node [centered, rotate=0.] at (4.6,9.) {\scriptsize $N_2$};
\node [centered, rotate=0.] at (6.9,7.8) {\scriptsize $N_3$};
\node [centered, rotate=0.] at (6.9,5.) {\scriptsize $N_4$};
\node [centered, rotate=0.] at (4.6,3.7) {\scriptsize $N_5$};
\node [centered, rotate=0.] at (2.3,5.) {\scriptsize $N_6$};
\end{tikzpicture}\qquad\Longrightarrow\qquad\frac{1}{d_\mathcal{A}^{9}}\sum_{x_i, y_i\in L_{M_i}}\sum_{e_i\in L_{N_i}}\sum_{p_\alpha,q_\beta\in L_M}\nonumber\\
\T\qquad
\begin{tikzpicture}[baseline={([yshift=-.8ex]current bounding box.center)},scale=1.5]
\draw [blue, ->-] (3.2,5.6) -- (3.2,5.8);
\draw [blue, ultra thick, snake it] (3.2,5.8) -- (3.2,6.1);
\draw [blue, ->-] (3.2,6.1) -- (3.2,6.4);
\node [centered, rotate=0.] at (3.05,5.7) {\scriptsize $x_6$};
\node [centered, rotate=0.] at (3.02,5.95) {\scriptsize $M_6$};
\node [centered, rotate=0.] at (3.05,6.25) {\scriptsize $y_6$};
\draw [red, thick, densely dashed] (3.2, 5.8) -- (2.65, 5.8);
\draw [red, thick, densely dashed] (3.2, 6.1) -- (4.1, 6.1);
\fill [red] (4.05, 6.1) circle (0.05);
\draw [blue, ->-] (3.2,6.4) -- (3.2,6.7);
\draw [blue, ultra thick, snake it] (3.2,6.7) -- (3.2,7.);
\draw [blue, ->-] (3.2,7.) -- (3.2,7.2);
\node [centered, rotate=0.] at (3.05,6.55) {\scriptsize $x_0$};
\node [centered, rotate=0.] at (3.02,6.85) {\scriptsize $M_0$};
\node [centered, rotate=0.] at (3.05,7.1) {\scriptsize $y_0$};
\draw [red, thick, densely dashed] (3.2, 6.7) -- (2.65, 6.7);
\draw [red, thick, densely dashed] (3.2, 7.) -- (3.8, 7.);
\fill [red] (3.8, 7.) circle (0.05);
\draw [blue, ->-] (3.2,7.2) -- (3.55,7.4);
\draw [blue, ultra thick, snake it] (3.55,7.4) -- (4.25,7.8);
\draw [blue, ->-] (4.25,7.8) -- (4.6,8.);
\node [centered, rotate=0.] at (3.45,7.2) {\scriptsize $x_1$};
\node [centered, rotate=0.] at (3.8,7.72) {\scriptsize $M_1$};
\node [centered, rotate=0.] at (4.35,8.) {\scriptsize $y_1$};
\draw [red, thick, densely dashed] (3.55, 7.4) -- (3.25, 7.9);
\draw [red, thick, densely dashed] (4.25, 7.8) -- (4.45, 7.45);
\fill [red] (4.47, 7.4) circle (0.05);
\draw [blue, ->-] (4.6,8.) -- (4.95,7.8);
\draw [blue, ultra thick, snake it] (4.95,7.8) -- (5.65,7.4);
\draw [blue, ->-] (5.65,7.4) -- (6.,7.2);
\node [centered, rotate=0.] at (5.9,7.4) {\scriptsize $y_2$};
\node [centered, rotate=0.] at (5.4,7.72) {\scriptsize $M_2$};
\node [centered, rotate=0.] at (4.77,7.8) {\scriptsize $x_2$};
\draw [red, thick, densely dashed] (4.95, 7.8) -- (5.25, 8.3);
\draw [red, thick, densely dashed] (5.65, 7.4) -- (5.38, 7.);
\fill [red] (5.39, 7.02) circle (0.05);
\draw [blue, ->-] (6.,7.2) -- (6.,6.7);
\draw [blue, ultra thick, snake it] (6.,6.7) -- (6.,6.1);
\draw [blue, ->-] (6.,6.1) -- (6.,5.6);
\draw [red, thick, densely dashed] (6., 6.7) -- (6.5, 6.7);
\draw [red, thick, densely dashed] (5.6, 6.1) -- (6., 6.1);
\fill [red] (5.56, 6.1) circle (0.05);
\node [centered, rotate=0.] at (5.85,6.9) {\scriptsize $x_3$};
\node [centered, rotate=0.] at (6.2,6.4) {\scriptsize $M_3$};
\node [centered, rotate=0.] at (6.15,5.85) {\scriptsize $y_3$};
\draw [blue, ->-] (6.,5.6) -- (5.65,5.4);
\draw [blue, ultra thick, snake it] (5.65,5.4) -- (4.95,5.);
\draw [blue, ->-] (4.95,5.) -- (4.6,4.8);
\node [centered, rotate=0.] at (5.95,5.35) {\scriptsize $x_4$};
\node [centered, rotate=0.] at (5.4,5.05) {\scriptsize $M_4$};
\node [centered, rotate=0.] at (4.9,4.78) {\scriptsize $y_4$};
\draw [red, thick, densely dashed] (5.65, 5.4) -- (5.95, 4.9);
\draw [red, thick, densely dashed] (4.95, 5.) -- (4.55, 5.7);
\fill [red] (4.58, 5.64) circle (0.05);
\draw [blue, ->-] (4.6,4.8) -- (4.25,5.);
\draw [blue, ultra thick, snake it] (4.25,5.) -- (3.55,5.4);
\draw [blue, ->-] (3.55,5.4) -- (3.2,5.6);
\node [centered, rotate=0.] at (3.48,5.6) {\scriptsize $y_5$};
\node [centered, rotate=0.] at (3.75,5.05) {\scriptsize $M_5$};
\node [centered, rotate=0.] at (4.5,5.03) {\scriptsize $x_5$};
\draw [red, thick, densely dashed] (4.25, 5.) -- (3.95, 4.5);
\draw [red, thick, densely dashed] (3.55, 5.4) -- (4.25, 5.8);
\fill [red] (4.25, 5.8) circle (0.05);
\draw [blue, ->-] (3.2,6.4) -- (3.6,6.4);
\draw [blue, ultra thick, snake it] (3.6,6.4) -- (4.,6.4);
\draw [blue, ->] (4.,6.4) -- (4.5,6.4);
\fill [blue] (3.2,6.4) circle (0.07);
\draw [red, thick, densely dashed] (3.6, 6.4) arc (180:0:1.);
\draw [red, thick, densely dashed] (4., 6.4) arc (-180:0:.8);
\node [centered, rotate=0.] at (3.4,6.5) {\scriptsize $p_\alpha$};
\node [centered, rotate=0.] at (3.8,6.53) {\scriptsize $M$};
\node [centered, rotate=0.] at (4.6,6.4) {\scriptsize $q_\beta$};
\node [centered, rotate=0.] at (3.4,6.8) {\color{red} $\times$};
\node [centered, rotate=0.] at (3.8,7.3) {\color{red} $\times$};
\node [centered, rotate=0.] at (4.9,7.5) {\color{red} $\times$};
\node [centered, rotate=0.] at (5.75,6.7) {\color{red} $\times$};
\node [centered, rotate=0.] at (5.4,5.6) {\color{red} $\times$};
\node [centered, rotate=0.] at (4.3,5.4) {\color{red} $\times$};
\node [centered, rotate=0.] at (3.6,5.8) {\color{red} $\times$};
\node [centered, rotate=0.] at (3.6,6.2) {\color{red} $\times$};

\draw [blue, ultra thick, snake it] (2.6,7.6) -- (2.9,7.4);
\draw [blue, ->-] (2.9,7.4) -- (3.2,7.2);
\fill [blue] (3.2,7.2) circle (0.07);
\draw [red, thick, densely dashed] (2.9,7.4) -- (2.68,7.07);
\node [centered, rotate=0.] at (2.5,7.7) {\scriptsize $N_1$};
\node [centered, rotate=0.] at (3.1,7.43) {\scriptsize $e_1$};
\draw [blue, ultra thick, snake it] (4.6,8.6) -- (4.6,8.3);
\draw [blue, ->-] (4.6,8.3) -- (4.6,8.);
\fill [blue] (4.6,8.) circle (0.07);
\draw [red, thick, densely dashed] (4.6,8.3) -- (4.2,8.3);
\node [centered, rotate=0.] at (4.6,8.7) {\scriptsize $N_2$};
\node [centered, rotate=0.] at (4.72,8.1) {\scriptsize $e_2$};
\draw [blue, ultra thick, snake it] (6.6,7.6) -- (6.3,7.4);
\draw [blue, ->-] (6.3,7.4) -- (6.,7.2);
\fill [blue] (6.,7.2) circle (0.07);
\draw [red, thick, densely dashed] (6.3,7.4) -- (6.08,7.73);
\node [centered, rotate=0.] at (6.7,7.7) {\scriptsize $N_3$};
\node [centered, rotate=0.] at (6.2,7.19) {\scriptsize $e_3$};
\draw [blue, ultra thick, snake it] (6.6,5.2) -- (6.3,5.4);
\draw [blue, ->-] (6.3,5.4) -- (6.,5.6);
\fill [blue] (6.,5.6) circle (0.07);
\draw [red, thick, densely dashed] (6.3,5.4) -- (6.7,5.73);
\node [centered, rotate=0.] at (6.8,5.1) {\scriptsize $N_4$};
\node [centered, rotate=0.] at (6.3,5.6) {\scriptsize $e_4$};
\draw [blue, ultra thick, snake it] (4.6,4.2) -- (4.6,4.5);
\draw [blue, ->-] (4.6,4.5) -- (4.6,4.8);
\fill [blue] (4.6,4.8) circle (0.07);
\draw [red, thick, densely dashed] (4.6,4.5) -- (5.,4.5);
\node [centered, rotate=0.] at (4.65,4.1) {\scriptsize $N_5$};
\node [centered, rotate=0.] at (4.38,4.67) {\scriptsize $e_5$};
\draw [blue, ultra thick, snake it] (2.6,5.2) -- (2.9,5.4);
\draw [blue, ->-] (2.9,5.4) -- (3.2,5.6);
\fill [blue] (3.2,5.6) circle (0.07);
\draw [red, thick, densely dashed] (2.9,5.4) -- (3.12,5.07);
\node [centered, rotate=0.] at (2.5,5.1) {\scriptsize $N_6$};
\node [centered, rotate=0.] at (3.12,5.38) {\scriptsize $e_6$};
\end{tikzpicture}
}

\newcommand{\Edge}[1]{
\begin{tikzpicture}[baseline={([yshift=-.8ex]current bounding box.center)},scale=1.]
\draw [] (0, 0) -- (0, 1.5);
\node [right, rotate=0.] at (0.,.75) {\scriptsize ${#1}$};
\end{tikzpicture}
}

\newcommand{\Tail}[1]{
\begin{tikzpicture}[baseline={([yshift=-.8ex]current bounding box.center)},scale=1.]
\draw [decorate,decoration={snake,segment length=.12cm, amplitude=.04cm}] (0, 0) -- (0, 1.5);
\node [right, rotate=0.] at (0.,.75) {\scriptsize ${#1}$};
\end{tikzpicture}
}

\newcommand{\HalfBraidingA}{
\begin{tikzpicture}[baseline={([yshift=-.8ex]current bounding box.center)},scale=1.]
\draw [->-] (0, 0) -- (0, .7);
\draw [->-] (.5, 0) -- (.5, .7);
\draw [] (-.2, .7) -- (.7, .7) -- (.7, 1.3) -- (-.2, 1.3) -- (-.2, .7);
\draw [->-] (0, 1.3) -- (0, 2.);
\draw [->-] (.5, 1.3) -- (.5, 2.);
\node [centered, rotate=0.] at (-.3, .2) {\scriptsize $X_J$};
\node [centered, rotate=0.] at (.7, .2) {\scriptsize $y$};
\node [centered, rotate=0.] at (-.2, 1.8) {\scriptsize $y$};
\node [centered, rotate=0.] at (.8, 1.8) {\scriptsize $X_J$};
\node [centered, rotate=0.] at (.25, 1.) {\scriptsize $c_{X_J, y}$};
\end{tikzpicture}\qquad =\qquad
\begin{tikzpicture}[baseline={([yshift=-.8ex]current bounding box.center)},scale=1.]
\draw [->-] (1, 0) -- (1, 1);
\draw [->-] (1, 1) -- (1, 2);
\draw [->-] (.5, 0) arc [start angle = 180, end angle = 100, x radius = .5, y radius = 1.];
\draw [-<-] (1.5, 2) arc [start angle = 0, end angle = -80, x radius = .5, y radius = 1.];
\node [left, rotate=0.] at (0.5, .25) {\scriptsize $X_J$};
\node [right, rotate=0.] at (1.5, 1.75) {\scriptsize $X_J$};
\node [centered, rotate=0.] at (1.2, .2) {\scriptsize $y$};
\node [centered, rotate=0.] at (.8, 1.8) {\scriptsize $y$};
\end{tikzpicture}
}

\newcommand{\HalfBraidingB}{
\begin{tikzpicture}[baseline={([yshift=-.8ex]current bounding box.center)},scale=.8]
\draw [->-] (1, 0) -- (1, 1);
\draw [->-] (1, 1) -- (1, 2);
\draw [->-] (.5, 0) arc [start angle = 180, end angle = 100, x radius = .5, y radius = 1.];
\draw [-<-] (1.5, 2) arc [start angle = 0, end angle = -80, x radius = .5, y radius = 1.];
\node [left, rotate=0.] at (0.5, .25) {\scriptsize $X_J$};
\node [right, rotate=0.] at (1.5, 1.75) {\scriptsize $X_J$};
\node [centered, rotate=0.] at (1.2, .2) {\scriptsize $y$};
\node [centered, rotate=0.] at (.8, 1.8) {\scriptsize $y$};
\end{tikzpicture}\ = \ \bigoplus_{p,q\in L_J}\ \bigoplus_{k\in L_\Fus}\ \sqrt{\frac{d_k}{d_y}}\ z_{pqy}^{J;k}\quad
\begin{tikzpicture}[baseline={([yshift=-.8ex]current bounding box.center)},scale=.8]
\draw [->-] (1, 0) -- (1, .8);
\draw [->-] (1, 1.2) -- (1, 2);
\draw [->-] (1, .8) -- (1, 1.2);
\draw [->-] (.5, 0) arc [start angle = 180, end angle = 90, x radius = .5, y radius = .7];
\draw [-<-] (1.5, 2) arc [start angle = 0, end angle = -90, x radius = .5, y radius = .7];
\node [left, rotate=0.] at (0.5, .25) {\scriptsize $p$};
\node [right, rotate=0.] at (1.5, 1.75) {\scriptsize $q$};
\node [centered, rotate=0.] at (1.2, .2) {\scriptsize $y$};
\node [centered, rotate=0.] at (.8, 1.8) {\scriptsize $y$};
\node [centered, rotate=0.] at (1.2, 1.) {\scriptsize $k$};
\end{tikzpicture}
}

\newcommand{\HalfBraidingC}{
\begin{tikzpicture}[baseline={([yshift=-.8ex]current bounding box.center)},scale=1]
\draw [->-] (1.15, .4) arc (-60:30:.5);
\draw [->-] (.6, .1) -- (.9, .3);
\draw [->-] (1.2, 1.3) -- (1., 1.6);

\draw [->-] (1, 0) -- (1, 1);
\draw [->-] (.2, 1.5) -- (1, 1);
\draw [->-] (1.8, 1.5) -- (1, 1);
\node [centered, rotate=0.] at (1.15, 0.) {\scriptsize $y$};
\node [centered, rotate=0.] at (.2, 1.3) {\scriptsize $u$};
\node [centered, rotate=0.] at (1.8, 1.3) {\scriptsize $v$};
\node [centered, rotate=0.] at (.4, 0.) {\scriptsize $p$};
\node [centered, rotate=0.] at (.8, 1.7) {\scriptsize $q$};
\end{tikzpicture}\qquad = \qquad 
\begin{tikzpicture}[baseline={([yshift=-.8ex]current bounding box.center)},scale=1]
\draw [->] (.5, .5) arc (-170:-240:1.);
\draw [->] (.5, .5) arc (-170:-190:1.);
\fill [white] (.55, 1.3) circle (.2);

\draw [->-] (1, 0) -- (1, 1);
\draw [->-] (.2, 1.5) -- (1, 1);
\draw [->-] (1.8, 1.5) -- (1, 1);
\node [centered, rotate=0.] at (1.15, 0.) {\scriptsize $y$};
\node [centered, rotate=0.] at (.2, 1.3) {\scriptsize $u$};
\node [centered, rotate=0.] at (1.8, 1.3) {\scriptsize $v$};
\node [centered, rotate=0.] at (.5, .3) {\scriptsize $p$};
\node [centered, rotate=0.] at (1.1, 1.6) {\scriptsize $q$};
\end{tikzpicture}
}

\newcommand{\HalfBraidingD}{
\begin{tikzpicture}[baseline={([yshift=-.8ex]current bounding box.center)},scale=.8]
\draw [->-] (1, 0) -- (1, 1);
\draw [->-] (1, 1) -- (1, 2);
\draw [->-] (.5, 0) arc [start angle = 180, end angle = 100, x radius = .5, y radius = 1.];
\draw [-<-] (1.5, 2) arc [start angle = 0, end angle = -80, x radius = .5, y radius = 1.];
\node [left, rotate=0.] at (0.5, .25) {\scriptsize $p$};
\node [right, rotate=0.] at (1.5, 1.75) {\scriptsize $q$};
\node [centered, rotate=0.] at (1.2, .2) {\scriptsize $j_E$};
\node [centered, rotate=0.] at (.8, 1.8) {\scriptsize $j_E$};
\end{tikzpicture} =
\sum_{k\in L_\Fus} \sqrt{\frac{d_k}{d_{j_E}}}\ {z_{pqj_E}^{J;k}} 
\begin{tikzpicture}[baseline={([yshift=-.8ex]current bounding box.center)},scale=.8]
\draw [->-] (1, 0) -- (1, .8);
\draw [->-] (1, 1.2) -- (1, 2);
\draw [->-] (1, .8) -- (1, 1.2);
\draw [->-] (.5, 0) arc [start angle = 180, end angle = 90, x radius = .5, y radius = .7];
\draw [-<-] (1.5, 2) arc [start angle = 0, end angle = -90, x radius = .5, y radius = .7];
\node [left, rotate=0.] at (0.5, .25) {\scriptsize $p$};
\node [right, rotate=0.] at (1.5, 1.75) {\scriptsize $q$};
\node [centered, rotate=0.] at (1.2, .2) {\scriptsize $j_E$};
\node [centered, rotate=0.] at (.8, 1.8) {\scriptsize $j_E$};
\node [centered, rotate=0.] at (1.2, 1.) {\scriptsize $k$};
\end{tikzpicture} =
W_E^{J;pq}\
\begin{tikzpicture}[baseline={([yshift=-.8ex]current bounding box.center)},scale=.8]
\draw [->-] (1, 0) -- (1, 2);
\node [centered, rotate=0.] at (1.3, 1.) {\scriptsize $j_E$};
\end{tikzpicture}\ .
}

\newcommand{\NonAbA}{
\begin{tikzpicture}[baseline={([yshift=-.8ex]current bounding box.center)},scale=.8]
\draw [->-] (3.2,5.6) -- (3.2,6.1);
\draw [->-] (3.2,6.1) -- (3.2,6.7);
\draw [->-] (3.2,6.7) -- (3.2,7.2);
\draw [->-] (3.2,7.2) -- (3.9,7.6);
\draw [->-] (2.5,7.6) -- (3.2,7.2);
\draw [->-] (2.5,5.2) -- (3.2,5.6);
\draw [->-] (3.2,5.6) -- (3.9,5.2);
\draw [decorate,decoration={snake,segment length=.12cm, amplitude=.04cm}, ->] (3.2,6.7) -- (4.,6.7);
\draw [decorate,decoration={snake,segment length=.12cm, amplitude=.04cm}, ->] (3.2,6.1) -- (2.4,6.1);
\node [centered, rotate=0.] at (2.4,5.) {\scriptsize $e_n$};
\node [centered, rotate=0.] at (4.1,5.) {\scriptsize $i_n$};
\node [centered, rotate=0.] at (4.1,7.8) {\scriptsize $i_2$};
\node [centered, rotate=0.] at (2.4,7.8) {\scriptsize $e_1$};
\node [centered, rotate=0.] at (3.,5.8) {\scriptsize $j$};
\node [centered, rotate=0.] at (3.,6.4) {\scriptsize $j$};
\node [centered, rotate=0.] at (3.4,7.) {\scriptsize $j$};
\node [centered, rotate=0.] at (4.4,6.7) {\scriptsize $1_{\mathcal{I}, 1}$};
\node [centered, rotate=0.] at (2.,6.1) {\scriptsize $1_{\mathcal{I},1}$};
\end{tikzpicture}
}

\newcommand{\NonAbB}{
\begin{tikzpicture}[baseline={([yshift=-.8ex]current bounding box.center)},scale=.8]
\draw [->-] (3.2,5.6) -- (3.2,6.1);
\draw [->-] (3.2,6.1) -- (3.2,6.7);
\draw [->-] (3.2,6.7) -- (3.2,7.2);
\draw [->-] (3.2,7.2) -- (3.9,7.6);
\draw [->-] (2.5,7.6) -- (3.2,7.2);
\draw [->-] (2.5,5.2) -- (3.2,5.6);
\draw [->-] (3.2,5.6) -- (3.9,5.2);
\draw [decorate,decoration={snake,segment length=.12cm, amplitude=.04cm}, ->] (3.2,6.7) -- (4.,6.7);
\draw [decorate,decoration={snake,segment length=.12cm, amplitude=.04cm}, ->] (3.2,6.1) -- (2.4,6.1);
\node [centered, rotate=0.] at (2.4,5.) {\scriptsize $e_n$};
\node [centered, rotate=0.] at (4.1,5.) {\scriptsize $i_n$};
\node [centered, rotate=0.] at (4.1,7.8) {\scriptsize $i_2$};
\node [centered, rotate=0.] at (2.4,7.8) {\scriptsize $e_1$};
\node [centered, rotate=0.] at (3.,5.8) {\scriptsize $j$};
\node [centered, rotate=0.] at (3.,6.4) {\scriptsize $k$};
\node [centered, rotate=0.] at (3.4,7.) {\scriptsize $j$};
\node [centered, rotate=0.] at (4.5,6.7) {\scriptsize $q_{J,\beta}$};
\node [centered, rotate=0.] at (1.9,6.) {\scriptsize $p^\ast_{J^\ast,\alpha}$};
\end{tikzpicture}
}

\begin{document}

\title{Symmetry-Enriched Topological Phases and Their Gauging: A String-Net Model Realization}

\date{\today}
\author[a]{Nianrui Fu}
\author[a]{Yu Zhao\footnote{Corresponding author}}
\author[a,b,c]{Yidun Wan\footnote{Corresponding author}}
\affiliation[a]{State Key Laboratory of Surface Physics, Department of Physics, Center for Field Theory and Particle Physics, and Institute for Nanoelectronic devices and Quantum computing, Fudan University, Shanghai 200433, China}
\affiliation[b]{Shanghai Research Center for Quantum Sciences, 99 Xiupu Road, Shanghai 201315, China}
\affiliation[c]{Hefei National Laboratory, Hefei 230088, China}
\emailAdd{nrfu25@m.fudan.edu.cn, yuzhao20@fudan.edu.cn, ydwan@fudan.edu.cn}

\abstract{We present a systematic framework for constructing exactly-solvable lattice models of symmetry-enriched topological (SET) phases based on an enlarged version of the string-net model. We also gauge the global symmetries of our SET models to obtain string-net models of pure topological phases. Without invoking externally imposed onsite symmetry actions, our approach promotes the string-net model of a pure topological order, specified by an input unitary fusion category $\Fus$, to an SET model, specified by a multifusion category together with a set of isomorphisms. Two complementary construction strategies are developed in the main text: (i) promotion via outer automorphisms of $\Fus$ and (ii) promotion via the Frobenius algebras of $\Fus$. The global symmetries derived via these two strategies are intrinsic to topological phases and are thus termed blood symmetries, as opposed to adopted symmetries, which can be arbitrarily imposed on topological phases. We propose the concept of symmetry-gauging family of topological phases, which are related by gauging their blood symmetries. With our approach, we construct the first explicit lattice realization of a nonabelian-symmetry-enriched topological phase -- the $S_3$ symmetry-enriched $\mathbb{Z}_2 \times \mathbb{Z}_2$ quantum-double phase. The approach further reveals the role of local excitations in SET phases and establishes their symmetry constraints.}

\maketitle

\flushbottom

\section{Introduction}\label{sec:intro}

When a pure topological phase\cite{Wen1990a,Kitaev2003a,Levin2004,Hu2013} is endowed with a global symmetry, it becomes a symmetry-enriched topological (SET) phase\cite{Mesaros2011,Hung2013,Huang2013,Chang2015,heinrich2016symmetry,cheng2017exactly,williamson2017a,lee2018}, which possesses richer physics and mathematics than the pure topological phase. While pure topological orders can be effectively described by lattice models -- such as the string-net\cite{Levin2004,Hung2012,Kitaev2012,schulz2013,Lan2014b,Lin2014,Hu2017,Hu2018,Bridgeman2020,zhao2022,Wang2022} and twisted quantum-double (TQD) models\cite{Hu2012a,Mesaros2011,Bravyi1998,Kitaev2003a,Bullivant2017,Cong2017,Wang2020,Wang2020b} -- constructing lattice models for SET phases presents additional challenge, although a few attempts have been made\cite{Mesaros2011,Chang2015,heinrich2016symmetry,cheng2017exactly,lee2018}. A natural question follows if this challenge is tackled: If we have such a lattice model of an SET phase, how can we possibly gauge the global symmetry and obtain a lattice model of a pure topological order? 

In this paper, we tackle this challenge and answer the question in the framework of the enlarged Hu-Geer-Wu (HGW) string-net model\footnote{Note that we shall not use the original version of the string-net (Levin-Wen) model due to Levin and Wen\cite{Levin2004} because of its limited Hilbert space. Instead, we shall adopt an enlarged version\cite{zhao2025c} of the extended string-net model due to Hu, Geer, and Wu\cite{Hu2018}, to be referred to as the enlarged Hu-Geer-Wu (HGW) model\cite{zhao2025c}. A subclass of this model was also obtained by Fourier-transforming the quantum-double model\cite{Wang2020b}. The HGW model was further extended to include gapped boundaries\cite{Wang2022} and gapped domain walls\cite{chen2025classification}.}\cite{Hu2018,zhao2025c}.

\begin{itemize}
\item We present a recipe to promote an enlarged HGW model of a pure topological phase to a model of an SET phase. Our approach relies solely on the input data of the enlarged HGW model of the underlying pure topological order of the SET phase, naturally extracting its intrinsically admitted global symmetries (to be termed \textit{blood symmetries}), thereby eliminating the need for any externally imposed explicit onsite symmetry actions on the model, as used in previous works. Our model makes computing the symmetry transformations of the anyon spectrum in SET phases convenient.

\item We further provide a procedure to gauge the SET phase's global symmetry, yielding a pure topological phase -- known as the \emph{parent phase} of the SET phase  --  in which the global symmetry of the SET phase manifests as a gauge invariance. 

\item We find \emph{symmetry-gauging families} of topological phases. In each such family, the topological phases are related by gauging their blood symmetries. 

\item Our approach clarifies the definition and reveals the role of local excitations in SET phases and establishes their symmetry constraints.

\end{itemize}

\textbf{Background} -- The string-net model with a given input unitary fusion category (UFC) $\Fus$ describes a pure topological phase, whose topological properties are captured by the Drinfeld center $\Cent(\Fus)$ of $\Fus$\cite{Muger2003,Levin2004,Kitaev2006,kong2022invitationtopologicalorderscategory}. Previous works modified the string-net model to describe SET phases by replacing the input UFC with a $G$-graded fusion category\cite{heinrich2016symmetry,cheng2017exactly} or a multifusion category\cite{Chang2015}. 

These approaches either begin with a parent pure topological order\cite{Chang2015,heinrich2016symmetry,lee2018} or rely on an \textit{ad hoc} given symmetry group $G$ except in the case of group extension\cite{cheng2017exactly}. These works either elevate $\Fus$ to be a multifusion category\cite{Chang2015} that encodes $G$ or extend $\Fus$ to be a $G$-crossed braided fusion category\cite{heinrich2016symmetry,cheng2017exactly}. The multifusion category and the $G$-crossed braided fusion category so obtained are equivalent in serving as the input data of the string-net model that describes the desired SET phase.

It is noteworthy that a generic multifusion category does not contain any symmetry information and cannot serve as the input data of an SET model. Any multifusion category considered in these approaches is in fact a multifusion category that is equipped with an extra structure encapsulating either a symmetry $G$ due to the given parent topological phase or an \textit{ad hoc} symmetry $G$.

These approaches require identifying complicated multifusion categories or resolving intricate categorical extension problems, making computation and physical interpretation challenging. More importantly, these approaches have a hard time in obtaining the model's spectrum and how the spectrum is transformed under global symmetry transformations, except in certain simple cases. 

Our approaches to modeling SET phases, although also define an SET model as a string-net model with an input multifusion category, overcomes the hurdles in previous works. 

\textbf{Our approach} -- We notice that a pure topological phase $\Cat$ described by the string-net model with input UFC $\Fus$ allows global symmetries that can be derived from $\Fus$ and those unrelated to $\Fus$, i.e., those symmetries that can be imposed in an \textit{ad hoc} way. Note that we are trying to construct string-net models of SET phases without spontaneous symmetry breaking (SSB). Such an SET phase usually contain multiple sectors related by global symmetry transformations. To encompass the sectors of an SET phase, our model's input data can no longer be merely a UFC but as to be seen would have to be a richer structure: a multifusion category equipped with a group $G$ of isomorphisms, whose $G$-actions characterize the symmetry transformations among the sectors.

Our SET construction approach has an advantage in explaining the origin of the symmetry of an SET phase. To be specific, we classify global symmetries of SET phases based on their actions on the input UFC $\Fus$ of the string-net model describing the underlying topological phase: The symmetry of an SET phase is (1) a \textbf{blood symmetry} if it is due to the intrinsic endo-equivalences of $\Fus$, e.g., the EM-exchange symmetry or the charge-conjugation symmetry; it is (2) an \textbf{adopted symmetry} if it stems from an externally imposed symmetry that acts trivially on $\Fus$, such as the global symmetries in SPT phases or the symmetry fractionalizations of the $\Z_2$ toric code anyons. This clear distinction eliminates \textit{ad hoc} extension constructions or the need of a given parent topological phase, enhances computational tractability, simplifies the classification of symmetry enrichments, and clarifies the physical interpretation of SET phases.

For the case of blood symmetries, we propose two strategies of promoting an $\Fus$ to be multifusion categories equipped with $G$-actions:
\begin{itemize}
    \item Strategy I\footnote{When $\Fus=\Vec(G)$ for a finite group $G$, this strategy aligns with the group extension approach mentioned in Ref.\cite{cheng2017exactly}.}: A UFC $\Fus$ may have outer automorphism\footnote{Inner automorphisms can only yield gauge invariances of the corresponding topological phase.}, which always form a group $G$ that naturally promotes $\Fus$ to be a multifusion category equipped with one or more $G$-actions, each defining an SET phase. For example, the the charge-conjugation symmetry enriched $\Z_3$ quantum-double phase arises from the $\Z_2$ outer automorphism group of the underlying $\Z_3$ UFC.
    \item Strategy II: A UFC $\Fus$ contains Frobenius algebra objects. It's known that such Frobenius algebras can be used to generate global symmetry transformations on the topological phase described by the string-net model with input $\Fus$ \cite{zhao2024a}. This result motivates us to construct a multifusion category in terms of the bimodules of the Frobenius algebra objects in $\Fus$. Such a multifusion category automatically encodes a global symmetry $G$\footnote{What's encoded in the multifusion category may be a gauge invariance instead of a global symmetry. Later, we shall give a criteria judging whether a global symmetry or gauge invariance appears.}. Nevertheless, at our disposal, there may be multiple compatible $G$-actions, yielding respectively different SET phases. An archetypal example is the EM-exchange-symmetry-enriched \(\Z_2\) toric code phase, derived from the isomorphism between \(\Vec(\Z_2)\) and \(\Rep(\Z_2)\).
\end{itemize}

For the case of adopted symmetries, we have a third strategy that can handle any adopted global symmetry $G$ one wishes to impose on a given topological phase $\Cat$, such that the $G$-action transforms the anyon spectrum of $\Cat$ trivially:\footnote{The $G$-action may generate nontrivial symmetry fractionalization of anyons, which we do not discuss in this paper.}
\begin{itemize}
    \item Strategy III: We can promote $\Fus$ to be a multifusion category by replicating $\Fus$ $|G|\times |G|$ times. There may be multiple permissible $G$-actions, each producing an SET phase. An extreme example is provided by symmetry-protected topological (SPT) orders, where the underlying topological order is trivial and admits no nontrivial symmetry actions. In such cases, the global symmetry must be imposed by hand.
\end{itemize}

We shall detail the first two strategies separately in subsequent sections and provide concrete examples respectively. Nevertheless, strategy III in fact coincides with the SET constructing method in Ref.\cite{cheng2017exactly}, so we put it in Appendix \ref{sec:symmfrac}. Note that our three strategies can be combined; however, we shall not discuss this in this paper.

Rather than working with the original Levin-Wen string-net model -- which, because of its restricted Hilbert space, cannot capture the full anyon spectrum -- we adopt the enlarged HGW string-net model\cite{zhao2025c}. This richer model not only realizes the complete anyon spectrum (recovering the Fourier-transformed quantum-double subclass\cite{Wang2020b}) but also makes transparent how the entire state space transforms under a global symmetry group \(G\). In particular, the enlarged HGW formalism provides a direct description of local excitations. The enlarged HGW model also exposes the internal spaces of nonabelian anyons and the nontrivial \(G\)-actions on these internal spaces. By examining these nontrivial $G$-actions, we discover a new phenomenon we call symmetry fragmentation, which lies well beyond the reach of symmetry fractionalization. We detailed symmetry fragmentation in our companion paper \cite{fu2025nonlinear}. 

Since all SET phases originate from anyon condensation\cite{Bais2009a,Barkeshli2010,you2012synthetic,Hung2013,HungWan2013a,Kong2013,You2013,eliens2014diagrammatics,Wan2017,Neupert2016,williamson2017a,lan2018classification,Burnell2018,lee2018,Hung2019,Aasen2019,Hu2021,zhao2022,lin2023,zhao2024b} in pure topological phases, we would like to reckon the picture of anyon condensation in our construction of SET-phase models. Anyon condensation in a (parent) pure topological phase can result in a child SET phase if there is no SSB\cite{Hung2013,williamson2017a,Hu2021}. The condensed anyons of the parent phase would become the local excitations in the child SET phase. Local excitations in an SET phase would generally deviate a topological vacuum state (a state free of nontrivial anyons) from a ground state of the SET phase, unlike in a pure topological order, where a topological vacuum state is also a ground state.  In fact, all the topological vacuum states in an SET phase occupy a Hilbert subspace that is spanned by all the states with solely local excitations. We shall dwell on these points in Section \ref{sec:local excitation}.

In what follows, we shall first establish our SET lattice model and then anatomize our three strategies for constructing the input multifusion categories and their associated $G$-actions of the model. Our analysis will be accompanied by plenty of illustrative examples. Notably, in Section \ref{sec:S3symmetry}, as in the literature to date, we shall offer the very first explicit lattice realization of a genuinely nonabelian symmetry-enriched topological phase: an $S_3$-symmetry-enriched $\Z_2\times\Z_2$ quantum double phase. Furthermore, we show that gauging this SET model leads to the $S_4$ string-net model.

\section{SET Lattice Model}
In this section, we shall establish our lattice model describing SET phases based on the HGW string-net model. 

\begin{figure}[t]
    \centering
    \begin{minipage}[t]{0.5\textwidth}
        \centering
        \Lattice
        \caption{The lattice of HGW model.}
        \label{fig:lat}
    \end{minipage}
    \hfill
    \begin{minipage}[t]{0.4\textwidth}
        \centering
        \FatLattice
        \caption{The fattened lattice of one plaquette.}
        \label{fig:fatlat}
    \end{minipage}
\end{figure}

\subsection{Input data: A Multifusion Category with Associated Isomorphisms}\label{subsec:multifusion}

A topological phase $\mathcal{C}$, when endowed with a global symmetry $G$, turns out to be an SET phase $\mathcal{C}_G$. The SET phase $\mathcal{C}_G$ differs from $\mathcal{C}$ in that it has a number of $G$-graded sectors, each of which is topologically the same as $\mathcal{C}$ albeit labeled with a distinctive symmetry group element. The $G$-action maps the $G$-graded sectors to one another. This fact motivates us to construct a lattice model of $\Cat_G$ by replicating the HGW string-net model of $\Cat$ a commensurate number of times, while the replicas are related by the group elements of $G$, such that in any basis state of the string-net model, each plaquette is labeled by a particular G-graded sector. Since the input data of the HGW model of the phase $\mathcal{C}$ is a UFC $\Fus$, a natural input data of the corresponding lattice model of the $\Cat_G$ phase should contain a commensurate number of copies of $\Fus$, with copies being related by the symmetry $G$. The mathematical structure of this natural input data is called a multifusion category $\M$ equipped with a $G$-action. 

As briefly described in the introduction, we have mainly three strategies for constructing such input data for our SET lattice model. Different strategies have their pros in promoting a UFC $\Fus$ into a multifusion category in different situations pertinent to the origins of the symmetries. Note that these strategies are not exclusive to one another and can be combined in handling more general situations. To get the gist of what is a multifusion category, let's consider a simple case scenario, where a multifusion category $\M$ based on an $\Fus$ is typically formulated as a matrix
\begin{equation}\label{eq:multifusionmatrix}
\M = \begin{pmatrix}
\Fus_{g_1g_1} & \Fus_{g_1g_2} & \cdots & \Fus_{g_1g_n} \\
\Fus_{g_2g_1} & \Fus_{g_2g_2} & \cdots & \Fus_{g_2g_n} \\
\vdots & \vdots & \ddots & \vdots \\
\Fus_{g_ng_1} & \Fus_{g_ng_2}& \cdots & \Fus_{g_ng_n}
\end{pmatrix},
\end{equation}
with each $\Fus_{g_ig_j}$ being a copy of $\Fus$ as a UFC. The size $n$ is typically equal to $|G|$ when $G$ is a finite group. A simple object of $\M$ is denoted by $x_{g_ig_j}$ (also a simple object of $\Fus_{g_ig_j}$), where $x$ is a simple object in \(\Fus\). The simple objects respect the following fusion rules:
\eqn[eq:mulfusrule]{
x_{g_ig_j}\otimes y_{g_k,g_l} = \begin{cases}(x\otimes y)_{g_ig_l}\quad(g_j =g_ k),\\ \nulm\qquad\qquad\ \ (g_j\ne g_k)\end{cases},
}
where $\nulm$ is the zero object of \(\M\), and $x\otimes y$ is a fusion in $\Fus$. 

While we shall detail a rigorous mathematical definition of multifusion category Appendix \ref{sec:multifusion}, a practical explanation and construction of multifusion categories with associated $G$-actions will be unfolded step by step in the subsequent sections pertaining to our strategies.

In light of the input data structure, the lattice of our SET model of phase $\mathcal{C}_G$ is exactly the same as that of the HGW string-net model of phase $\mathcal{C}$. It is a honeycomb lattice, where each plaquette carries a tail (wiggly dangling edge), as protraited in Figure \ref{fig:lat}. Each edge and tail in the lattice carries a degree of freedom taking value in the simple objects of $\M$, such that the fusion rules of $M$ is obeyed at each vertex of the lattice. 

\begin{figure}
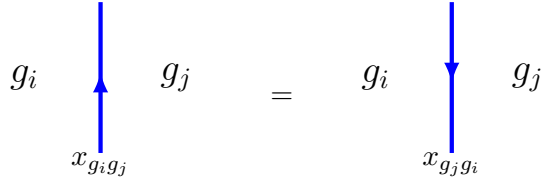

    \centering
    \DW
    \caption{Domain wall between two $G$-graded sectors, $g_i$-sector and $g_j$-sector. When moving along the domain-wall $x_{g_i g_j}$ in the indicated direction, $g_i$-sector is always on the left.}
    \label{fig:dw}
\end{figure}

From the fusion rules \eqref{eq:mulfusrule}, we can fatten the lattice Figure \ref{fig:lat} by extending each edge or tail to double-line, where the two group element indices lie one the double lines\cite{Chang2015}, as depicted in Figure \ref{fig:fatlat}. We can see that now each plaquette can be labeled uniquely by a group element $g$, as $g$ also labels the internal loop of the plaquette. This procedure is equivalent to assign spins into the plaquettes\cite{heinrich2016symmetry}. We say that a plaquette labeled by $g$ (in fact the corresponding phase) lies in a $g$-sector, which is a $G$-graded sector.  

The simple objects from the diagonal elements $\Fus_{g,g}$ of the multifusion matrix \eqref{eq:multifusionmatrix} describe the degrees of freedoms in sector-$g$, while the simple objects in the off-diagonal elements describe the domain wall degrees of freedoms, such as the simple objects in $\Fus_{g_ig_j}$ describes the domain wall between sector-$g_i$ and sector-$g_j$, as depicted in Figure \ref{fig:dw}.

\subsection{Hamiltonian}

The Hamiltonian takes the form
\eqn[eq:hamiltonian]{
H=-\sum_{p}Q_p=-\sum_{p}\sum_{s}Q_p^{s},
}
where $p$ runs over all plaquettes and $s$ over all simple objects of the multifusion category $\mathcal{M}$. This Hamiltonian is exactly the same as that in Ref. \cite{zhao2024b}, which redefines the Hamiltonian of the HGW string-net model, such that all elementary anyon excitations have the same energy.  The definition of the operators $Q_p^{s}$ is reviewed in Appendix \ref{sec:review}.

Although Hamiltonian \eqref{eq:hamiltonian} formally agrees with that of the HGW string-net model describing a pure topological phase, the physics it describes has a significant distinction with that of the HGW model: The system now has not only topological excitations -- anyons -- but also local excitations, which is absent from the HGW model. The topological excitations in the SET phase described by Hamiltonian \eqref{eq:hamiltonian} are defined with respect to the local excitations, which should be regarded as part of the vacuum. This gives way to the phenomenon known as symmetry fractionalization. Namely, since the topological vacuum contains local excitations, it is acted on by the global symmetry nontrivially and yields different $G$-graded sectors. The fusion rules of anyons then allow certain types of anyons in a $G$-graded sector for carrying fractionalized symmetry charges. Note however that the ground-state degeneracy of an SET phase $\Cat_G$ coincides with that of pure topological phase $\Cat$ on any topology, despite the existence of $G$-graded sectors. In Section \ref{sec:symmfrac}, we shall prove what is claimed here in this paragraph and show concrete examples. 

Now, let's start to explicitly construct the input data of our SET model, i.e., the multifusion categories with their associated sets of isomorphisms.

\section{Strategy I: Constructing SET Input Data from Automorphisms of \eqs{\Fus}, Symmetry Transformation, And Gauging}\label{sec:strategyI}

In this section, we elaborate on our first strategy of constructing the input data of the SET model describing the phase $\Cat_G$. The input data enables us to present the global-symmetry transformations on the SET model and the anyons in $\Cat$. Even at the level of the input data of such a model, we will be able to find out the input and output data of a pure topological phase -- a parent phase of $\Cat$ -- obtained by gauging the global symmetry $G$ of the SET phase $\Cat_G$. 

We shall introduce this strategy in three steps, accompanied by a concrete simple example -- $\Z_3$ quantum-double phase enriched by the charge-conjugation symmetry.

\subsection{Step 1: Find the Outer Automorphism Group \eqs{\Out(\Fus)} of \eqs{\Fus}}\label{subsec:case1step1}

First, we need to find all the outer automorphisms\footnote{Inner automorphisms only yield gauge invariance.} of $\Fus$. These outer automorphisms form a group $\Out(\Fus)$, which turns out to be a global symmetry of the topological phase described by the HGW model with input UFC $\Fus$. We denote each outer automorphism by $\phi_{g},\ g\in\Out(\Fus)$. The outer automorphism group $\Out(\Fus)$ would enrich the pure topological phase $\Cent(\Fus)$ to be an SET phase $\Cent(\Fus)_{\Out(\Fus)}$. The order $|\Out(\Fus)|$ of the outer automorphism group is the number of $G$-graded sectors in the SET phase $\Cent(\Fus)_{\Out(\Fus)}$. Each $G$-graded sector is uniquely labeled by a group element $g\in \Out(\Fus)$.

\begin{example}{Charge-Conjugation Symmetry-Enriched $\Z_3$ Quantum-Double Phase}{1}
The input UFC of $\Z_3$ quantum-double phase is $\Vec(\Z_3)$, which is defined in Appendix \ref{sec:review}.In essence, the three simple objects of $\Vec(\Z_3)$ are $\{0,1,2\}$, subject to fusion rules as the group multiplication rules. 

There are 9 anyon types of the $\Z_3$ quantum-double phase $\Cent(\Vec(\Z_3))$, denoted by $1$, $e$, $e^2$, $m$, $m^2$, $em$, $e^2m$, $em^2$, and $e^2m^2$, where $e$ ($m$) refers to the basic charge (flux). They are defined in Appendix \ref{subsec:Z3 ztensor}.

UFC $\Vec(\Z_3)$ has a trivial outer automorphism $\phi_+$ and a nontrivial one $\phi_-$:
\eqn{\label{eq:Z3outerautomorphism}
    \phi_+:\ 0\to0,\ 1\to1,\ 2\to2;\\
    \phi_-:\ 0\to0,\ 1\to2,\ 2\to1.
    }
We have $\Out(\Vec(\Z_3))=\Z_2 =\{+,-\}$ because $\phi_-\circ\phi_-=\phi_+$. Thus, the pure topological phase $\Cent(\Vec(\Z_3))$ can be enriched by this $\Z_2$ to be the SET phase $\Cent(\Vec(\Z_3))_{\Out(\Vec(\Z_3))}$, which has two $G$-graded sectors, $+$ and $-$.
\end{example}

\subsection{Step 2: Promote \eqs{\Fus} to A Multifusion Category by \eqs{\Out(\Fus)}}

After finding the $\Out(\Fus)$, we can promote $\Fus$ to a multifusion category $\mathcal{M}_{\Out(\Fus)}$ by $\Out(\Fus)$. The multifusion matrix takes the form
\eqn{\mathcal{M}_{\Out(\Fus)}=\begin{pmatrix}
  \Fus_{g_1g_1} & \Fus_{g_1g_2} & \cdots\\
  \Fus_{g_2g_1} & \Fus_{g_2g_2} & \cdots\\
  \vdots & \vdots & \ddots
\end{pmatrix},
}
which is a $|\Out(\Fus)|\times |\Out(\Fus)|$ matrix with $\Fus_{g_ig_j}$ being a copy of $\Fus$. Here, $g_ig_j\in\Out(\Fus)$. An isomorphism map $\D_{g_ig_j}$ can be defined by
\begin{equation}
\begin{aligned}
\D_{g_ig_j}:\ &\Fus_{g_ig_j}\to\Fus\\
 & x_{g_ig_j} \mapsto x. 
\end{aligned}
\end{equation}
Each diagonal element $\Fus_{g_ig_i}$ is equipped with the outer automorphism $\phi_{g_i}$. The fusion rules are given by
\begin{equation}\label{eq:fusion rules}
    x_{g_ig_j}\otimes y_{g_k,g_l}=\delta_{g_jg_k}(x\otimes y)_{g_ig_l}.
\end{equation}

The anyon types in the SET phase $\Cent(\Fus)_{\Out(\Fus)}$ are in one-to-one correspondence with the simple objects of the Drinfeld center of the multifusion category $\mathcal{M}_{\Out(\Fus)}$. Nevertheless, to identify the anyon types and name them properly would require knowing the symmetry transformations in the SET phase, which is to be constructed in the next subsection.

\begin{example}{Charge-Conjugation Symmetry-Enriched $\Z_3$ Quantum-Double Phase}{2}

The group $\Out(\Vec(\Z_3))$ promotes $\Vec(\Z_3)$ to be a multifusion category:
\eqn{\mathcal{M}_{\Out(\Vec(\Z_3))}=\begin{pmatrix}\label{eq:multifusionZ3}
    \{0_{++},1_{++},2_{++}\} & \{0_{+-},1_{+-},2_{+-}\}\\
    \{0_{-+},1_{-+},2_{-+}\} & \{0_{--},1_{--},2_{--}\}\\
    \end{pmatrix}.
    }
    
    An example of the generic fusion rules \eqref{eq:fusion rules} in this case is
    \begin{equation}
        1_{+-}\otimes 2_{-+}=0_{++}.
    \end{equation}

Using this multifusion category equipped with outer automorphisms \eqref{eq:Z3outerautomorphism} as the input data of the HGW model, we will be able to describe the charge-conjugation symmetry-enriched $\Z_3$ quantum-double phase $\Cent(\Vec(\Z_3))_{\Out(\Vec(\Z_3))}$.

In \eqref{eq:multifusionZ3}, the two diagonal elements give the degrees of freedoms in two $G$-graded sectors respectively: $\{0_{++},1_{++},2_{++}\}$ are the degrees of freedoms in the $+$ sector, and $\{0_{--},1_{--},2_{--}\}$ are the degrees of freedoms in the $-$ sector. The off-diagonal elements supply the degrees of freedoms on the domain wall between two different sectors: $\{0_{+-},1_{+-},2_{+-}\}$ are the degrees of freedoms on the $+\dash-$ domain wall ($+$-sector on the left) while $\{0_{-+},1_{-+},2_{-+}\}$ are the degrees of freedoms on the $-\dash+$ domain wall ($-$-sector on the left). See Fig. \ref{fig:dw}, where one sets $g_i=+$ and $g_j=-$. A single plaquette could happen to be in a $+$-sector or $-$-sector.

\end{example}

\subsection{Step 3: Construct the Symmetry Transformation}

From the isomorphisms, we can construct the symmetry operator $\G_{g}$ of the SET model by
\begin{equation}
    \G_g=\bigotimes_{l}\G^l_{g},
\end{equation}
where $l$ goes over all the edges and tails. Here, $\G^l_{g}$ is an onsite operator acting on a single edge or tail $l$ only:
\begin{equation}\label{eq:symoperator}
    \G^l_{g}(x_{g_ig_j})=[\phi_{g}(x)]_{(gg_i)(gg_j)}:= \D^{-1}_{(gg_i)(gg_j)}(\phi_{g}(x)).
\end{equation}
Onsite operators $\G^l_{g'}$ and $\G^l_g$ are composed as
\eqn{
&\G^l_{g'}\circ\G^l_{g}(x_{g_ig_j})=\G^l_{g'}([\phi_{g}(x)]_{(gg_i)(gg_j)})=[\phi_{g'}(\phi_{g}(x))]_{(g'gg_i)(g'gg_j)}=[\phi_{g'g}(x)]_{(g'gg_i)(g'gg_j)}\\
=&\G^l_{g'g}(x_{g_ig_j}),
}
such that 
\eqn{
\G_{g'}\circ\G_{g}=\G_{g'g}.
}
Therefore, the set of symmetry operators $\{\G_{g}| g\in \Out(\Fus)\}$ is a representation of the symmetry group $\Out(\Fus)$.

\begin{example}{Charge-Conjugation Symmetry-Enriched $\Z_3$ Quantum-Double Phase}{3}
The only nontrivial onsite symmetry operator is $\G^l_{-}$, as defined generally in \eqref{eq:symoperator}. In the current case, for example, we have 
\eqn{
&\G^l_{-}(1_{-+})=2_{+-},\ \G^l_{-}(1_{++})=2_{--},\\
&\G^l_-\circ\G^l_-=\G^l_+.
}
It's easy to see that $\G_-$ exchanges $+$ and $-$ sectors.
\newline

Now let's see how anyons change under $\G_-$. Since $1_{++}$ and $2_{--}$ are exchangeable under $\G^l_-$, they form a symmetry doublet and can be renamed as $r_{++}$ and $r_{--}$. Likewise, we rename $2_{++}$ ($1_{--}$) as $r^2_{++}$ ($r^2_{--}$), $0_{+-}$ ($0_{-+}$) as $s_{+-}$ ($s_{-+}$), $1_{+-}$ ($2_{-+}$) as $rs_{+-}$ ($rs_{-+}$), and $2_{+-}$ ($1_{-+}$) as $sr_{+-}$ ($sr_{-+}$). From the $z$-tensors (Appendix \ref{subsec:Z2Z3 ztensor}), we can see that
\begin{equation}\label{eq:ztensor example}
    z^{(m,m^2);s_{+-}}_{r_{++}r^2_{--}s_{+-}}=z^{(m,m^2);rs_{+-}}_{r_{++}r^2_{--}rs_{+-}}=z^{(m,m^2);sr_{+-}}_{r_{++}r^2_{--}sr_{+-}}=1\neq0.
\end{equation}
Thus, the anyon $(m,m^2)$ in the SET model looks like an $m$ anyon in sector $+$, while looks like an $m^2$ anyon in sector $-$. For this reason, we denote this type of anyon by $(m,m^2)$.

Generally, the original anyon types in the pure $\Z_3$ quantum-double phase are not completely well-defined because a $\Z_3$ quantum-double anyon in the $+$ sector would become its charge-congjugation in the $-$ sector. This implies that the original 9 anyons types shoud be regrouped into 9 pairs $(a,b)$, where $a$ and $b$ are each other's charge conjugation. Such a pair $(a,b)$ appears to be $a$ ($b$) in the $+$ ($-$) sector. For example, $(m,m^2)$ appears to be $m$ ($m^2$) in the $+$ ($-$) sector. These 9 pairs are in fact the simple objects in the Drinfeld center of the multifusion category \eqref{eq:multifusionZ3}, their half-braiding $z$-tensors can be found in Appendix \ref{subsec:Z2Z3 ztensor}. The general form of anyons in the SET phase is
\begin{equation}\label{eq:Z3z2Spectrum}
\{(e^i m^j,e^{(3-i)\mod 3} m^{(3-j) \mod 3})| i,j=0,1,2\},
\end{equation}
where $e$ ($m$) is the basic charge (flux) of the pure topological phase $\Cent(\Vec(\Z_3))$. Such an anyon, say, e.g., $(m,m^2)$, appears in the $+$ ($-$) sector as an anyon $m$ ($m^2$) of the $\Z_3$ quantum-double phase.

From onsite operator $\G_-^l$, we can construct the symmetry operator $\G_-$, which will transform the 9 anyon types defined in \eqref{eq:Z3z2Spectrum} as
\begin{equation}
    (e^i m^j,e^{(3-i)\mod 3} m^{(3-j) \mod 3}) \leftrightarrow (e^{(3-i)\mod 3} m^{(3-j) \mod 3},e^i m^j).
\end{equation}

\end{example}

\subsection{Step 4: Gauging}\label{sec:gauging}

To gauge the symmetry, we need to identify different sectors before and after symmetry transformation. Thus, for any $x_{g_ig_j}\in\mathcal{M}$, we need to identify all the elements in the set $\{\G^l_g(x_{g_ig_j})|g\in\Out(\Z_3)\}$, rendering the set an equivalence class, to be denoted by the pair $[g,x]$, where $g\in\Out(\Fus),\ x\in\Fus$. These equivalence classes should be regarded as the simple objects of the input UFC $\Fus_p$ of the parent phase. Thus, the identification process can be understood as an $|\Out(\Fus)|$-to-1 map:
\eqn{\label{eq:identification}
\mathcal{I}:\ &\mathcal{M}\to\Fus_p\\
&x_{g_ig_j}\mapsto \left[(g_i^{-1}g_j),\phi_{g_i^{-1}}(x)\right].
}
The fusion rules between the simple objects of $\Fus_p$ can be obtained from the fusion rules of $\M$:
\eqn{\label{eq:identification_fusion}
[g_i,x]\otimes[g_j,y]=[g_ig_j,x\otimes\phi_{g_{i}}(y)].
}
The $6j$-symbols of $\Fus_p$ can also be obtained from those of $\mathcal{M}$.

In particular, when the input UFC $\Fus$ of the original topological phase is a group algebra $\Vec(G)$, such as in the quantum-double phase introduced in Appendix \ref{sec:review}, the identification process can be viewed as group extension: fusion rules \eqref{eq:identification_fusion} reduce to
\eqn{
gx=&[g,e]\otimes[e,x]=[g,\phi_{g}(x)]=[e,\phi_{g}(x)]\otimes[g,e]\\
=&\phi_{g}(x)g,\qquad\forall g\in\Out(G),\ x\in G.
}
Thus, the input UFC $\Fus_p$ after gauging is the extension of $G$ by $\Out(G)$:
\eqn{
\Fus_p=G\rtimes\Out(G).
}

In fact, after identification, the outer automorphism of $\Vec(G)$ now becomes inner automorphism of $\Fus_p$. As pointed out earlier, inner automorphism only yields gauge invariances, while outer automorphisms generate global symmetries, our identification process described above is indeed a gauging process.

\begin{example}{Charge-Conjugation Symmetry-Enriched $\Z_3$ quantum-double Phase}{4}

    For the $\Z_3$ quantum-double phase, the identification \eqref{eq:identification} is showed in Table \ref{tab:Z3 identification}.
    \begin{table}[H]
        \centering
        \begin{adjustbox}{{width=\linewidth}}
        \begin{tabular}{||c||c|c|c|c|c|c||}
           \hline
           \textbf{Before}  & $0_{++},0_{--}$ & $1_{++},2_{--}$ & $2_{++},1_{--}$ & $0_{+-},0_{-+}$ & $1_{+-},2_{-+}$ & $2_{+-},1_{-+}$ \\
           \hline
           \textbf{After}  & $1:=[+,0]$ & $r:=[+,1]$ & $r^2:=[+,2]$ & $s:=[-,0]$ & $rs:=[-,1]$ & $sr:=[-,2]$\\
           \hline
        \end{tabular}
        \end{adjustbox}
        \caption{The identification when gauging the charge conjugation symmetry of $\Z_3$ quantum-double phase}
        \label{tab:Z3 identification}
    \end{table}

    The fusion rules \eqref{eq:identification_fusion} become
    \eqn{
    r^3=s^2=1,\ sr=r^2s,
    }
   which are just $S_3$ group multiplication rules. Truly, we have
    \begin{equation}
    S_3 = \Z_3\rtimes \Z_2.
    \end{equation}

    Hence, gauging the charge-conjugation symmetry of $\Z_3$ quantum-double phase results in the $S_3$ quantum-double phase. This result agrees with the literature\cite{cheng2017exactly}.

\end{example}

\section{Strategy II: Constructing SET Input Data from Frobenius algebras in \eqs{\Fus}, Symmetry Transformation, and Gauging}\label{sec:bimodule}

In this section, we introduce our second strategy of constructing the input data of and gauging an SET model.

This strategy consists of $7$ steps, to be elaborated one by one, accompanied by a simple example -- the $\Z_2$ toric code model enriched by the EM-exchange symmetry.

\subsection{Step 1: Find two Frobenius algebras}

First, we need to find $2$ Frobenius algebras of an input UFC $\Fus$.

A Frobenius algebra $\A$ in a fusion category $\Fus$ is a subset $L_{\A}$ of simple objects, equipped with product \(f: L^3_\A \to \C\), satisfying:
\eqn{\sum_{t_\tau \in L_\A} f_{r_\rho s_\sigma t_\tau} f_{a_\alpha b_\beta t_\tau^*} G^{rst}_{abc} \sqrt{d_c d_t} &= \sum_{\gamma = 1}^{n_c} f_{a_\alpha c_\gamma s_\sigma} f_{r_\rho c_\gamma^* b_\beta}\ ,\\ \\
\sum_{a_\alpha b_\beta \in L_\A} f_{a_\alpha b_\beta c_\gamma} f_{b_\beta^* a_\alpha^* c_\gamma^*} \sqrt{d_a d_b} = d_\A \sqrt{d_c},\qquad f_{a_\alpha b_\beta c_\gamma} &= f_{b_\beta c_\gamma a_\alpha}, \qquad f_{0 a_\alpha b_\beta} = \delta_{ab^*} \delta_{\alpha\beta},}
where
\eqn{
d_\A := \sum_{a \in L_\Fus} n_a d_a
}
is the \emph{quantum dimension} of the Frobenius algebra \(\A\).

Any given $\Fus$ always has a trivial Frobenius algebra $\A_e$: $L_{\A_e}=\{1\},\ f_{111}=1$, where $1$ is the unit object in $\Fus$. We shall need $2$ Frobenius algebras of $\Fus$, one of which must be the trivial one. Let $\A_e$ be the trivial Frobenius algebra. The other Frobenius algebra $\A_g$ could be any nontrivial one.

\begin{example}{EM-Exchange Symmetry Enriched Toric Code}{1}
    As an example, we consider the $\Z_2$ toric code model, which describes the $\Z_2$ quantum-double phase. The input UFC is $\Vec(\Z_2)$, having two simple objects $\{0,1\}$. There are $4$ anyon types, denoted by $1,e,m$ and $f$, as shown in Appendix \ref{subsec:toric ztensor}. The trivial Frobenius algebra is $\A_e=0$. The only nontrivial Frobenius algebra $\A_g$ in $\Vec(\Z_2)$ is defined by
    \eqn{
    L_{\A_g} = \{0, 1\}, \qquad f_{000} = f_{011} = f_{101} = f_{110} = 1.
    }
\end{example}

\subsection{Step 2: Calculate all the Bimodules}

First we take a brief review of the definition of bimodules. For $2$ arbitrary Frobenius algebras ${\A_1}$ and ${\A_1}$ in a UFC \(\Fus\), an ${\A_1}\dash{\A_2}$ bimodule \(M\) is defined by a pair of functions \((n^M, P_M)\), where the function \(n^M: L_\Fus \to \NN\) returns the \emph{multiplicity} \(n^M_a\) of \(a \in L_\Fus\) appearing in bimodule \(M\), satisfying \(n^M_a = n^M_{a^*}\), and the basis elements of \(M\) are labeled by \(a_i\), where \(a \in L_\Fus\) satisfies \(n_a^M > 0\), and \(i = 1, 2, \ldots, n^M_a\) labels the multiplicity index. We denote the set of all basis elements in bimodule \(M\) as \(L_M\). 

An ${\A_1}\dash{\A_2}$ bimodule is acted on by Frobenius algebra ${\A_1}$ (${\A_2}$) from its left (right), which is recorded by function \(P_M: L_{\A_1}\times L_{\A_2} \times L_M \times L_\Fus \times L_M \to \C\), satisfying the following defining equations:
\eqn{&\sum_{uv\in L_\Fus}\ \sum_{y_\upsilon\in L_M}\ [P_M]^{a_\alpha r_\rho}_{x_\chi u y_\upsilon}\ [P_M]^{b_\beta s_\sigma}_{y_\upsilon v z_\zeta}\ G^{v^* by}_{urw}\ G^{w^* bu}_{axc}\ G^{sz^* v}_{wrt^*}\ \sqrt{d_ud_vd_wd_yd_cd_t}\\ 
=\ &\sum_{\gamma = 1}^{n_c}\ \sum_{\tau = 1}^{n_t}\ P^{c_\gamma t_\tau}_{x_\chi w z_\zeta}f^{\A_1}_{a_\alpha c_\gamma^* b_\beta}\ f^{\A_2}_{r_\rho s_\sigma t_\tau},
}
\eqn{\ [P_M]^{00}_{x_\chi y z_\zeta} = \delta_{xy}\delta_{yz}\delta_{\chi\upsilon}\delta_{\upsilon\zeta}.
}
Here, $f^{\A_1}$ and $f^{\A_2}$ are the defining algebra products respectively for Frobeniua algebras ${\A_1}$ and ${\A_2}$.

The quantum dimension of bimodule $M$ is defined by
\eqn{
d_M = \frac{1}{\sqrt{d_{\A_1} d_{\A_2}}}\sum_{a\in L_\Fus}n^M_ad_a.
}

We can fuse bimodules. Such fusion usually involves more than two Frobenius algebras in general. Let ${\A_1}$, ${\A_2}$, and ${\A_3}$ be three (not necessarily different) Frobenius algebras of an UFC $\Fus$. Then let $M^1_{{\A_1}{\A_2}},M^2_{{\A_2}{\A_3}}$, and $M^3_{{\A_3}{\A_1}}$ be respectively ${\A_1}\dash{\A_2}$, ${\A_2}\dash{\A_3}$, and ${\A_3}\dash{\A_1}$ bimodules. They can fuse and comprise a fusion vertex only if in any of the three plaquettes meeting at the fusion vertex, the Frobenius algebra acting on the bimodules is unique. (For an explicit illustration, see Appendix \ref{sec:theory}.) The fusion rules are determined by the fusion matrix $\Delta_{M^1_{{\A_1}{\A_2}}M^2_{{\A_2}{\A_3}}M^3_{{\A_3}{\A_1}}}$, which is defined by
\eqn{\ [\Delta_{M^1_{{\A_1}{\A_2}}M^2_{{\A_2}{\A_3}}M^3_{{\A_3}{\A_1}}}]_{r_\rho s_\sigma t_\tau}^{x_\chi y_\upsilon z_\zeta} := \frac{1}{d_{\A_1} d_{\A_2} d_{\A_3}}\sum_{b_\beta\in L_{\A_1}}\sum_{c_\gamma \in L_{\A_2} }\sum_{a_\alpha\in L_{\A_3}}\sum_{p\in L_\Fus}\sum_{u_\rho\in L_{M_{{\A_1}{\A_2}}^1}}\sum_{v_\sigma\in L_{M_{{\A_2}{\A_3}}^2}}\\
\sum_{w_\lambda\in L_{M_{{\A_3}{\A_1}}^3}}
[P_1]^{b_\beta c_\gamma^*}_{x_\chi u r_\rho}\ [P_2]^{c_\gamma a_\alpha^*}_{y_\upsilon v s_\sigma}\ 
\ [P_3]^{a_\alpha b_\beta^*}_{z_\zeta w t_\tau}\ G^{bxu^*}_{c^* r^* p}\ G^{swp}_{br^* t^*}\ G^{pvz}_{aw^* s^*}\ G^{xyz}_{vpc}\ \sqrt{d_ud_vd_wd_ad_bd_cd_rd_sd_t}\ d_p\ &,}
where the triple \((x_\chi, y_\upsilon, z_\zeta)\) is the row index, and \((r_\rho, s_\sigma, t_\tau)\) is the column index. See Appendix \ref{sec:theory} for more detail. With this understanding, we may neglect the Frobenius algebra labels in studying the fusion of bimodules if no confusion would be caused.

The fusion rules\footnote{We assume in this paper that the bimodule fusion rules are devoid of multiplicities. Otherwise, ${\rm Tr}[\Delta_{M_1M_2M_3}]=n$, where $n$ is the multiplicity. Generalization to such cases is straightforward.} of three bimodules \(M_1, M_2, M_3\) can be read off directly from the fusion matrix as
\eqn{\delta_{M_1M_2M_3} = {\rm Tr}[\Delta_{M_1M_2M_3}] = \sum_{r_\rho, s_\sigma, t_\tau\in L_M}[\Delta_{M_1M_2M_3}]_{r_\rho s_\sigma t_\tau}^{r_\rho s_\sigma t_\tau}.}

The $6j$-symbols can be calculated through the $+1$ eigenvector $\V_{M_1M_2M_3}$ of fusion matrix $\Delta_{M_1M_2M_3}$, which is normalized by
\eqn{\label{normalization}
\sum_{x_\chi\in L_{M_1}}\sum_{y_\upsilon\in L_{M_2}}\sum_{z_\zeta\in L_{M_3}}|\V_{M_1M_2M_3}^{x_\chi y_\upsilon z_\zeta}|^2\ \sqrt{d_xd_yd_z} = N_{M_1M_2M_3}\sqrt{d_{M_1}d_{M_2}d_{M_3}}\ ,}
where $N_{M_1M_2M_3}$ is a normalization constant which depends on the Frobenius algebras acting on the bimodules. The vector elements $\V_{M_1M_2M_3}^{x_\chi y_\upsilon z_\zeta}$ are called the \emph{vertex coefficients}. 

For our purpose in this step, only two Frobenius algebras ${\A_1}$ and ${\A_2}$ of $\Fus$ are relevant, giving rise to only $4$ possible fusion patterns:
\begin{equation}
    \begin{split}
        &N_{M^1_{{\A_1}{\A_1}}M^2_{{\A_1}{\A_1}}M^3_{{\A_1}{\A_1}}}=d_{\A_1}^2,\ N_{M^1_{{\A_2}{\A_2}}M^2_{{\A_2}{\A_2}}M^3_{{\A_2}{\A_2}}}=d_{\A_2}^2,\\
        &N_{M^1_{{\A_1}{\A_1}}M^2_{{\A_1}{\A_2}}M^3_{{\A_2}{\A_1}}}=d_{\A_1}^{\frac{3}{2}}d_{\A_2}^{\frac{1}{2}},\ N_{M^1_{{\A_2}{\A_2}}M^2_{{\A_2}{\A_1}}M^3_{{\A_1}{\A_2}}}=d_{\A_1}^{\frac{1}{2}}d_{\A_2}^{\frac{3}{2}}.\\
    \end{split}
\end{equation}
Using the vertex coefficients, the bimodule $6j$ symbols can be expressed as
\eqn{\label{6jgeneral}
G^{M_1M_2M}_{M_3M_4M'}=\sum_{x_\chi\in L_{M_1}}\sum_{y_\upsilon\in L_{M_2}}\sum_{z_\zeta\in L_{M}}\sum_{r_\rho\in L_{M_3}}\sum_{s_\sigma\in L_{M_4}}\sum_{t_\tau\in L_{M'}}\\ \frac{\V_{M_1M_2M}^{x_\chi y_\upsilon z_\zeta}\V_{M_3M_4M^*}^{r_\rho s_\sigma z^*_\zeta}\V_{M^*_1M^*_4M'^*}^{x^*_\chi s^*_\sigma t^*_\tau}\V_{M^*_3M^*_2M'}^{r^*_\rho y^*_\upsilon t_\tau}}{d_{\A_1}^{N_{\A_1}-\frac{1}{2}}d_{\A_2}^{N_{\A_2}-\frac{1}{2}}\sqrt{d_{M_1}d_{M_2}d_{M_3}d_{M_4}d_{M}d_{M'}}}G^{xyz}_{rst},
}
where $N_{\A_1}$ is the number of the plaquettes with Frobenius algebra ${\A_1}$ in it, $N_{\A_2}$ is the number of the one with ${\A_2}$ in it. For detailed information, see appendix \ref{sec:theory}.

It is worth mentioning that since our normalization \eqref{normalization} condition only constrains the vertex coefficients $\V_{M_1M_2M_3}^{x_\chi y_\upsilon z_\zeta}$ up to a phase, leaving room for a gauge redundancy:
\eqn{\label{gauge}
\V_{M_1M_2M_3}^{x_\chi y_\upsilon z_\zeta}\to e^{i\theta_{M_1M_2M_3}}\V_{M_1M_2M_3}^{x_\chi y_\upsilon z_\zeta},\ \theta_{M_1M_2M_3}\in\mathbb{R}.
}
Here $\theta_{M_1M_2M_3}$ is anti-symmetric. Hence, the $6j$ symbols are changed by
\eqn{
G^{M_1M_2M}_{M_3M_4M'}\to \frac{e^{i\theta_{M_1M_2M}}e^{i\theta_{M_3M_4M^*}}}{e^{i\theta_{M_4M_1M'}}e^{i\theta_{M_2M_3M'^*}}}G^{M_1M_2M}_{M_3M_4M'},
}
which corresponds to the scaling symmetry discussed in Ref.\cite{Hung2012}. We shall fix this gauge in Section \ref{subsec:isomorphism}.

It happens that all $\A\dash\A$ bimodules comprise a UFC ${_\A\Bimod_\A(\Fus)}$, i.e., they are closed under fusion. It is known that ${_\A\Bimod_\A(\Fus)}$ is Morita equivalent to $\Fus$, in the sense that their Drinfeld centers are isomorphic\cite{etingof2016,zhao2024a}. When $\A=\{1\}$, i.e., the trivial Frobenius algebra, we have ${_\A\Bimod_\A(\Fus)} = \Fus$.

In this step, we need to calculate all the ${\A_e}\dash{\A_e}$, ${\A_e}\dash{\A_g}$, ${\A_g}\dash{\A_e}$ and ${\A_g}\dash{\A_g}$ bimodules, as well as their fusion rules and $6j$-symbols.

\begin{example}{EM-Exchange Symmetry Enriched Toric Code}{2}
    For $\Vec(\Z_2)$, we list all the simple (i.e., irreducible) bimodules over the trivial Frobenius algebra $\A$ and the nontrivial one $\B$ here:
    \begin{enumerate}
        \item ${\A_e}\dash{\A_e}$ simple bimodules are just the simple objects of $\Fus$ as $\A_e$ is trivial, we denote them by $M_0,M_1$. $d_{M_0}=d_{M_1}=1$.
        \item There are $2$ simple ${\A_g}\dash{\A_g}$ bimodules, denoted by $M_+,M_-$. $d_{M_+}=d_{M_-}=1$.
        \item There is only $1$ simple ${\A_e}\dash{\A_g}$ bimodule denoted by $M_{\sigma}$, $d_{M_{\sigma}}=\sqrt{2}$.
        \item There is only $1$ simple ${\A_g}\dash{\A_e}$ bimodule denoted by $M^*_\sigma$, which is the opposite bimodule of $M_{\sigma}$, $d_{M^*_{\sigma}}=\sqrt{2}$.
    \end{enumerate}
    The data of these bimodules and the fusion rules can $6j$-symbols between them can be found in Appendix \ref{Z2}.
\end{example}

\subsection{Step 3: Construct the Transformation \eqs{\T}}\label{subsec:isomorphism}

The UFCs ${_{\A_e}\Bimod_{\A_e}(\Fus)}$ and ${_{\A_g}\Bimod_{\A_g}(\Fus)}$ are Morita equivalent to the original UFC \(\Fus\)\cite{etingof2016,zhao2024a}, which means that they can describe the same topological phase as \(\Fus\) does. Nevertheless, the Morita equivalence can only give us a duality, not yet a symmetry, generated by transformations from $\Fus$ to $\Fus$. To this end, we need to establish isomorphisms among ${_{\A_e}\Bimod_{\A_e}(\Fus)}$, ${_{\A_g}\Bimod_{\A_g}(\Fus)}$, and $\Fus$.

While leaving more details to Appendix \ref{sec:trans}, here we briefly describe how to construct symmetry (or gauge) transformation through bimodules.

For any Frobenius algebra $\A$ of $\Fus$ satisfying ${_\A\Bimod_\A(\Fus)}\cong\Fus$, we can choose an isomorphism $\varphi:\Fus\to {_\A\Bimod_\A(\Fus)}$. There is a duality map $\D:{_\A\Bimod_\A(\Fus)}\to\Fus$ induced by the corresponding bimodule functions $P$. The duality map $\D$ maps every fusion vertex of ${_\A\Bimod_\A(\Fus)}$ to a linear combination of the fusion vertices in $\Fus$. Namely, for any fusion vertex \(\phi_{M_1 M_2}^{M_3} : M_1 \otimes M_2 \to M_3\), 
\eqn{
\D(\phi_{M_1 M_2}^{M_3}) = \sum_{z_\zeta \in L_{M_3}} \Bigg[ \sum_{x_\chi \in L_{M_1}} \sum_{y_\upsilon \in L_{M_2}} \mathcal{V}_{M_1 M_2 M_3^*}^{x_\chi y_\upsilon z_\zeta^*} \ \phi_{x_\chi y_\upsilon}^{z_\zeta} \Bigg],
}
where \(\phi_{xy}^z : x \otimes y \to z\) are fusion vertices of $\Fus$.

Note that the isomorphism $\varphi$ is not unique in general. Composing the duality map $\D$ with a chosen isomorphism $\varphi$ yields a transformation $\T$ over the pure topological order:
\eqn[eq:compositionmap]{
\T := \D \circ \varphi : \Fus\to\Fus.
}
One can show that such a transformation $\T$ preserves the modular $S$ and $T$ matrices of the model and hence of the output topological phase. So, $\T$ appears to be a symmetry transformation; however, it may be a gauge transformation. As far as strategy II is concerned, a global symmetry transformation differs from a gauge transformation in that it permutes anyon types\cite{zhao2024a}. Now that $\Fus$ has multiple Frobenius algebras $\A_1,\A_2,\A_3,\dots $, we can construct respectively transformations $\T_1,\T_2,\T_3,\dots $. These transformations can be composed arbitrarily.

Now come back to our SET construction. Since $\A_e$ is trivial, there is a canonical isomorphism from $\Fus$ to ${_{\A_e}\Bimod_{\A_e}(\Fus)}$:
\begin{equation}\label{eq:trivial isomorphism}
    \begin{split}
        \varphi_e:\ &\Fus\to{_{\A_e}\Bimod_{\A_e}(\Fus)}\\
        &a\mapsto M_a,
    \end{split}
\end{equation}
where
\begin{equation}\label{eq:trivial bimodule}
    L_{M_a}=\{a\},\ [P_{M_a}]^{11}_{aaa}=1.
\end{equation}
In this case, we have
\begin{equation}
    \D_e(\phi_{M_a M_b}^{M_c}) = \phi_{ab}^{c}.
\end{equation}
It's obvious that $\T_e:=\D_e \circ \varphi_e$ is the identity transformation.

For the Frobenius algebra $\A_g$, if it happens that ${_{\A_g}\Bimod_{\A_g}(\Fus)}\overset{\varphi_g}{\cong} \Fus$, we can also construct a transformation $\T_g:\D_g\circ\varphi_g$. 

It is worth noting that the transformations $\T_g$ constructed here acts solely on the pure topological order $\Cent(\Fus)$, so it is not yet the symmetry transformation of the SET model. In subsequent steps, one will see that after constructing the input multifusion category $\M$ of the model of SET phase $\Cent(\Fus)_G$, we can lift a $\T_g$ to a global symmetry transformation $\G_g$ of the SET model.

Recall that the fusion vertices of $\Fus$ and those of $_\A\Bimod_\A(\Fus)$ are related by the vertex coefficients $\V_{M_1M_2M_3}^{x_\chi y_\upsilon z_\zeta}$. Now a chosen isomorphism $\varphi_g:\Fus \to {_{\A}\Bimod_{\A}(\Fus)}$ also relates the fusion vertices of the two parties. Would these two relations be in conflict? They could. But such a conflict can always be resolved by a gauge fixing because the vertex coefficients have a gauge redundancy \eqref{gauge}. Hence, for any chosen isomorphism $\varphi_g$, compatible vertex coefficients can be chosen by a gauge fixing. This gauge fixing is crucial for the gauging procedure to be discussed in Section \ref{subsec:gauging}, as it will allow us to consistently obtain from the $6j$-symbols of an SET model those of the parent model. Hereafter, we will always have done the necessary gauge fixing implicitly.

\begin{example}{EM-Exchange Symmetry Enriched Toric Code}{3}
    Consider our toric code example. 
    \begin{enumerate}
        \item For ${\A_e}\dash{\A_e}$ bimodules, the isomorphism $\varphi_e$ can be read off directly from \eqref{eq:trivial isomorphism} as
        \begin{equation}\label{eq:toric iso 1}
            \begin{split}
                \varphi_e:\ &0\mapsto M_0,\\
                &1\mapsto M_1.
            \end{split}
        \end{equation} 
        The corresponding transformation $\T_e=\idm$.
        \item For ${\A_g}\dash{\A_g}$ bimodules, we have only one choice for the isomorphism $\varphi_g$:
        \begin{equation}\label{eq:toric iso 2}
            \begin{split}
                \varphi_g:\ &0\mapsto M_+,\\
                &1\mapsto M_-.
            \end{split}
        \end{equation}
    \end{enumerate}
    Plug $\varphi_g$ into \eqref{eq:compositionmap}, we can construct the transformation $\mathcal{T}_g$, which transforms the $\Vec(\Z_2)$ fusion vertices as:
    \begin{equation}
        \begin{split}
            \T_g:\ &\Vertex{0}{0}{0}\ \to\ \Vertex{0}{0}{0}+\Vertex{0}{1}{1}+\Vertex{1}{0}{1}+\Vertex{1}{1}{0},\\
            &\Vertex{0}{1}{1}\ \to\ \Vertex{0}{0}{0}+\Vertex{0}{1}{1}+i\Vertex{1}{0}{1}-i\Vertex{1}{1}{0}.
        \end{split}
    \end{equation}
Note that here we only show how $\T_g$ transforms fusion vertices, but the symmetry transformation is defined how $\T_g$ transforms the states on the entire lattice. In the current example, we can show that $\T_g$ is indeed a global symmetry transformation because it exchanges the toric code anyon types $e$ and $m$ while retains $1$ and $f$.
\end{example}

\subsection{Step 4: Find all the Transformations}\label{subsec:composition}

In the previous step, we constructed a nontrivial transformation operator $\T_g$ on the pure topological order $\Cent(\Fus)$. We can compose $\T_g$ with itself repeatedly to obtain a set of transformations closed under composition. We deal with the case where such a set forms a group $G$ under composition. We then use group elements, such as $g_i$, to label such transformation operators, $\T_{g_i}$.

For each such transformation $\T_{g_i}$, we need to find the corresponding Frobenius algebra $\A_{g_i}$, which gives rise to a duality map $\D_{g_i}$, whose composition with an isomorphism $\varphi_{g_i}$ yields the $\T_{g_i}$. Only when we find all the $(\A_{g_i},\varphi_{g_i})$ for all $g_i\in G$, can we proceed to the next step.

\begin{example}{EM-Exchange Symmetry Enriched Toric Code}{4}
    For the toric code example, we have
    \begin{equation}
        \T_g^2=\idm.
    \end{equation}
    Hence, $\{\T_e,\T_g\}=\Z_2$. For convenience, we use $\Z_2$ group elements to label them: $\T_+:=\T_e$, $\T_-:=\T_g$.
\end{example}

\subsection{Step 5: Construct the Input Multifusion Category of SET}\label{sec:construct multifusion}

For a set of transformations, $\{\T_{g_i}|g_i\in G\}$, we can use the corresponding set of Frobenius algebras $\{\A_{g_i}|g_i\in G\}$ to construct the input multifusion category for the SET model.

We can calculate all $\A_{g_i}\dash A_{g_j}$ simple bimodules for any $g_i,g_j\in G$ and use them to build a multifusion category. Generally, such a multifusion category takes the form:
\eqn{\M=\mat{
    {_{\A_e}\Bimod_{\A_e}(\Fus)} & {_{\A_e}\Bimod_{\A_{g_i}}(\Fus)} & {_{\A_e}\Bimod_{\A_{g_j}}(\Fus)} & \cdots \\
    {_{\A_{g_i}}\Bimod_{\A_e}(\Fus)} & {_{\A_{g_i}}\Bimod_{\A_{g_i}}(\Fus)} & {_{\A_{g_i}}\Bimod_{\A_{g_j}}(\Fus)} & \cdots \\
    {_{\A_{g_j}}\Bimod_{\A_e}(\Fus)} & {_{\A_{g_j}}\Bimod_{\A_{g_i}}(\Fus)} & {_{\A_{g_j}}\Bimod_{\A_{g_j}}(\Fus)} & \cdots \\
    \vdots & \vdots & \vdots & \ddots
    },\label{multifusion-matrix}}
which serves as the input of the SET string-net model. Here, each diagonal element, such as ${_{\A_{g_i}}\Bimod_{\A_{g_i}}(\Fus)}$, characterizes a $G$-graded sector.

\begin{example}{EM-Exchange Symmetry Enriched Toric Code}{5}
    Since $\T_+,\T_-$ form a $\Z_2$ group, there are only $2$ $G$-graded sectors in the toric code example. Thus, the multifusion category takes the form
    \eqn{\label{eq:Z2multifusion}
    \M=\mat{
    \{M_0,M_1\} & \{M_{\sigma}\}\\
    \{M^*_{\sigma}\} & \{M_+,M_-\}\\
    },
    }    
    which, together with the isomorphisms \eqref{eq:toric iso 1} and \eqref{eq:toric iso 2}, is the input data of the EM-exchange symmetry-enriched toric-code model.
\end{example}

\subsection{Step 6: Global Symmetry Transformation of SET}

After constructing the input multifusion category $\M$ \eqref{multifusion-matrix}, we need to establish the global symmetry transformation $\G_g$($g\in G$) of the model:
\begin{equation}\label{eq:symm}
    \G_g:=\bigotimes_{l\in P}\G_g^l,
\end{equation}
where $\G_g^l$ is an onsite operator acting on the d.o.f on edge (tail) $l$. Local actions $\G_g^l$ can be obtained from $\varphi_g$ via the following procedure:
\begin{enumerate}
    \item Local actions $\G_g^l$ should preserve the fusion rules of $\M$. Namely,
    \begin{equation}\label{eq:symconstraint}
        \G_g^l(x)\otimes\G_g^l(y)=\G_g^l(x\otimes y),\ \forall x,y\in\M.
    \end{equation}   
    \item The degrees of freedom on an edge/tail is a bimodule and can take value in all diagonal/off-diagonal sets of the multifusion matrix. The action of $\G_g^l$ on a diagonal bimodule is completely determined by the isomorphism $\varphi_g$. Let's denote a diagonal bimodule by an object $a\in\Fus$ indexed by a group element $g\in G$:
    \begin{equation}
        a_{gg}:=\varphi_g(a).
    \end{equation}
    Then, we have
    \begin{equation}\label{eq:sym on diag}
        \G_g^l:\ a_{g_ig_i}\mapsto a_{(gg_i)(gg_i)},\ \forall a\in\Fus.
    \end{equation}
    \item The $\G_g^l$'s action on the off-diagonal elements can be obtained through the constraint \eqref{eq:symconstraint} together with the symmetry action on diagonal bimodules \eqref{eq:sym on diag}. It is worth noting that the $\G_g^l$'s action on the off-diagonal elements may not be unique.
\end{enumerate}


\begin{example}{EM-Exchange Symmetry Enriched Toric Code}{6}
    In this example, there is only one solution of the nontrivial onsite operator $\G^l_-$:
    \begin{equation}\label{eq:Z2 symm toric}
        \begin{split}
            \G^l_-:\ &M_0\mapsto M_+,\ M_1\mapsto M_-,\ M_{\sigma}\mapsto M^*_{\sigma},\\
            &M_+\mapsto M_0,\ M_-\mapsto M_1,\ M^*_{\sigma}\mapsto M_{\sigma}.
        \end{split}
    \end{equation}
    The nontrivial global symmetry operator $\G_-$ is then the tensor product \eqref{eq:symm} of $\G^l_-$ for all edges/tails $l$.
\newline

    This SET phase has 4 types of anyons: $(1,1)$, $(e,m)$, $(m,e)$, $(f,f)$, each being a pair of the toric code anyons. Detailed information can be found in Appendix \ref{subsec:Z2toric ztensor}. As the example in Strategy I, the pairs show how the anyon appears to be in different $G$-graded sectors. For example, $(e,m)$ appears to be $e$ ($m$) in the $+$ ($-$) sector. It is obvious that the global symmetry transformation $\G_-$ exchanges $(e,m)$ and $(m,e)$ while preserving $(1,1)$ and $(f,f)$.
\end{example}

\subsection{Step 7: Gauging}\label{subsec:gauging}

To gauge the global symmetry, the $G$-graded sectors related by global symmetry transformations should be regarded physically equivalent, i.e., they should be identified. To this end, we need to identify all the simple objects of the multifusion category $\M$ related by symmetry transformations. That is, we ought to identify all the elements in the set $\{\G_g^l(x)|g\in G\}$. In this way, the multifusion category would be converted into a UFC $\Fus_p$ of the gauged model -- the model of the parent pure topological order. We can obtain the fusion rules and $6j$-symbols of $\Fus_p$ from those of the multifusion category.

\begin{example}{EM-Exchange Symmetry Enriched Toric Code}{7}

For our EM-exchange toric code SET, according to \eqref{eq:Z2 symm toric}, $M_0$ ($M_1$) and $M_+$ ($M_-$) are related by the symmetry transformation $\G_-$. So, we shall identify $M_0$ ($M_1$) and $M_+$ ($M_-$) and name the identified object as $1$ ($\psi$). Besides, $M_{\sigma}$ and $M^*_{\sigma}$ should also be identified and can be named as $\sigma$. As such, $\Fus_p=\{1,\sigma,\psi\}$, whose fusion rules of are shown in Table \ref{tab:Z2SETfusion}. 

\begin{table}[H]
    \centering
    \begin{tabular}{|c|c|c|c|}
    \hline
        $\times$ & $1$ & $\sigma$ & $\psi$ \\
        \hline
        $1$ & $1$ & $\sigma$ & $\psi$ \\
        \hline
        $\sigma$ & $\sigma$ & $1+\psi$ & $\sigma$ \\
        \hline
        $\psi$ & $\psi$ & $\sigma$ & $1$ \\
        \hline
    \end{tabular}
    \caption{The fusion table of input fusion category of parent phase}
    \label{tab:Z2SETfusion}
\end{table}

The nonzero $6j$-symbols of $\Fus_p$ can be obtained from the $6j$-symbol of $\M$ \eqref{SET_toric_6j}:
\begin{equation}
    \begin{split}
        &G^{111}_{111}=G^{\psi\psi1}_{\psi\psi1}=G^{111}_{\psi\psi\psi}=1,\ G^{111}_{\sigma\sigma\sigma}=G^{1\psi\psi}_{\sigma\sigma\sigma}=\frac{1}{\sqrt[4]{2}},\\
        &G^{\sigma\sigma1}_{\sigma\sigma1}=G^{\sigma\sigma1}_{\sigma\sigma\psi}=\frac{1}{\sqrt{2}},\ G^{\sigma\sigma1}_{\sigma\sigma1}=-\frac{1}{\sqrt{2}}.
    \end{split}
\end{equation}

Clearly, $\Fus_p$ is precisely the Ising UFC. Therefore, the parent phase is the doubled Ising phase. This result agrees with the literature\cite{cheng2017exactly,heinrich2016symmetry}.

\end{example}

\section{Local Excitations}\label{sec:local excitation}
Our model hosts local excitations, on top of which the topological excitations, i.e., anyons, are defined. Such local excitations are a generic feature of SET phases. A \emph{local excitation} in an SET phase is one that retains the system's state when it completes any loop path -- be it noncontractible, crossing symmetry domain walls, or encircling other excitations.

In an SET phase, not only the nontrivial anyon types but also the topological vacuum (trivial anyon type) must be defined modulo the local excitations. So to understand the local excitations in our SET model, we need to examine how the topological vacuum in our model is defined. 

Let's briefly recall how the topological vacuum in a pure topological phase $\Cat$ described by the HGW model with input UFC $\Fus$ is defined. The anyon types in $\Cat$ are the simple objects in the Drinfeld center $\Cent(\Fus)=:\Cat$ of $\Fus$. Such a simple object is an object (not necessarily simple) in $\Fus$ equipped with a number of $z$-tensors. The topological vacuum is a simple object of $\Cent(\Fus)$ that braids trivially with all the simple objects of $\Cent(\Fus)$. In particular, the $z$-tensors associated with the topological vacuum have all their tensor components being fixed $\C$-numbers. 

In contrast, consider an SET phase $\Cat_G$ described by our model with an input multifusion category $\M$. The anyon types in $\Cat_G$ are the simple objects in the Drinfeld center $\Cent(\M)$ of $\M$. Such a simple object is an object (not necessarily simple) in $\M$ equipped with a $z$-tensor. The topological vacuum in the SET phase however comes with a $z$-tensor whose tensor components are not completely fixed $\C$-numbers but contain certain continuous degrees of freedom. As we will show, these degrees of freedom precisely characterize the local excitations. The $z$-tensors associated with nontrivial anyons types also contain these degrees of freedom. One can imagine that when an SET phase anyon hops in the system, it always carries certain local excitations. Nevertheless, the topological properties, such as braiding and fusion, of the anyons are not affected by the local excitations at all.

We claim that the creation operator of any type of local excitation in an SET phase commutes with the global symmetry for the following reason. To reduce the clutter, it suffices to consider SET phases on the sphere because such an SET phase has a unique ground state, i.e., a topological vacuum state free of local excitations. A universal property of such SET phases is that any of its topological vacuum states should be invariant under the global symmetry transformations. Now that acting on a topological vacuum state by the creation operator $W$ of any type of local excitation on a topological vacuum state is again a topological vacuum state, we can conclude that
\begin{equation}\label{eq:WcommuteG}
    [W,\mathcal{G}] =0,
\end{equation}
where $\mathcal{G}$ collectively denotes the global symmetry transformations in the SET phase. This commutativity holds on any topology where an SET phase may be defined. 

The commutativity \eqref{eq:WcommuteG} in fact plays a constraint on the vacuum $z$-tensors' degrees of freedom characterizing local excitations: As will be seen, each such degree of freedom can take only a set of discrete values, each characterized by an irreducible representation of symmetry group $G$.

To facilitate our understanding of local excitations in general cases, let us study the following simple example.

\begin{exampleb}
Consider again the EM-exchange symmetry-enriched toric-code model. The $z$-tensors for the topological vacuum (trivial anyon) $(1,1)$ are
\begin{equation}\label{eq:vacuum}
    \begin{split}
        &z^{(1,1);M_0}_{M_0M_0M_0}=z^{(1,1);M_1}_{M_0M_0M_1}=z^{(1,1);M_+}_{M_+M_+M_+}=z^{(1,1);M_-}_{M_+M_+M_-}=1,\\
        &z^{(1,1);M_{\sigma}}_{M_0M_+M_{\sigma}}=e^{i\theta},\ z^{(1,1);M^*_{\sigma}}_{M_+M_0M^*_{\sigma}}=e^{-i\theta},\ \theta\in[0,2\pi),
    \end{split}
\end{equation}
where a $U(1)$ degree of freedom $e^{i\theta}$ clearly exists. Using these $z$-tensors, we can construct the trivial-anyon creation operator $W^{(1,1)}$ (see Appendix \ref{sec:spec}), which also contains the $U(1)$ degree of freedom $e^{i\theta}$. Operator $W^{(1,1)}$ creates no anyons at all when acting on a topological vacuum or ground state; however, it can excite a ground state unless $\theta=0$. Hence, $W^{(1,1)}$ creates a pair of excitations when $\theta\neq 0$, and since no anyons are created, such these excitations are \textit{local excitations}, which are characterized by $\theta$. To emphasize, we redenote $W^{(1,1)}$ by $W^{L,\theta}_P$, and it takes the form 
\begin{equation}\label{eq:Z2 local excitation}
    W^{L,\theta}_P=\prod_{l\in P}e^{i\theta(\delta_{M_{\sigma},l} -\delta_{M^*_{\sigma},l})}.
\end{equation}
The superscript $L$ of $W^{L,\theta}_P$ indicates ``local'', and subscript $P$ labels the path along which $W^{L,\theta}_P$ is defined. When $\theta\neq 0$, $W^{L,\theta}_P$ creates a pair of local excitations at the two ends of path $P$, which intersects a number of lattice edges $l$. The delta-function $\delta_{M_{\sigma},l}$ ($\delta_{M^*_{\sigma},l}$) returns 1 if the degree of freedom on edge $l$ is $M_{\sigma}$ ($M^*_{\sigma}$) and otherwise returns $0$. In other words, $\delta_{M_{\sigma},l}$ detects whether edge $l$ is a domain wall. 
\newline

The EM-exchange transformation $\mathcal{G}$ acts on the local excitation creation operator $W^{L,\theta}_P$ as
\begin{equation}
    \mathcal{G} W^{L,\theta}_P \mathcal{G}^{-1} = W^{L,-\theta}_P.
\end{equation}
According to Eq.~\eqref{eq:WcommuteG}, the operator $W^{L,\theta}_P$ must commute with $\G$, which imposes a constraint that only the values
\begin{equation}
    \theta = 0, \pi
\end{equation}
are allowed in SET phases.
\newline

Clearly, $W^{L,0}_P = \idm$, which is trivial; however, $W^{L,\pi}_P$ creates at the two endpoints of the path $P$ a pair of nontrivial excitations, which are indeed excitations according to Hamiltonian in Eq. \eqref{eq:hamiltonian}. Given any basis state of the Hilbert space, i.e., a lattice state in which all edge/tail degree of freedom are fixed, and a $P$ starts in the $+$ ($-$) sector and end in the $-$ ($+$) sector, acting $W^{L,\pi}_P$ on the state multiplies a $-1$ sign to the state. If path $P$ lies completely within the same $G$-graded sector, no matter a $+$ or $-$ sector, $W^{L,\pi}_P$ does not affect the state. On the other hand, if $P$ is a loop, $W^{L,\pi}_P$ acts trivially on any state, complying with our definition of local excitations.
\newline

In this case, there are only two types of local excitations, respectively characterized by $\theta=0$ (trivial type) and $\theta=\pi$ (nontrivial type). The two permitted values of $\theta$ also label two irreducible representations of $\Z_2$.
\end{exampleb}

Now we consider general cases. Note that in the example above and in fact in any SET phase whose global symmetry is Abelian, all local excitations are one-dimensional. Nevertheless, when in an SET phase that has a nonabelian global symmetry, local excitations may be multi-dimensional. 

We leave the detailed derivation in Appendix \ref{sec:local excitation derivation} and show only the results here. For an SET model with a generic input multifusion category $\M$ and a set of isomorphisms that yield a symmetry group $G$, the local-excitation types in the SET model are characterized by irreducible representations $\rho$ of group $G$. The creation operators for local excitation $\rho$ take the form
\begin{equation}
    [W^{L,\rho}_P]_{a_0a_n}=\sum_{a_1,a_2,\cdots ,a_{n-1}=1}^{N_\rho}\prod_{k=0}^{n-1}[W^{L,\rho}_{l_{k}}]_{a_ka_{k+1}},
\end{equation}
\begin{equation}
    [W^{L,\rho}_l]_{ab}\ExcitedC\ :=\rho_{ab}(g_i^{-1}g_j)\ExcitedD,
\end{equation}
where Path $P$ intersects $n$ lattice edges, $l_0$ through $l_{n-1}$. Here, $N_{\rho}$ is the dimension of representation $\rho$ and hence of local excitation $\rho$'s internal space, spanned by $\{(1_{g_i g_i})_a| a = 1, 2,\dots, N_\rho\}$. 

In an SET phase, the topological properties of anyons, such as anyons' quantum dimensions, fusion rules, and braiding statistics, are independent of local excitations. In our SET model with an input multifusion category $\M$, however, to define anyon types, we need to find the minimal solutions of the $z$-tensors for the Drinfeld center of $\M$. These $z$-tensors inevitably carry the degrees of freedom that characterize the local excitations. To define the anyon types unambigously and extract their topological properties, we can fix these degrees of freedom to certain definite values. For the trivial anyon type, there is a canonical way to do the fixing by setting all the local degrees of freedom to be trivial. For example, we set $\theta=0$ in defining the trivial anyon in the EM-exchange symmetry-enriched toric-code model. For nontrivial anyons, we can freely choose the values of the degrees of freedom. We have done so in all the $z$-tensors in Appendix \ref{subsec:Z2Z3 ztensor} and \ref{subsec:Z2S3 ztensor}.

\section{Other Examples}\label{sec:examples}

\subsection{Strategy I Example: Constructing and Gauging the \eqs{S_3}-Symmetry-Enriched \eqs{\Z_2 \times \Z_2} Quantum-Double Phase}\label{sec:S3symmetry}

In this section, we provide a more nontrivial example for strategy I: constructing and gauging the $S_3$ symmetry enriched $\Z_2\times\Z_2$ quantum-double phase. In the literature to date, this example is the first explicitly constructed and studied nonabelian-symmetry-enriched topological phase.

\subsubsection{Constructing SET Model}

The input UFC of $\Z_2 \times \Z_2$ quantum-double phase is $\Vec(\Z_2 \times \Z_2)$, whose four simple objects are $\{1,a,b,c\}$, fusing as $c=ab$ and $a^2=b^2=c^2=1$. There are $16$ anyon types in this topological phase: $(a,b),\ a,b\in\{1,e,m,f\}$, where $1,e,m$, and $f$ are the $\Z_2$ toric code anyon types. In fact, the $\Z_2\times\Z_2$ quantum-double phase can be viewed as two decoupled layers of the $\Z_2$ toric code phase. More details of this phase can be found in Appendix \ref{subsec:Z2*Z2 ztensor}.

The UFC $\Vec(\Z_2 \times \Z_2)$ has $6$ outer automorphisms:
\eqn[eq:automophismS3]{
\phi_1:&\ &1\to1,\ a\to a,\ b\to b,\ c\to c;\\
\phi_r:&\ &1\to1,\ a\to b,\ b\to c,\ c\to a;\\
\phi_{r^2}:&\ &1\to1,\ a\to c,\ b\to a,\ c\to b;\\
\phi_s:&\ &1\to1,\ a\to b,\ b\to a,\ c\to c;\\
\phi_{rs}:&\ &1\to1,\ a\to c,\ b\to b,\ c\to a;\\
\phi_{sr}:&\ &1\to1,\ a\to a,\ b\to c,\ c\to b.
}
They form a $S_3$ group: $\Out(\Vec(\Z_2\times\Z_2))=S_3$, which can be the symmetry group of $\Z_2 \times \Z_2$ quantum-double phase. This symmetry permutes the $\Z_2\times\Z_2$ quantum-double anyons types as in Table \ref{tab:Z2timesZ2_S3symmetry}.
\begin{table}[ht]
    \centering
    \begin{tabular}{|c|c|c|c|c|c|}
    \hline\hline
        $1$ & $r$ & $r^2$ & $s$ & $rs$ & $sr$\\
        \hline\hline
        $(1,1)$ & $(1,1)$ & $(1,1)$ & $(1,1)$ & $(1,1)$ & $(1,1)$ \\
        \hline
        $(e,1)$ & $(e,e)$ & $(1,e)$ & $(1,e)$ & $(e,1)$ & $(e,e)$ \\
        \hline
        $(1,e)$ & $(e,1)$ & $(e,e)$ & $(e,1)$ & $(e,e)$ & $(1,e)$ \\
        \hline
        $(e,e)$ & $(1,e)$ & $(e,1)$ & $(e,e)$ & $(1,e)$ & $(e,1)$ \\
        \hline
        $(m,1)$ & $(1,m)$ & $(m,m)$ & $(1,m)$ & $(m,m)$ & $(m,1)$ \\
        \hline
        $(f,1)$ & $(e,f)$ & $(m,f)$ & $(1,f)$ & $(f,m)$ & $(f,e)$ \\
        \hline
        $(m,e)$ & $(e,m)$ & $(f,f)$ & $(e,m)$ & $(f,f)$ & $(m,e)$ \\
        \hline
        $(f,e)$ & $(1,f)$ & $(f,m)$ & $(e,f)$ & $(m,f)$ & $(f,1)$ \\
        \hline
        $(1,m)$ & $(m,m)$ & $(m,1)$ & $(m,1)$ & $(1,m)$ & $(m,m)$ \\
        \hline
        $(e,m)$ & $(f,f)$ & $(m,e)$ & $(m,e)$ & $(e,m)$ & $(f,f)$ \\
        \hline
        $(1,f)$ & $(f,m)$ & $(f,e)$ & $(f,1)$ & $(e,f)$ & $(m,f)$ \\
        \hline
        $(e,f)$ & $(m,f)$ & $(f,1)$ & $(f,e)$ & $(1,f)$ & $(f,m)$ \\
        \hline
        $(m,m)$ & $(m,1)$ & $(1,m)$ & $(m,m)$ & $(m,1)$ & $(1,m)$ \\
        \hline
        $(f,m)$ & $(f,e)$ & $(1,f)$ & $(m,f)$ & $(f,1)$ & $(e,f)$ \\
        \hline
        $(m,f)$ & $(f,1)$ & $(e,f)$ & $(f,m)$ & $(f,e)$ & $(1,f)$ \\
        \hline
        $(f,f)$ & $(m,e)$ & $(e,m)$ & $(f,f)$ & $(m,e)$ & $(e,m)$ \\
        \hline\hline
    \end{tabular}
    \caption{Permutation pattern of the $\Z_2\times\Z_2$ quantum-double anyons under the $S_3$ symmetry.}
    \label{tab:Z2timesZ2_S3symmetry}
\end{table}

The outer automorphisms \eqref{eq:automophismS3} allows us to promote $\Vec(\Z_2\times\Z_2)$ into a multifusion category $\mathcal{M}$:
\begin{equation}
    \M=\bigoplus_{g_i,g_j\in S_3}\Vec(\Z_2\times\Z_2)_{g_ig_j},
\end{equation}
which has $36$ copies of $\Vec(\Z_2\times\Z_2)$.

Use $\mathcal{M}$ as the input multifusion category, we can realize the $S_3$ symmetry-enriched $\Vec(\Z_2\times\Z_2)$ string-net model. The symmetry operators $\{\G_g|g\in S_3\}$ are instances of the general ones defined in \eqref{eq:symoperator}.

\subsubsection{Gauging the SET Model}

Following the gauging procedure in Section \ref{sec:gauging}, we can gauge the global symmetry $S_3$ of the $S_3$-symmetry-enriched $\Z_2\times \Z_2$ quantum double phase and obtain the parent enlarged HGW model of the parent pure topological phase. We find that the parent model's input UFC $\Fus_p$ has $24$ simple objects, presented as
\begin{equation}\label{eq:gaugedS4}
\left(a,b,r,s | a^2,b^2,s^2,r^3,aba^{-1}b^{-1},rsrs,sasb,rar^2b,rbr^2ab\right),
\end{equation}
where $(a,b)$ are the $\Z_2\times\Z_2$ generators, and $r,s$ are the $S_3$ generators. One can readily see that the UFC described in \eqref{eq:gaugedS4} is the group algebra $\Vec(S_4)$. Indeed, we can make the following identification.
\begin{equation}
    \begin{split}
        & 1=(1)(2)(3)(4);\\
        & a=(14)(23),\ b=(13)(24),\ c=(12)(34);\\
        & r=(123)(4),\ r^2=(132)(4);\\
        & s=(12)(3)(4),\ rs=(13)(2)(4),\ sr=(23)(1)(4).
    \end{split}
\end{equation}
In fact, mathematically,
\begin{equation}
    (\mathbb{Z}_2\times\mathbb{Z}_2)\rtimes S_3=S_4.
\end{equation}
Therefore, gauging the $S_3$-symmetry-enriched $\Z_2 \times \Z_2$ quantum-double phase results in the $S_4$ quantum-double phase.

\subsection{Strategy II Example: Constructing and Gauging the \eqs{\mathbb{Z}_2}-Symmetry-Enriched \eqs{\Z_2\times\Z_2} Quantum-Double Phase}\label{sec:Z2timesZ2}

The input UFC of the $\Z_2\times\Z_2$ quantum-double model is $\Vec(\Z_2\times\Z_2)$, as introduced in \ref{sec:S3symmetry}.

\subsubsection{Constructing SET Model}

Following the procedure in Section \ref{sec:bimodule}, we can construct the $\Z_2$-symmetry-enriched $\Z_2\times\Z_2$ quantum-double model. We first need two frobenius algebras in $\Vec(\Z_2\times\Z_2)$. We always choose the trivial Frobenius algebra $\A_+$ with $L_{\A_+}=\{1\}$. We choose the other Frobenius algebra we need as $\A_-$, whose basis elements are $L_{\A_-}=\{1,a,b,c\}$, with $f$-functions being $f_{xyz}=\delta_{xyz},\forall x,y,z\in \Vec(\Z_2\times\Z_2)$.

The simple bimodules of $\A_+$ and $\A_-$ are listed as follows.
\begin{enumerate}
    \item Since $\A_+$ is the trivial Frobenius algebra in $\Vec(\Z_2\times \Z_2)$, the ${\A_+}\dash{\A_+}$ bimodules form a UFC ${_{\A_+}\Bimod_{\A_+}}(\Vec(\Z_2\times \Z_2))$ that is identified with $\Vec(\Z_2\times \Z_2)$. We denote the $4$ simple bimodules in ${_{\A_+}\Bimod_{\A_+}}(\Vec(\Z_2\times \Z_2))$ as $M_1,M_a,M_b,M_c$, which are respectively identified with the $\Vec(\Z_2\times \Z_2)$ simple objects $1,a,b,c$ according to \eqref{eq:trivial bimodule}.
    \item There are $4$ simple ${\A_-}\dash{\A_-}$ bimodules: $M_I,M_A,M_B,M_C$, which form a UFC that is isomorphic to $\Vec(\Z_2\times \Z_2)$. We denote this UFC by ${_{\A_-}\Bimod_{\A_-}}(\Vec(\Z_2\times \Z_2))$, whose categorical data is detailed in Appendix \ref{Z2timesZ2}. Here, we note that $d_{M_I}=d_{M_A}=d_{M_B}=d_{M_C}=1$.
    \item There is only one simple ${\A_+}\dash{\A_-}$ bimodule $M_{\chi}$, with $d_{M_{\chi}}=2$.
    \item There is only one simple ${\A_-}\dash{\A_+}$ bimodule $M^*_{\chi}$, which is just the opposite bimodule (see Eq. \eqref{eq:opposite bimodule}) of $M_{\chi}$. We have $d_{M_{\chi}^*}=2$.
\end{enumerate}

The fusion rules and $6j$-symbols between all the bimodules listed above can be found in Appendix \ref{Z2timesZ2}.

The isomorphism $\varphi_+$ from $\Vec(\Z_2\times \Z_2)$ to $_{A_+}\Bimod_{\A_+}(\Vec(\Z_2\times \Z_2))$ comes from \eqref{eq:trivial isomorphism}:
\begin{equation}\label{eq:iso1}
    \varphi_+:\ 1\mapsto M_1,\ a\mapsto M_a,\ b\mapsto M_b,\ c\mapsto M_c.
\end{equation}
The isomorphism $\varphi_-$ from $\Vec(\Z_2\times \Z_2)$ to $_{A_-}\Bimod_{\A_-}(\Vec(\Z_2\times \Z_2))$ can be chosen as
\begin{equation}\label{eq:iso2}
    \varphi_-:\ 1\mapsto M_I,\ a\mapsto M_A,\ b\mapsto M_B,\ c\mapsto M_C.
\end{equation}

From the bimodules and the isomorphisms, we can construct the transformations $\T_+$ and $\T_-$. $\T_-$ do the EM-exchange transformation on each $\Z_2$ toric-code layer of the $\Z_2\times\Z_2$ quantum-double phase while at the same time exchanging the two layers. The composition rules are $\T_-^2=\T_+=\idm$. Since $\{\T_+,\T_-\}$ form a $\Z_2$, according to the discussion in Section \eqref{subsec:composition}, the global symmetry of the SET phase being constructed should be a $\Z_2$ symmetry.

The multifusion category of these bimodules takes the form:
\begin{equation}
    \mathcal{M}=\left(\begin{array}{cc}
        \{M_1,M_a,M_b,M_c\} & \{M_{\chi}\} \\
        \{M^*_{\chi}\} & \{M_I,M_A,M_B,M_C\}
    \end{array}\right),
\end{equation}
which, together with isomorphisms \eqref{eq:iso1} and \eqref{eq:iso2}, is the input data of the $\Z_2$-symmetry-enriched $\Z_2\times\Z_2$ quantum-double model.

The nontrivial symmetry transformation operator $\G_-$ acts on the multifusion simple objects as
\begin{align*}
    \G_{-}:\ &M_1\mapsto M_I,\ M_a\mapsto M_A,\ M_b\mapsto M_B,\ M_c\mapsto M_C,\ M_{\chi}\mapsto M_{\chi}^*,\\
    &M_I\mapsto M_1,\ M_A\mapsto M_a,\ M_B\mapsto M_b,\ M_C\mapsto M_c,\ M^*_{\chi}\mapsto M_{\chi}.
\end{align*}

\subsubsection{Gauging this SET Model}

According to Section \ref{subsec:gauging}, to gauge this SET model, we need to identify $M_1$ with $M_I$, $M_a$ with $M_A$, $M_b$ with $M_B$, $M_c$ with $M_C$, and $M_{\chi}$ with $M^*_{\chi}$. We name the five pairs of identified objects respectively as $0,1,2,3$, and $4$, which are the simple objects of the parent input UFC $\Fus_p=\{0,1,2,3,4\}$, whose fusion rules are recorded in Table \ref{tab:Z2timesZ2bimod}. The nonzero $6j$-symbols of $\Fus_p$ read:
\begin{table}[ht]
    \centering
    \begin{tabular}{|c|c|c|c|c|c|}
    \hline
        $\times$ & $0$ & $1$ & $2$ & $3$ & $4$ \\
        \hline
        $0$ & $0$ & $1$ & $2$ & $3$ & $4$ \\
        \hline
        $1$ & $1$ & $0$ & $3$ & $2$ & $4$ \\
        \hline
        $2$ & $2$ & $3$ & $0$ & $1$ & $4$ \\
        \hline
        $3$ & $3$ & $2$ & $1$ & $0$ & $4$ \\
        \hline
        $4$ & $4$ & $4$ & $4$ & $4$ & $0+1+2+3$ \\
        \hline
    \end{tabular}
    \caption{Fusion table of $\Rep(D_4)$}
    \label{tab:Z2timesZ2bimod}
\end{table}
\begin{equation}
    \begin{split}
        &G^{000}_{000}=G^{000}_{111}=G^{000}_{222}=G^{000}_{333}=G^{011}_{011}=G^{022}_{022}=G^{033}_{033}=1,\\
        &G^{011}_{322}=G^{011}_{233}=G^{022}_{133}=G^{123}_{123}=G^{132}_{132}=1,\\
        &G^{000}_{444}=G^{011}_{444}=G^{022}_{444}=G^{033}_{444}=G^{123}_{444}=\frac{1}{\sqrt{2}},\\
        &G^{044}_{044}=G^{044}_{144}=G^{044}_{244}=G^{044}_{344}=G^{144}_{144}=G^{244}_{244}=G^{344}_{344}=\frac{1}{2},\\
        &G^{144}_{244}=G^{144}_{344}=G^{244}_{344}=-\frac{1}{2}.
    \end{split}
\end{equation}
Hence, we can see that $\Fus_p=\Rep(D_4)$.

\subsection{Strategy II Example: Constructing and Gauging the EM-Exchange Symmetry-Enriched \eqs{S_3} Quantum-Double Phase}\label{sec:S3}

The input UFC of $S_3$ quantum-double phase is $\Vec(S_3)$, whose $6$ simple objects are the group elements of $S_3$: $\{1,r,r^2,s,rs,sr\}$, fusing as $r^3=1$ and $rs=sr^2$.

There are $8$ kinds of anyons in $\Vec(S_3)$ string-net model, denoted by $A,B,C,D,E,F,G$ and $H$ respectively. More details of this phase can be found in Appendix \ref{subsec:S3 ztensor}.

\subsubsection{Constructing SET Model}

Following the procedure in Section \ref{sec:bimodule}, we can construct the EM-exchange symmetry-enriched $S_3$ quantum-double model. We first need two Frobenius algebras in $\Vec(S_3)$. We always choose the trivial Frobenius algebra $\A_+$ with $L_{\A_+}=\{1\}$. We choose the other Frobenius algebra we need as $\A_-$, whose basis elements are $L_{\A_-}=\{1,r,r^2\}$, with $f$-functions being $f_{abc}=\delta_{abc},\ \forall a,b,c\in L_{\mathcal{\A_-}}$.

The simple bimodules of $\A_+$ and $\A_-$ are listed as follows.
\begin{enumerate}

    \item Since $\A_+$ is the trivial Frobenius algebra in $\Vec(S_3)$, the ${\A_+}\dash{\A_+}$ bimodules form a UFC ${_{\A_+}\Bimod_{\A_+}}(S_3)$ that is identified with $\Vec(S_3)$. We denote the $6$ simple bimodules in ${_{\A_+}\Bimod_{\A_+}}(\Vec(S_3))$ as $M_1,M_r,M_{r^2},M_{s},M_{rs},M_{sr}$, which are respectively identified with the $\Vec(S_3)$ simple objects $1,r,r^2,s,rs,sr$ according to \eqref{eq:trivial bimodule}.
    \item There are $6$ simple ${\A_-}\dash{\A_-}$ bimodules: $M_I,M_R,M_{R^2},M_S,M_{RS},M_{SR}$, which form a UFC that is isomorphic to $\Vec(S_3)$. We denote this UFC by ${_{\A_-}\Bimod_{\A_-}}(\Vec(S_3))$, whose categorical data is detailed in Appendix \ref{Z2timesZ2}. Here, we note that $d_{M_I}=d_{M_R}=d_{M_{R^2}}=d_{M_S}=d_{M_{RS}}=d_{M_{SR}}=1$.
    \item There are $2$ simple ${\A_+}\dash{\A_-}$ bimodules, $M_\alpha$ and $M_\beta$, with $d_{M_\alpha}=d_{M_\beta}=\sqrt{3}$.
    \item There is $2$ simple ${\A_-}\dash{\A_+}$ bimodules $M^*_\alpha$ and $M^*_\beta$, which are the opposite bimodules (see Eq. \eqref{eq:opposite bimodule}) of $M_\alpha$ and $M_\beta$ respectively. We have $d_{M^*_\alpha}=d_{M^*_\beta}=\sqrt{3}$.
\end{enumerate}

The fusion rules and $6j$-symbols between all the bimodules listed above can be found in Appendix \ref{S3}.

The isomorphism $\varphi_+$ from $\Vec(S_3)$ to $_{A_+}\Bimod_{\A_+}(\Vec(S_3))$ comes from \eqref{eq:trivial isomorphism}:
\begin{equation}\label{eq:iso1S3}
    \varphi_-:\ 1\mapsto M_1,\ r\mapsto M_r,\ r^2\mapsto M_{r^2},\ s\mapsto M_s,\ rs\mapsto M_{rs},\ sr\mapsto M_{sr}.
\end{equation}
The isomorphism $\varphi_-$ from $\Vec(S_3)$ to $_{A_-}\Bimod_{\A_-}(\Vec(S_3))$ can be chosen as
\begin{equation}\label{eq:iso2S3}
    \varphi_-:\ 1\mapsto M_I,\ r\mapsto M_R,\ r^2\mapsto M_{R^2},\ s\mapsto M_S,\ rs\mapsto M_{RS},\ sr\mapsto M_{SR}.
\end{equation}

From the bimodules and the isomorphisms, we can construct the transformations $\T_+$ and $\T_-$. $\T_-$ exchange anyon $C$ and $F$ in the $S_3$ quantum-double phase. The composition rules are $\T_-^2=\T_+=\idm$. Since $\{\T_+,\T_-\}$ form a $\Z_2$, according to the discussion in Section \eqref{subsec:composition}, the global symmetry of the SET phase being constructed should be a $\Z_2$ symmetry. Such a symmetry can fragment the internal space of the nonabelian anyons into symmetry definite states, such details will be discussed in a companion paper \cite{fu2025nonlinear}.

The multifusion category of these bimodules takes the form:
\begin{equation}
    \mathcal{M}=\left(\begin{array}{cc}
        \{M_1,M_r,M_{r^2},M_{s},M_{rs},M_{sr}\} & \{M_{\alpha},\ M_{\beta}\} \\
        \{M^*_{\alpha},\ M^*_{\beta}\} & \{M_I,M_R,M_{R^2},M_{S},M_{RS},M_{SR}\}
    \end{array}\right),
\end{equation}
which, together with the isomorphism \eqref{eq:iso2}, is the input data of the EM-exchange symmetry-enriched $S_3$ quantum-double phase.

The nontrivial symmetry transformation operator $\G_-$ acts on the multifusion simple objects as:
\begin{align*}
    \G_{-}:\ &M_1\mapsto M_I,\ M_r\mapsto M_R,\ M_{r^2}\mapsto M_{R^2},\ M_s\mapsto M_S,\\
    &M_{rs}\mapsto M_{RS},\ M_{sr}\mapsto M_{SR},\ M_{\alpha}\mapsto M_{\alpha}^*,\ M_{\beta}\mapsto M_{\beta}^*,\\
    &M_I\mapsto M_1,\ M_R\mapsto M_r,\ M_{R^2}\mapsto M_{r^2},\ M_S\mapsto M_s,\\
    &M_{RS}\mapsto M_{rs},\ M_{SR}\mapsto M_{sr},\ M_{\alpha}^*\mapsto M_{\alpha},\ M_{\beta}^*\mapsto M_{\beta}.\\
\end{align*}

\subsubsection{Gauging the SET Model}

According to Section \ref{subsec:gauging}, to gauge this SET model, we need to identify $M_1$ with $M_I$, $M_r$ with $M_R$, $M_{r^2}$ with $M_{R^2}$, $M_s$ with $M_S$, $M_{rs}$ with $M_{RS}$, $M_{sr}$ with $M_{SR}$, $M_{\alpha}$ with $M^*_{\alpha}$, and $M_{\beta}$ with $M^*_{\beta}$. We name the eight pairs of identified objects respectively as $1,r,r^2,s,rs,sr,\alpha$ and $\beta$, which are the simple objects of the parent input UFC $\Fus_p=\{1,r,r^2,s,rs,sr,\alpha,\beta\}$, whose fusion rules are recorded in Table \ref{tab:S3CF}. The nonzero $6j$-symbols of $\Fus_p$ read:
\begin{table}[ht]
    \centering
    \begin{tabular}{|c|c|c|c|c|c|c|c|c|}
    \hline
        $\times$ & $1$ & $r$ & $r^2$ & $s$ & $rs$ & $sr$ & $\alpha$ & $\beta$ \\
        \hline
        $1$ & $1$ & $r$ & $r^2$ & $s$ & $rs$ & $sr$ & $\alpha$ & $\beta$ \\
        \hline
        $r$ & $r$ & $r^2$ & $1$ & $rs$ & $sr$ & $s$ & $\alpha$ & $\beta$ \\
        \hline
        $r^2$ & $r^2$ & $1$ & $r$ & $sr$ & $s$ & $rs$ & $\alpha$ & $\beta$ \\
        \hline
        $s$ & $s$ & $sr$ & $rs$ & $1$ & $r^2$ & $r$ & $\beta$ & $\alpha$ \\
        \hline
        $rs$ & $rs$ & $s$ & $sr$ & $r$ & $1$ & $r^2$ & $\beta$ & $\alpha$ \\
        \hline
        $sr$ & $sr$ & $rs$ & $s$ & $r^2$ & $r$ & $1$ & $\beta$ & $\alpha$ \\
        \hline
        $\alpha$ & $\alpha$ & $\alpha$ & $\alpha$ & $\beta$ & $\beta$ & $\beta$ & $1+r+r^2$ & $s+rs+sr$ \\
        \hline
        $\beta$ & $\beta$ & $\beta$ & $\beta$ & $\alpha$ & $\alpha$ & $\alpha$ & $s+rs+sr$ & $1+r+r^2$ \\
        \hline
    \end{tabular}
    \caption{Fusion table of $\MA$}
    \label{tab:S3CF}
\end{table}
\begin{align*}
        &G^{a,b,c}_{d,e,f}=\delta_{abc^*}\delta_{cde}\delta_{a^*e^*f^*}\delta_{d^*b^*f},\ \forall a,b,c,d,e,f\in\{1,r,r^2,s,rs,sr\},\\
        &G^{a,b,c}_{\mu,\nu,\gamma}=\frac{1}{\sqrt[4]{3}}\delta_{abc}\delta_{\nu c\mu}\delta_{\gamma a\nu}\delta_{\mu b\gamma},\ \forall a,b,c\in\{1,r,r^2,s,rs,sr\},\ \mu,\nu,\gamma\in \{\alpha,\beta\},\\
        &G^{\alpha,\alpha,1}_{\alpha,\alpha,1}=G^{\alpha,\alpha,1}_{\alpha,\alpha,r}=G^{\alpha,\alpha,1}_{\alpha,\alpha,r^2}=\frac{1}{\sqrt{3}},\ G^{\alpha,\alpha,r}_{\alpha,\alpha,r}=G^{\alpha,\alpha,r^2}_{\alpha,\alpha,r^2}=\frac{e^{-i\frac{2\pi}{3}}}{\sqrt{3}},\ G^{\alpha,\alpha,r}_{\alpha,\alpha,r^2}=\frac{e^{i\frac{2\pi}{3}}}{\sqrt{3}},\\
        &G^{\beta,\beta,1}_{\beta,\beta,1}=G^{\beta,\beta,1}_{\beta,\beta,r}=G^{\beta,\beta,1}_{\beta,\beta,r^2}=\frac{1}{\sqrt{3}},\ G^{\beta,\beta,r}_{\beta,\beta,r}=G^{\beta,\beta,r^2}_{\beta,\beta,r^2}=\frac{e^{i\frac{2\pi}{3}}}{\sqrt{3}},\ G^{\beta,\beta,r}_{\beta,\beta,r^2}=\frac{e^{-i\frac{2\pi}{3}}}{\sqrt{3}},\\
        &G^{\alpha,\alpha,1}_{\beta,\beta,s}=G^{\alpha,\alpha,1}_{\beta,\beta,rs}=G^{\alpha,\alpha,1}_{\beta,\beta,sr}=G^{\alpha,\alpha,r}_{\beta,\beta,s}=G^{\alpha,\alpha,r^2}_{\beta,\beta,s}=\frac{1}{\sqrt{3}},\\
        &G^{\alpha,\alpha,r}_{\beta,\beta,rs}=G^{\alpha,\alpha,r^2}_{\beta,\beta,sr}=\frac{e^{-i\frac{2\pi}{3}}}{\sqrt{3}},\ G^{\alpha,\alpha,r}_{\beta,\beta,sr}=G^{\alpha,\alpha,r^2}_{\beta,\beta,rs}=\frac{e^{i\frac{2\pi}{3}}}{\sqrt{3}},\\
        &G^{\alpha,\beta,s}_{\alpha,\beta,s}=G^{\alpha,\beta,s}_{\alpha,\beta,rs}=G^{\alpha,\beta,s}_{\alpha,\beta,sr}=\frac{1}{\sqrt{3}},\ G^{\alpha,\beta,rs}_{\alpha,\beta,rs}=G^{\alpha,\beta,sr}_{\alpha,\beta,sr}=\frac{e^{i\frac{2\pi}{3}}}{\sqrt{3}},\ G^{\alpha,\beta,rs}_{\alpha,\beta,sr}=\frac{e^{-i\frac{2\pi}{3}}}{\sqrt{3}}.\\
\end{align*}

The two UFCs $\Fus_p$ and $A_5$ are Morita equivalent. This Morita equivalence is proven in Appendix \ref{Morita_equivalence}). As such, we shall denote $\Fus_p$ by $\MA$ to emphasize this Morita equivalence.

\section{Symmetry-Gauging Family}\label{family}

Our strategies I and II of constructing and gauging SET phases appear to differ in mathematics, they are in fact akin to each other by their nature in physics, in the sense that they do not impose an ad hoc global symmetry to a topological phase but instead extract the possible global symmetries intrinsically allowed by the defining data of the topological phase. In contrast, our strategy III (Appendix \ref{sec:symmfrac}) directly endows a topological phase with an ad hoc symmetry, which generally has nothing to do with the defining data of the phase. Therefore, it would be better to define a few important physical concepts that can help grasp the physics in and better distinguish the three strategies. This is what we are going to do as follows.  

\begin{enumerate}
    \item \textbf{Blood symmetry} and \textbf{Blood parent phase}. In Strategy I and II, we construct an SET phase solely from the input UFC $\Fus$ of an enlarged HGW model describing a pure topological phase. The global symmetry in such an SET phase arises merely from the corresponding $\Fus$; that is, it is intrinsic to the pure topological phase we begin with. For this reason, we term a global symmetry arising via these two strategies a \textbf{blood symmetry}. We use the \textbf{blood parent phase} as the term for the parent pure topological phase of this SET phase.
    
    \item \textbf{Adopted symmetry} and \textbf{adopted parent phase}. In strategy III (Appendix \ref{sec:symmfrac}), to construct an SET phase, we must impose on a child pure topological phase a global symmetry that may have nothing to do with the child phase. We call such a global symmetry an \textbf{adopted symmetry}. We term the the parent phase of such an SET phase a \textbf{step parent} phase and the child phase an \textbf{adopted child phase}.

    \item \textbf{Terminal topological phase} and \textbf{initial topological phase}. We may continue to apply strategy I or II to a parent UFC $\Fus_p$ obtained via strategy I or II from a given UFC $\Fus$. This way we may obtain a grand parent UFC $\Fus_{gp}$. We may iterate this procedure again and again until we cannot do this anymore, in the sense that the resultant ancestor UFC does not have any blood symmetry. The corresponding ancestor topological phase is a \textbf{terminal topological phase}, which occurs when the following two conditions hold.
    \begin{itemize}
    \item All the inequivalent, connected Frobenius algebras in the input UFC of the corresponding string-net model are all Morita equivalent.
    \item The input UFC has no nontrivial outer automorphism.  
    \end{itemize}

    Opposite to terminal topological phases, there exist initial topological phases. An \textbf{initial topological phase} is one that cannot be the parent phase obtainable via Strategy I or II. In other words, if an initial topological phase is the parent phase obtained by gauging the global symmetry of any topological phase, the global symmetry must be an adopted symmetry. 
    
    Consequently, since a trivial phase bears no blood symmetry at all or be the parent phase of any other phase, it is both a terminal and an initial topological (trivial) phase. 
    
    The doubled Ising phase is an example of terminal topological phase: the Ising UFC of the corresponding string-net model has only two inequivalence, connected Frobenius algebras, which happen to be Morita equivalent. The doubled Ising phase can come from gauging the EM-exchange-symmetry-enriched $\Z_2$ toric code phase (see Section \ref{sec:bimodule}). Here, the $\Z_2$ toric code is an initial topological phase. In fact, the $\Z_n$ quantum-double phase for any $n\geq 2$ is an initial topological phase. 

    \item \textbf{Symmetry-gauging family}. Via strategy I and/or strategy II, it is possible to connect a certain initial topological phase with a certain terminal topological phase by an oriented web of topological phases. This web has two special nodes, one being the initial topological phase while the other the terminal topological phase. In between these two special nodes are other nodes as pure topological phases. A directed link connecting two nodes represents constructing and gauging an SET phase to obtain the topological phase at the head node of the link from the topological phase at the tail node of the link. See Figure \ref{fig:family} for an illustration. We term such a web of topological phases a \textbf{symmetry-gauging family}, which forms a directed acyclic graph.
    
    As a more involved example, let us consider $\Z_3$ quantum-double phase as the initial topological phase, which can have two blood symmetries: the charge-conjugation symmetry $s$ due to the outer automorphism of $\Vec(\Z_3)$ and the EM-exchange symmetry $\sigma$ due to the two Frobenius algebras in $\Vec(\Z_3)$. The symmetry-gauging family with this initial topological phase comprises two paths as follows.
    \begin{enumerate}
        \item Path 1: First, construct the charge-conjugation symmetry-enriched $\Z_3$ quantum-double phase and gauge the symmetry, obtaining $S_3$ quantum-double phase as shown in Section \ref{sec:strategyI}. The $S_3$ quantum-double phase also has a blood symmetry -- the EM-exchange symmetry, which exchanges the anyon types $C$ and $F$. This symmetry is in fact due to that of the $\Z_3$ quantum double phase. Then, construct the EM-exchange symmetry-enriched $S_3$ quantum-double phase and gauge the symmetry, obtaining the doubled $\MA$ phase as shown in Section \ref{sec:bimodule}. The doubled $\MA$ phase is a terminal topological phase because it has no more blood symmetry.
        
        \item Path 2: First, construct the EM-exchange symmetry-enriched $\Z_3$ quantum-double phase and gauge the symmetry, obtaining doubled $\Z_3$-Ising phase. The doubled $\Z_3$-Ising phase also has a blood symmetry -- the charge-conjugation symmetry, which is due to that of the $\Z_3$ quantum double phase. Then, construct the charge-conjugation symmetry-enriched $S_3$ quantum-double phase and gauge the symmetry, obtaining the doubled $\MA$ phase.
    \end{enumerate}
    These two paths comprise a symmetry-gauging family -- $\Z_3$ symmetry-gauging family, as depicted in Figure \ref{fig:family}.
    \begin{figure}
        \centering
        \begin{tikzpicture}
        \node at (0.2, 1) {$D({\tt Vec}(\mathbb{Z}_3))$};
        \draw [->] (1.2, 1.1) -- (3.5, 1.5);
        \draw [->] (1.2, .9) -- (3.3, 0.6);
        \node at (4.5, 1.5) {$D({\tt Vec}(S_3))$};
        \node at (4.5, .5) {$D(\mathbb{Z}_3{\tt -Ising})$};
        \node at (2.1, 1.5) {$s$};
        \node at (2.1, .5) {$\sigma$};
        \draw [->] (5.5, 1.5) -- (7.5, 1.1);
        \draw [->] (5.7, .5) -- (7.5, .9);
        \node [right] at (6.4, 1.6) {$\alpha = \sigma, \beta = s\sigma$};
        \node [right] at (6.4, .4) {$s, \sigma s$};
        \node [right] at (7.5, 1.) {$D(\MA)$};
        \end{tikzpicture}
        \caption{$\Z_3$ symmetry-gauging family. Here, $s$ and $\sigma$ respectively represent the charge-conjugation symmetry and the EM-exchange symmetry of $\Z_3$ quantum-double phase.}
        \label{fig:family}
    \end{figure}
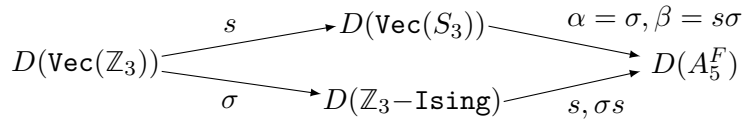
\end{enumerate}

\section{Conclusion and Discussion}
We have developed a unified and constructive framework for modeling and gauging SET phases based on the enlarged HGW string-net model, by providing three concrete strategies.  Strategies I and II extract global symmetries directly from the input unitary fusion category $\Fus$ of the enlarged HGW model that describes the $\Cent(\Fus)$ topological phase. These two strategies avoid ad hoc symmetry assignments and offer a natural classification into intrinsic (``blood'') symmetries arising from categorical structure, as opposed to adopted symmetries imposed externally in strategry III. 

The main text presents Strategy I, which promotes $\Fus$ via its outer automorphism group, and Strategy II, which promotes $\Fus$ via its Frobenius algebras and their bimodules. These strategies allow us to not only construct multifusion-category inputs and explicit onsite symmetry operators for building SET models but also gauge the SET models by identifying the relevant $G$-graded sectors to produce a parent model of pure topological phase. Strategy III, presented in Appendix~\ref{sec:symmfrac}, handles imposed (adopted) symmetries by replicating $\Fus$ and constructing permitted $G$-actions; this appendix strategy is well suited for SPT-like constructions. We have shown a number of concrete examples. Notably, we provided in the literature to date the first example of a nonabelian-symmetry-enriched topological phase -- the $S_3$-symmetry-enriched $\Z_2\times \Z_2$ quantum double phase.

Our strategies I and II focus on the blood symmetries of topological phases. We find that a family of topological phases are related by gauging their blood symmetries. We termed such a family a symmetry-gauging family, which always has an initial topological phase (the smallest topological phase, i.e., with fewest anyon types, in the family) and a terminal topological phase (the largest topological phase in the family). We used a web of nodes and oriented links to present such a family. While each node is a topological phase, each oriented link points from a node (child topological phase) to another node (parent topological phase), which results from gauging a certain blood symmetry (labeled on the link) of the child phase. Due to the duality between gauging and condensation, if we reverse all the oriented links in a symmetry-gauging family of topological phases, we obtain a \textbf{condensation family} of topological phases, where now each oriented link points from a parent topological phase to a child topological phase that results from certain anyon condensation in the parent phase. 

It's important to note that any condensations in a condensation family of topological phases always exhibit anyon splitting -- a distinctive phenomenon of anyon condensation\footnote{For the original idea of anyon condensation, see Ref.\cite{Bais2009a}. For anyon condensation in the string-net model, see Ref.\cite{zhao2022}.}. Conversely, if the anyon condensation in a parent topological phase exhibit anyon splitting, the corresponding child topological phase must have a blood symmetry. The gauging procedure in strategy III however is dual to anyon condensation that does not involve anyon splitting at all.

Our SET model also manifests the local excitations and clarifies their definition in SET phases, their interplay with global symmetries, and the discrete symmetry constraints on their internal degrees of freedom. Given an SET phase and its parent topological phase, which via certain anyon condensation therein becomes the SET phase if the global symmetry is not spontaneously broken. The local excitation types of the SET phase precisely correspond to the condensed anyon sectors of the parent phase. We stress on condensed anyon sectors because when anyon splitting occurs in anyon condensation, certain parent phase nonabelian anyons may split into multiple sectors that may not all condense. The condensed anyon sectors become the local excitations of the child SET phase. For example, the doubled Ising phase is the parent of the EM-exchange symmetry-enriched $\Z_2$ toric-code model. The nontrivial local excitation in this SET phase is precisely the condensed anyon $\psi\bar{\psi}$ in doubled Ising phase. 

Our construction of SET models leads us to discover a novel phenomenon -- symmetry fragmentation -- of SET phases involving nonabelian anyons: In an SET phase, the internal space of a nonabelian anyon type and/or the combined internal space of more than one anyon type may fragment into subspaces respectively carrying definite global symmetry charges. This phenomenon is worth of its own story and was reported in a companion paper\cite{fu2025nonlinear}.

A few problems deserve future exploration. In all examples we have studied (including those not shown in the paper) for strategy II, we have not witnessed blood symmetries larger than $\Z_2$. Perhaps one would have to look into more complex UFCs to find larger blood symmetries. In this paper, we have focused on model construction but not the classifications of SET phases. It is interesting to see whether our model would give rise to new classifications of SET phases. Another intriguing question to study is whether and how the local excitations in our SET model are related to symmetry fractionalization; both parties are likely to be related because they all involve domain-wall crossing. Another interesting topic is the symmetry sectors of an SET phase. A symmetry sector of an SET phase is a subspace of the phase's Hilbert space that is an eigenspace of the global symmetry. Symmetry sectors are generally not the $G$-graded sectors discussed in this paper. Defining the symmetry sectors of our SET model is rather involved, especially in the case of nonabelian symmetries, where further enlargement of our model's Hilbert space may be required. We shall study symmetry sectors in future work.

\begin{acknowledgments}
The authors thank Hongguang Liu, Yuting Hu, Siyuan Wang, and Yifei Wang for inspiring and helpful discussions. YW is supported by NSFC Grant No. KRH1512711, the Shanghai Municipal Science and Technology Major Project (Grant No. 2019SHZDZX01), Science and Technology Commission of Shanghai Municipality (Grant No. 24LZ1400100), and the Innovation Program for Quantum Science and Technology (No. 2024ZD0300101). YW is grateful for the hospitality of the Perimeter Institute during his visit, where the main part of this work is done. This research was supported in part by the Perimeter Institute for Theoretical Physics. Research at Perimeter Institute is supported by the Government of Canada through the Department of Innovation, Science and Economic Development and by the Province of Ontario through the Ministry of Research, Innovation and Science. 
\end{acknowledgments}

\appendix

\section{Strategy III: Constructing SET Input Data from An Expected Symmetry, Symmetry Transformation, And Gauging}\label{sec:symmfrac}

In this section, we dwell on the case where a topological phase $\Cent(\Fus)$ is endowed with a global symmetry $G$ that may be unrelated to the input UFC $\Fus$ of the HGW model describing $\Cent(\Fus)$. Such a symmetry is referred to as an adopted symmetry in the introduction. For this, we shall need our strategy III.

Our strategy III coincides with the SET constructing method in Ref.\cite{cheng2017exactly}, and can realize symmetry fractionalization as shown in Ref.\cite{cheng2017exactly}.

We will elaborate the strategy in three steps, accompanied by a concrete example--imposed $\Z_2$ symmetry enriched toric code model.

\subsection{Step 1: Promote \eqs{\Fus} to a Multifusion Category by \eqs{G}}

First, we should promote $\Fus$ to a Multifusion category $\mathcal{M}$ by an arbitrary symmetry group $G$. $\mathcal{M}$ takes the form
\begin{equation}\label{eq:multifusionmatrix2}
\M = \begin{pmatrix}
\Fus_{g_1g_1} & \Fus_{g_1g_2} & \cdots & \Fus_{g_1g_n} \\
\Fus_{g_2g_1} & \Fus_{g_2g_2} & \cdots & \Fus_{g_2g_n} \\
\vdots & \vdots & \ddots & \vdots \\
\Fus_{g_ng_1} & \Fus_{g_ng_2}& \cdots & \Fus_{g_ng_n}
\end{pmatrix},
\end{equation}
which has already been introduced in Section \ref{subsec:multifusion}. Here, each element $\Fus_{g_ig_j}$ of the multifusion matrix is a copy of $\Fus$, where the matrix index $g_i,g_j$ are group elements of $G$. The fusion rules of $\mathcal{M}$ can be obtained directly from those of $\Fus$:
\begin{equation}\label{eq:fusion}
    x_{g_ig_j}\otimes y_{g_kg_l} = \begin{cases}(x\otimes y)_{g_ig_l}\quad(g_j =g_ k),\\ \nulm\qquad\qquad\ \ (g_j\ne g_k)\end{cases},
\end{equation}
while the nonzero $6j$ symbols of $\mathcal{M}$ can also be read off from those of $\Fus$:
\begin{equation}\label{eq:6j}
    G^{a_{g_ig_j}b_{g_jg_k}e_{g_ig_k}}_{c_{g_kg_l}d_{g_lg_i}f_{g_jg_l}}=G^{abe}_{cdf},\ \forall\ g_ig_jg_k,g_l\in G.
\end{equation}

\begin{example}{Imposed \eqs{\Z_2} Symmetry Enriched Toric Code}{1}

Consider toric code model as the child phase, whose input fusion category is $\Fus=\Vec(\Z_2)$. We choose $\Z_2$ as the symmetry group. Thus, the multifusion matrix \eqref{eq:multifusionmatrix2} is
\begin{equation}\label{eq:impose symm multifusion Z2}
    \M = \begin{pmatrix}
    \{0_{++},1_{++}\} & \{0_{+-},1_{+-}\} \\
    \{0_{-+},1_{-+}\} & \{0_{--},1_{--}\} \\
\end{pmatrix}.
\end{equation}

The fusion rules and $6j$-symbols of $\M$ can be obtained directly from those of $\Fus$, as shown in \eqref{eq:fusion} and \eqref{eq:6j}. For example, we have $0_{+-}\otimes1_{-+}=1_{++}$ and $G^{0_{++}1_{+-}1_{+-}}_{0_{--}1_{-+}1_{+-}}=G^{011}_{011}=1$.

\end{example}

\subsection{Step 2: Construct Global Symmetry Transformation}

After building the multifusion category $\M$, we need to construct a set of global symmetry transformations $\{\G_g|g\in G\}$. Each symmetry transformation $\G_g$ is defined by
\begin{equation}
    \G_g = \bigotimes_l \G^l_g,
\end{equation}
where $l$ goes over all the edges/tails $l$.

There are $2$ constraints on the symmetry transformations. First, the set of all symmetry operators must form a group under composition, which means that
\begin{equation}\label{eq:op compose}
    \G^l_g\circ\G^l_h=\G^l_{gh},\ \forall g,h\in G.
\end{equation}
Second, the symmetry action must preserve the fusion rules:
\begin{equation}\label{eq:preserve fusion}
    \G^l_g(x)\otimes\G^l_g(y)=\G^l_g(x\otimes y).
\end{equation}

The symmetry transformations can be determined as follows:
\begin{enumerate}
    \item Since the symmetry group $G$ acts trivially on the $G$-graded sectors without permuting anyons, the symmetry transformation on the degrees of freedoms of $G$-graded sectors (the diagonal elements) is given by:
    \begin{equation}\label{eq:diag trans}
        \G^l_{g}:\ x_{g_i,g_i}\mapsto x_{gg_i,gg_i}.
    \end{equation}
    \item The symmetry transformation on the domain wall degrees of freedoms should map objects in $\Fus_{g_ig_j}$ to objects in $\Fus_{(gg_i)(gg_j)}$. The exact form of the transformation can be derived from \eqref{eq:op compose}, \eqref{eq:preserve fusion} and \eqref{eq:diag trans}. It is worth noting that the solution is generally not unique: different choices correspond to distinct symmetry fractionalization patterns introduced in Ref.\cite{cheng2017exactly}.
\end{enumerate}

\begin{example}{Imposed \eqs{\Z_2} Symmetry Enriched Toric Code}{2}

For our $\Z_2$ symmetry enriched toric code example, let's consider the only nontrivial symmetry operator $\G_-$. There are two solutions for the onsite operator $\G^l_-$:
\begin{equation}\label{eq:iso Z2_1}
    \begin{split}
        (\G^l_-)_1:\ 0_{++}\mapsto0_{--},\ 1_{++}\mapsto1_{--},\\
        \ 0_{--}\mapsto0_{++},\ 1_{--}\mapsto1_{++},\\
        \ 0_{+-}\mapsto0_{-+},\ 1_{+-}\mapsto1_{-+},\\
        \ 0_{-+}\mapsto0_{+-},\ 1_{-+}\mapsto1_{+-};
    \end{split}
\end{equation}
\begin{equation}\label{eq:iso Z2_2}
    \begin{split}
        (\G^l_-)_2:\ 0_{++}\mapsto0_{--},\ 1_{++}\mapsto1_{--},\\
        \ 0_{--}\mapsto0_{++},\ 1_{--}\mapsto1_{++},\\
        \ 0_{+-}\mapsto1_{-+},\ 1_{+-}\mapsto0_{-+},\\
        \ 0_{-+}\mapsto1_{+-},\ 0_{-+}\mapsto1_{+-}.\\
    \end{split}
\end{equation}

They provide $2$ different symmetry transformations $(\G_-)_1$ and $(\G_-)_2$, giving $2$ different SET models, even though both models share the same input multifusion category \eqref{eq:impose symm multifusion Z2}.

\end{example}

\subsection{Step 3: Gauging}

Same as the previous two strategies, to gauge the SET model, we need to identify the simple objects related by the symmetry transformation -- identify all the elements in the set $\{\G^l_g(a)|g\in G\}$ as one simple object in the parent phase $\Fus_p$.

It is worth noting that, with the same input multifusion category while different symmetry transformations, after gauging we can get different $\Fus_p$. 

\begin{example}{Imposed \eqs{\Z_2} Symmetry Enriched Toric Code}{3}

For our $2$ different $\Z_2$ symmetry enriched toric code model, gauging them yields different parent phase $\Fus_p$:
\begin{enumerate}
    \item For SET model with symmetry $(\G_-)_1$ \eqref{eq:iso Z2_1}, we shall identify $0_{++}$ and $0_{--}$ as $1$, identify $1_{++}$ and $1_{--}$ as $a$, identify $0_{+-}$ and $0_{-+}$ as $b$, identify $1_{+-}$ and $1_{-+}$ as $c$. Thus, $\Fus_p=\{1,a,b,c\}$, with the nonzero fusion rules:
    \begin{equation}\label{eq:gauging Z2_1 fusion}
    \delta_{111}=\delta_{1aa}=\delta_{1bb}=\delta_{1cc}=\delta_{abc}=\delta_{acb}=1.
    \end{equation}
    Hence, $\Fus_p=\Vec(\Z_2\times\Z_2)$.
    \item For SET model with symmetry $(\G_-)_2$ \eqref{eq:iso Z2_2}, we shall identify $0_{++}$ and $0_{--}$ as $0$, identify $1_{++}$ and $1_{--}$ as $2$, identify $0_{+-}$ and $1_{-+}$ as $1$, identify $1_{+-}$ and $0_{-+}$ as $3$. Thus, $\Fus_p=\{0,1,2,3\}$, with the nonzero fusion rules:
    \begin{equation}\label{eq:gauging Z2_2 fusion}
    \delta_{000}=\delta_{022}=\delta_{013}=\delta_{031}=\delta_{211}=\delta_{211}=1.
    \end{equation}
    Hence, $\Fus_p=\Vec(\Z_4)$.
\end{enumerate}

\end{example}

\subsection{Example: Constructing and Gauging SPT Models}\label{sec:SPT}

An interesting application of Strategy III is constructing and gauging SPT models, as SPT models can be regarded as SET model with trivial underlying topological order ($\Fus=\Vec$)\cite{chen2013symmetry}.

To construct an SPT model, we need to choose a symmetry group $G$ first. Then, we can construct the input multifusion category:
\begin{equation}\label{eq:SPT Mn}
    \M = \begin{pmatrix}
    \{1\}_{g_1g_1} & \{1\}_{g_1g_2} & \cdots & \{1\}_{g_1g_n} \\
    \{1\}_{g_2g_1} & \{1\}_{g_2g_2} & \cdots & \{1\}_{g_2g_n} \\
    \vdots & \vdots & \ddots & \vdots \\
    \{1\}_{g_ng_1} & \{1\}_{g_ng_2}& \cdots & \{1\}_{g_ng_n}
\end{pmatrix},
\end{equation}
where $g_1,g_2,\cdots,g_n\in G,\ |G|=n$.

In the SPT case, as there is only $1$ simple object in $\Vec$, there is only one solution of symmetry transformation:
\begin{equation}\label{eq:SPT symmop}
    \G_g^l:\ 1_{g_ig_j}\mapsto1_{(gg_i)(gg_j)},\ \forall g,g_ig_j\in G.
\end{equation}
With \eqref{eq:SPT Mn} and \eqref{eq:SPT symmop}, we have the SPT model.

To gauge the SPT model, we need to do the identification:
\begin{equation}\label{eq:SPT identify}
    \begin{split}
        \mathcal{I}:\ &\M\ \ \to\Fus_p,\\
        &1_{g_ig_j}\mapsto g^{-1}_ig_j.
    \end{split}
\end{equation}
Thus, we have $\Fus_p=\Vec(G)$.

\section{String-net Model}\label{sec:review}

This section provides a concise overview of the string-net model as outlined in Ref. \cite{zhao2024}, which was adapted from that in \cite{Hu2018}. The string-net model is an exactly solvable model defined on a \(2\)-dimensional lattice. An example lattice is depicted in Fig. \ref{fig:lat}. Each plaquette within the lattice carries a tail linked to any selected edge, directed inward. In Appendix \ref{sec:pachner}, We will demonstrate that different choices of the edge to which the tail is attached are equivalent. Each edge and tail is oriented, but we'll show that different choices of directions are equivalent.

The input data of the string-net model is a unitary fusion category \(\Fus\), described by a finite set \(L_\Fus\), whose elements are called \emph{simple objects}, equipped with three functions \(N: L_\Fus^3 \to \NN\), \(d: L_\Fus\to\RR^+\), and \(G: L^6_\Fus\to\C\). The function \(N\) sets the \emph{fusion rules} of the simple objects, satisfying
\eq{
\sum_{e \in L_\Fus} N_{ab}^e N_{ec}^d = \sum_{f\in L_\Fus} N_{af}^d N_{bc}^f, \qquad\qquad N_{ab}^c = N_{c^* a}^{b^*}.
}
There exists a special simple object \(1 \in L_\Fus\), called the \emph{trivial object}, such that for any \(a, b \in L_\Fus\),
\eq{
N_{1a}^b = N_{1b}^a = \delta_{ab},
}
where \(\delta\) is the Kronecker symbol. For each \(a \in L_\Fus\), there exists a unique simple object \(a^* \in L_\Fus\), called the \emph{opposite object} of \(a\), such that
\eq{
N_{ab}^1 = N_{ba}^1 = \delta_{ba^*}.
}
We only consider the case where for any \(a, b, c \in L_\Fus\), \(N_{ab}^c = 0\) or \(1\). In this case, we define
\eq{
\delta_{abc} = N_{ab}^{c^*} \in \{0, 1\}.
}

The basic configuration of the string-net model is established by assigning each edge and tail with a simple object from \(L_\Fus\), subject to the vertex constraint \(\delta_{ijk} = 1\) for the three edges or tails converging at each vertex, which are oriented inward and labeled counterclockwise by \(i, j, k \in L_\Fus\). We can reverse the direction of any edge or tail and simultaneously conjugate its label as \(j \to j^*\), which keeps the configuration invariant. The Hilbert space \(\Hil\) of the model is spanned by all possible configurations of these labels on the edges and tails.

The function \(d: L_\Fus\to\RR^+\) gives the \emph{quantum dimensions} of the simple objects in \(L_\Fus\). It is the largest eigenvalues of the fusion matrix and forms the \(1\)-dimensional representation of the fusion rule:
\eq{d_ad_b = \sum_{c\in L_\Fus}N_{ab}^cd_c.}
In particular, \(d_1 = 1\), and for any \(a\in L_\Fus, d_a = d_{a^*}\ge 1\). 

The function \(G: L_\Fus^6\to\CC\) defines the \(6j\)-\emph{symbols} of the fusion algebra. It satisfies
\eqn[eq:sixj]{\sum_nd_nG^{pqn}_{v^*u^*a}G^{uvn}_{j^*i^*b}G^{ijn}_{q^*p^*c} = G^{abc}_{i^*pu^*}&G^{c^*b^*a^*}_{vq^*j},\qquad\sum_nd_nG^{ijp}_{kln}G^{j^*i^*q}_{l^*k^*n} = \frac{\delta_{pq^*}}{d_p}\delta_{ijp}\delta_{klq},\\
G^{ijm}_{kln} = G^{klm^*}_{ijn^*} = G^{jim}_{lkn^*}= G^{mij}_{nk^*l^*} &= \alpha_m\alpha_n\overline{G^{j^*i^*m^*}_{l^*k^*n^*}},\qquad \Big|G^{abc}_{1bc}\Big| = \frac{1}{\sqrt{d_bd_c}}\delta_{abc},
}
where \(\alpha_m = G^{1mm^*}_{1m^* m}\in\{\pm 1\}\) is the Frobenius-Schur indicator of simple object \(m\).

The Hamiltonian of the string-net model reads
\eqn{H := - \sum_{{\rm Plaquettes}\ P}Q_P,\qquad\qquad Q_P := \frac{1}{D}\sum_{s\in L_\Fus}d_sQ_P^s,}
where operator \(Q_P^s\) acts on edges surrounding plaquette \(P\) and has the following matrix elements on a hexagonal plaquette:
\begin{align*}Q_P^s\ \PlaquetteSrc\
:=\ \delta_{p,1}\delta_{j_1,j_7}\ \sum_{j_k\in L_\Fus}\ \prod_{k = 1}^{6}\ \Bigg(\sqrt{d_{i_k}d_{j_k}}G^{e_ki_ki_{k+1}^*}_{sj_{k+1}^* j_k}\Bigg)\PlaquetteTar\ ,
\end{align*}
and
\eq{D := \sum_{a\in L_\Fus}d_a^2}
is the total quantum dimension of UFC \(\Fus\). Here, we present only the action of the \(Q_P\) operator on a hexagonal plaquette. The matrix elements of \(Q_P\) operators on other types of plaquettes are defined similarly. For simplicity, we also omit the ``$\ket{\cdot}$'' labels surrounding all diagrams unless they are explicitly required.

It turns out that
\eq{
(Q_P^s)^\dagger = Q_P^{s^*},\qquad Q_P^rQ_P^s = \sum_{t\in L_\Fus} N_{rs}^tQ_P^t,\qquad Q_P^2 = Q_P,\qquad Q_{P_1}Q_{P_2} = Q_{P_2}Q_{P_1}.
}
The summands \(Q_P\) in Hamiltonian \(H\) are commuting projectors, so the Hamiltonian is exactly solvable. The ground-state subspace \(\Hil_0\) of the system is the projection
\eqn{
\Hil_0 = \Bigg[\prod_{{\rm Plaquettes\ } P}Q_P\Bigg]\Hil.
}
If the lattice has the sphere topology, the model has a unique ground state \(\ket\Phi\) up to scalar factors.

\subsection{Topological Features}\label{sec:pachner}

We briefly review the topological nature of the ground-state subspace of the string-net model defined in Ref. \cite{Hu2018}. Topologically, any two lattices with the same topology can be transformed into each other by the \emph{Pachner moves}. There are unitary linear maps between the Hilbert spaces of two string-net models with the same input UFC on different lattices related by the Pachner moves\cite{Hu2012}, formally denoted as operators \(\T\). The ground states are invariant under such linear transformations. There are three kinds of elementary Pachner moves, whose corresponding linear transformations are:
\eqn[eq:pachner]{
&\T \quad \PachnerOne\ ,\\
&\T \quad \PachnerTwo\ ,\\
&\T \quad \PachnerThree\ .}
Here, ``\({\color{red}\times}\)'' marks a plaquette to be contracted. These
three elementary Pachner moves and their corrresponding unitary transformations can be composed. Given initial and final lattices, there are multiple ways to compose these elementary Pachner moves; nevertheless, all such compositions yield the same transformation matrices on the ground-state Hilbert space.

We have also noted that for a given plaquette, different choices of edge to which the tail is attached are equivalent. These variations lead to distinct lattice configurations and, consequently, different Hilbert spaces for the lattice model. The equivalence between such two Hilbert spaces is established by the following linear transformation \(\T'\):
\eqn[eq:tailmove]{
\T'\quad\PachnerFour\ .
}
The states where tails attach to other edges can be obtained recursively in this manner.

For convenience, in certain cases, we will temporarily incorporate auxiliary states with multiple tails within a single plaquette. These states, despite having multiple tails in one plaquette, are all equivalent to states within the Hilbert space (with only one tail in each plaquette):
\eqn[eq:tailfuse]{
\PachnerFive\ .
}

\subsection{Excited States}\label{sec:spec}

An \emph{excited state} \(\ket\varphi\) of the string-net model is an eigenstate such that \(Q_P\ket\varphi = 0\) at some plaquettes \(P\). In such a state, \emph{anyons} are said to inhabit these plaquettes \(P\). The ground states are also described as trivial excited states, in which there are only \emph{trivial anyons} in all plaquettes. We assume the sphere topology, where the model has a unique ground state; nevertheless, the results in this section apply to other topologies.

We start with the simplest excited states with a pair of anyons situated in two \emph{adjacent} plaquettes sharing a common edge \(E\). This specific state is produced through the ribbon operator \(W_E^{J;pq}\), which is a composition of an auxiliary operator \(\mathcal{W}_E^{J; pq}\)
\eqn[eq:create]{
\mathcal{W}_E^{J; pq} \ExcitedA := \sum_{k \in L_\Fus} \sqrt{\frac{d_k}{d_{j_E}}} \ {z_{pqj_E}^{J;k}} \ \ \ExcitedB \ 
}
and Pachner moves \eqref{eq:tailmove} and \eqref{eq:tailfuse}, where $j_E$ is the degrees of freedom on edge $E$. For any state in the Hilbert space of the model, each plaquette contains exactly one tail attached to a specific edge (potentially edge $E$) of the plaquette.The operator \(\mathcal{W}_E^{J; pq}\) as described in equation \eqref{eq:create} introduces two additional tails on edge \(E\). These newly introduced tails are required to be moved and fused with the original tails of the plaquettes by the subsequent Pachner moves in the creation opeartor.

The coefficients in Eq. \eqref{eq:create},  \(z_{pqj}^{J; k}\), are called the \emph{half-braiding tensor} of anyon type \(J\), defined by the following equation:
\eqn[eq:halfA]{
\frac{\delta_{jt}N_{rs}^t}{d_t} z_{pqt}^{J;w} = \sum_{u,l,v\in L_\Fus} z_{lqr}^{J;v} z_{pls}^{J;u}d_u d_v G^{r^*s^*t}_{p^*wu^*} G^{srj^*}_{qw^*v} G^{s^*ul^*}_{rv^*w}.
}
We will discuss the physical significance of this equation in Appendix \ref{sec:halfbraid}. The label \(J\), referred to as the \emph{anyon type}, denotes distinct minimal solutions \(z^J\) of Eq. \eqref{eq:halfA} that cannot be expressed as the sum of any other nonzero solutions. Categorically, anyon type \(J\) are labeled by simple objects in the \emph{center} of UFC \(\Fus\), a modular tensor category whose categorical data encode all topological properties of the topological order that the string-net model describes, denoted as \(\Cent(\Fus)\). In particular, the trivial anyon $\mathcal{I}\in L_{\Cent(\Fus)}$ satisfies
$$z^{\mathcal{I};k}_{pqj} = \delta_{p1}\delta_{q1}\delta_{jk}.$$

The statistics of anyon \(J\) are recorded by the diagonal element of modular \(T\) matrix of UMTC \(\Cent(\Fus)\), where
\eq{
T_{JK} = \frac{1}{d_t}\delta_{JK}\sum_{p \in L_\Fus} d_p z_{ttt}^{J;p}.}
Here, \(t\) is an arbitrary charge of anyon \(J\). The braiding of two anyons \(J\) and \(K\) is recorded by the modular \(S\) matrix, whose matrix elements are
\eq{S_{JK} = \frac{1}{D}\ \sum_{p, q, k \in L_\Fus} d_k \bar{z}_{ppq}^{J;k} \bar{z}_{qqp}^{K;k}.
}
An anyon \(J\) has trivial self-statistics if \(\theta_J = 1\); two anyons \(J\) and \(K\) braid trivially if and only if \(S_{JK} = d_J d_K\), where \(d_J\) is the quantum dimension of anyon \(J\), defined as
\eq{
d_J = \sum_{J\text{'Charges } p} d_p.
}

States containing two quasiparticles situated in nonadjacent plaquettes are produced using ribbon operators that extend along more extensive paths. These extended ribbon operators are formed by concatenating shorter ribbon operators. As an illustration, to generate two quasiparticles \(J^*\) and \(J\) characterized by charges \(p^*_0\) and \(p_n\) within two nonadjacent plaquettes \(P_0\) and \(P_n\), one may select a series of plaquettes denoted as \((P_0, P_1, \ldots, P_n)\), where \(P_i\) and \(P_{i+1}\) are adjoining plaquettes sharing a common edge \(E_i\). The ribbon operator \(W_{P_0P_n}^{J; p_0p_n}\) is
\eq{
W_{P_0P_n}^{J; p_0p_n} := \Bigg[\sum_{p_1 p_2 \cdots p_{n-1} \in L_\Fus} \prod_{k = 1}^{n-1} \left(d_{p_k} B_{P_k} W_{E_k}^{J; p_k p_{k+1}}\right)\Bigg]W_{E_0}^{J; p_0 p_1}.
}
Different choices of plaquette paths \((P_0, P_1, \cdots, P_n)\) give the same operator \(W_{P_0P_n}^{J; p_0 p_n}\) provided that these sequences can be continuously deformed into one another. Following the same procedure, we can also define the creation operator of three or more anyons.

At the end of this section, we introduce the measurement operator \(M_P^J\), which measures the presence of an anyon \(J\) excited in the plaquette \(P\):
\eqn{M_P^J\ \PlaquetteSrc
:= \sum_{s,t\in L_\Fus} \frac{d_sd_tz_{pps}^{J; t}}{d_p} \PlaquetteMsr\ .}
The set of measurement operators are orthonormal and complete:
\eq{M_P^JM_P^K = \delta_{JK}M_P^J,\qquad \sum_{J\in L_{\mathcal{Z}(\Fus)}}M_P^J = \idm.}

\subsection{The Output UMTC is the Center of the Input UFC}\label{sec:halfbraid}

As previously discussed, the string-net model produces a unitary modular tensor category (UMTC), denoted as \(\Cent(\Fus)\), representing the center of its input unitary fusion category (UFC), \(\Fus\). This appendix aims to elucidate in detail how this representation is conceptualized.

Categorically, an object \(J\) in the center \(\Cent(\Fus)\) is denoted as a pair \(J = (X_J, c_{X_J, \cdot})\), where \(X_J\) is an object in UFC \(\Fus\), and \(c_{x_J, \cdot}\), referred to as a \emph{half-braiding}, is a set of morphisms
\eq{
\{ c_{X_J, y}: X_J\otimes y\to y\otimes X_J | y\in \Fus
\}. }
A morphism \(c_{X_J, y}\) braids object \(X_J\) with object \(y\) in \(\Fus\) and can be depicted as
\eq{\HalfBraidingA.}
In a UFC, all morphisms can be decomposed as direct sums of fusion of simple objects, and so can a half-braiding:
\eqn[eq:halfB]{\HalfBraidingB.}
Here, \(L_J\subseteq L_\Fus\), such that the direct sum of simple objects in \(L_J\) is \(X_J\):
\eq{X_J = \bigoplus_{p \in L_J} p.}
The expansion coefficients \(z_{pqk}^{J;y}\) are known as the \emph{half-braiding tensor} of \(J\). A characteristic of half-braiding is that it should be commutative with any fusion within \(\Fus\):
\eqn[eq:halfC]{\HalfBraidingC.}
Expanding Eqs. \eqref{eq:halfC} using Eq. \eqref{eq:halfB} leads to Eq. \eqref{eq:halfA}.

For a string-net model with input UFC $\Fus$, an anyon type \(J\) is a simple object in the output UMTC \(\Cent(\Fus)\), and $J$'s charges take value in \(L_J\). The action of creation operator \(W_E^{J;pq}\) represents the half-braiding morphism \(c_{X_J, j_E}\) of object \(X_J\) with \(j_E\in L_\Fus\), the degrees of freedom on edge \(E\):
\eq{\HalfBraidingD}

\subsection{The String-Net Model with Noncommutative Input UFCs}\label{appendix:noab}

In this appendix, we briefly introduce the string-net model, whose input UFC $\Fus$ has noncommutative fusion rules $\delta$, i.e., there exists $a, b, c\in L_\Fus$ for which $\delta_{abc}\ne\delta_{acb}$. Note that the states in the Hilbert space of such a string-net model are required to satisfy the constraint \(\delta_{abc} \ne 0\) for any three adjacent edges or tails converging at a vertex, with degrees of freedom \(a, b, c\) arranged in a \emph{counterclockwise} orientation, and all directed towards the vertex.

For these noncommutative string-net models, it is possible that Eq. \eqref{eq:halfA} for half-braiding tensors $z$ yield matrix solutions. In other words, a minimal solution \( z_{pqj}^{J;k} \) of the half-braiding tensor may not be a complex number but instead a unitary matrix. Here, $J$ indexes the anyon types that distinguish different minimal solutions, and $p, q, j, k$ belong to the set $L_\Fus$. Consequently, the tensor $z^J$ is expressed as:
\eq{[z_{pqj}^{J;k}]_{\alpha\beta} \in \mathbb{C}, \qquad 1 \leq \alpha \leq n_{(J, p)},\qquad 1\le \beta \leq n_{(J, q)},}
where $n_{(J, p)}\in\mathbb{N}$. For those anyons $J$ with $n_{(J, p)} > 1$ for charge $p$, a dyon type is not only labeled by the anyon types $J$ but also by the multiplicity index $1 \le \alpha\le n_{(J,p)}$:
$$(J, p, \alpha),\qquad J\in L_{\Cent(\Fus)},\qquad p\text{ is }J\text{'s charge},\qquad 1\le\alpha\le n_{(J, p)}.$$
For convenience, we refer to the pair \((J, p)\), consisting of anyon type \(J\) and charge type \(p\), as a \emph{dyon multiplet}. Each dyon multiplet \((J, p)\) contains \(n_{(J, p)}\) distinct dyon types \((J, p, \alpha)\), where \(\alpha\) is called the \emph{multiplet index}. The number \(n_{(J, p)}\) is referred to as the \emph{degeneracy} of the dyon multiplet \((J, p)\). The pair \((p, \alpha)\), consisting of the charge type \(p\) and the multiplet index \(\alpha\), is called the \emph{local degrees of freedom} of the anyon \(J\).

In the Hilbert space of a noncommutative string-net model, two excited states, each containing a dyon in the same plaquette with the same anyon type \(J\) and charge type \(p\) but different multiplet indices \(\alpha\), should be orthogonal excited states. However, the original Hilbert space, whose local Hilbert space on each tail is spanned solely by the charges \(p \in L_\Fus\), cannot accommodate such a large number of distinct excited states. To capture the full dyon spectrum, we must therefore \emph{enlarge} the local Hilbert space of each \emph{tail}, as in our previous work\cite{zhao2024a}. The degrees of freedom on edges do not need to be extended as the edges pertain to the ground states because any path along an edge forms a closed loop, where each vertex only respects the fusion rules, disregarding the multiplet indices.

For convenience, the local subspace on each tail is spanned by local degrees of freedoms of dyons, denoted as \( p_{J, \alpha} \), where \( J \) is an anyon types carrying charge \( p \), and \( 1 \leq \alpha \leq n_{(J,p)} \). Different basis states on a tail are orthogonal local states:
$$\Bigg\langle\quad\Tail{p_{J,\alpha}}\ \Bigg|\quad\Tail{q_{K,\beta}}\ \Bigg\rangle = \delta_{JK}\delta_{pq}\delta_{\alpha\beta}.$$
Here, we enlarge the local Hilbert spaces of tails even for commutative string-net models. This does not affect our discussion, as two excited states with different anyon types in the same plaquette must be orthogonal.

The simplest creation operator \(W_E^{J;(p,\alpha)(q,\beta)}\), which creates a pair of dyons \((J^*, p^*, \alpha)\) and \((J, q, \beta)\) in two adjacent plaquettes separated by edge \(E\), is defined as
\eqn{
W_E^{J; (p,\alpha)(q,\beta)} \NonAbA := \sum_{k \in L_\Fus} \sqrt{\frac{d_k}{d_j}}\ \ \overline{[{z_{pqj}^{J;k}}]_{\alpha\beta}} \ \ \NonAbB \ ,
}
Here, \(\mathcal{I} \in L_{\Cent(\Fus)}\) is the trivial anyon with the unique charge \(1 \in L_\Fus\) and no degeneracy, \(n_\mathcal{I} = 1\). The multiplication of creation operators depends on the CG coefficients of the tensor products of the half-braiding \(z\)-tensors, which we do not detail here.

A typical example of such a noncommutative input UFC is \({\tt Vec}(G)\), where \(G\) is a nonabelian group. The simple objects of UFC \({\tt Vec}(G)\) are labeled by group elements in $G$, and for any $a, b, c, d, e, f\in G$,
$$\delta_{abc} = 1\quad\text{if}\quad c^* = ab\quad\text{else}\quad 0,\qquad\qquad d_a = 1,\qquad\qquad G^{abe}_{cdf} = \delta_{abe}\delta_{bcf^*}\delta_{cde^*}\delta_{daf},$$
where \(c^*\) denotes the inverse element of \(c \in G\). Equation \eqref{eq:halfA} for the \(z\)-tensors becomes
$$z^{J;(pa)}_{p(a^* pa)a}z^{J;(a^* pab)}_{(a^* pa)(b^* a^* pab)b} = z^{J;(pab)}_{p(b^* a^* pab)(ab)}.$$
Consequently, the anyon types $J$ are labeled by a pair
$$J = (\bar p, \rho),$$
where $\bar p = \{g^* pg| g\in G\}$ for $p\in G$ is a conjugacy class of $G$ containing element $p$, and $\rho$ is an irreducible representation of the centralizer of conjugacy class $\bar p$:
$$Z(\bar p) = \{g\in G| g^* pg = p\}.$$
The multiplet degeneracy of each dyon multiplet $(J, p)$ of anyon type $J$ is the dimension of irrep $\rho$ of the centralizer $Z(\bar p)$:
$$n_{((\bar p, \rho), p)} = {\tt dim}\rho,$$
and the quantum dimension of anyon $J$ is $n_{((\bar p, \rho), p)}|\bar p|$, representing the total number of local degrees of freedom of \(J\).

In particular, the fluxons in the string-net model with input UFC ${\tt Vec}G$ are those dyons with anyon type $(\bar 1, \rho)$, whose trivial conjugacy class is $\bar 1 = \{1\}\subset G$ and $Z(\bar 1) = G$. The correspoinding half-braiding tensor is:
$$z_{1,1,r}^{(\bar 1, \rho);r} = D_{\alpha\beta}^\rho(g),\qquad 1\le\alpha, \beta\le{\tt dim}\rho,\qquad\forall r\in G.$$
where \(D^\rho\) is the representation matrix of the irrep \(\rho\) of the group \(G\).

\subsection{Quantum-Double Phase}

A special case of string-net model is the \emph{quantum-double phase}. For a quantum-double phase, the input UFC of the model is $\Vec(G)$ where $G$ is a group. Th simple objects are group elements of $G$, while the fusion rules are given by the group multiplication rules:
\eqn{\delta_{abc}=\delta_{ab,c^*},}
where $a^* =a^{-1}$. The $6j$-symbols reads:
\begin{equation}
    G^{abm}_{cdn}=\delta_{abm} \delta_{bcn^*} \delta_{cdm^*} \delta_{dan}.
\end{equation}

In this paper, the child phase in every example is a quantum-double phase.

\section{Multifusion Category}\label{sec:multifusion}

In this section we provide a brief introduction of multifusion category.

A multifusion category $\M$ is a finite semisimple rigid monoidal $\C$-linear category, with finitely many simple objects and finite-dimensional Hom spaces. In contrast to a fusion category where the tensor unit is required to be simple, the tensor unit $\idm$ in the multifusion category may be decomposed into a direct sum of simple objects:
\begin{equation}
    \idm=\bigoplus_{0\leq i\leq n}\idm_{ii}.
\end{equation}
Here, each $\idm_{ii}$ is a simple object satisfying $\idm_{ii} \otimes \idm_{ii} = \idm_{ii}$, and $\idm_{ii} \otimes \idm_{jj} = \nulm$ for $i \ne j$.

The set $\{\idm_{ii}\}$ defines a partitioning of $\M$ into blocks, such that each block $\M_{ij}$ consists of objects with $\idm_{ii}$ acting as a left unit and $\idm_{jj}$ as a right unit:
\begin{equation}\label{eq:partition}
    \M=\bigoplus_{0\leq i,j\leq n}\M_{ij},
\end{equation}
where
\begin{equation}
    \forall X_{ij}\in\M_{ij},\ \idm_{ii}\otimes X_{ij}=X_{ij}\otimes\idm_{jj}=X_{ij}.
\end{equation}

The fusion rules of $\M$ respect this block structure: $\forall X_{ij}\in\M_{ij},\ X_{kl}\in\M_{kl},$ 
\begin{equation}\label{eq:multifusion fusion}
    \begin{cases}
        X_{ij}\otimes X_{kl}\in \M_{il},\quad\ j=k\\
        X_{ij}\otimes X_{kl}=\nulm,\qquad j\neq k
    \end{cases}.
\end{equation}

The partition of $\M$ \eqref{eq:partition} can be shown in a multifusion matrix:
\begin{equation}
    \M=\begin{pmatrix}
        \M_{11} & \M_{12} & \cdots & \M_{1n} \\
        \M_{21} & \M_{22} & \cdots & \M_{2n} \\
        \vdots & \vdots & \ddots & \vdots \\
        \M_{n1} & \M_{n2} & \cdots & \M_{nn}
    \end{pmatrix}.
\end{equation}
Now \eqref{eq:multifusion fusion} could be viewed as the matrix product rules of the multifusion matrix.

\section{Frobenius Algebras and Bimodules}\label{sec:theory}

In this section, we briefly review the definition of Frobenius algebras in a given fusion category and their bimodules.

\subsection{Frobenius Algebra}\label{sec:frob}

A Frobenius algebra \(\A\) in a fusion category \(\Fus\) is characterized by a pair of functions \((n, f)\). The function \(n: L_\Fus \to \mathbb{N}\) returns the \emph{multiplicity} \(n_a\) of each \(a \in L_\Fus\) appearing in the Frobenius algebra \(\A\), satisfying \(n_1 = 1\) and \(n_a = n_{a^*}\). The simple objects of \(\A\) are labeled by \(a_\alpha\), where \(a \in L_\Fus\) satisfies \(n_a > 0\), and \(\alpha = 1, 2, \ldots, n_a\) is the \emph{multiplicity index}. The set of all simple objects in \(\A\) is denoted \(L_\A\).

The algebraic multiplication of \(\A\) is given by a function \(f: L^3_\A \to \C\), satisfying:
\eqn[eq:frob]{\sum_{t_\tau \in L_\A} f_{r_\rho s_\sigma t_\tau} f_{a_\alpha b_\beta t_\tau^*} G^{rst}_{abc} \sqrt{d_c d_t} &= \sum_{\gamma = 1}^{n_c} f_{a_\alpha c_\gamma s_\sigma} f_{r_\rho c_\gamma^* b_\beta}\ ,\\ \\
\sum_{a_\alpha b_\beta \in L_\A} f_{a_\alpha b_\beta c_\gamma} f_{b_\beta^* a_\alpha^* c_\gamma^*} \sqrt{d_a d_b} = d_\A \sqrt{d_c},\qquad f_{a_\alpha b_\beta c_\gamma} &= f_{b_\beta c_\gamma a_\alpha}, \qquad f_{0 a_\alpha b_\beta} = \delta_{ab^*} \delta_{\alpha\beta},}
where
\eqn{
d_\A := \sum_{a \in L_\Fus} n_a d_a
}
gives the \emph{quantum dimension} of the Frobenius algebra \(\A\).

For lattice model representations, we adopt the following conventions: 
Red edges or tails indicate labels from \(L_\A\), while a red dot at a vertex signifies multiplication by a coefficient \(f\). This is expressed graphically as
\eqn{
\FrobeniusA\ .
}
Dashed red edges or tails denote summation over all labels in \(L_\A\) for the corresponding edge. 

The defining relations \eqref{eq:frob} for \(\A\) are then represented graphically through Pachner moves:
\eq{\FrobeniusB\ ,}
\eq{\FrobeniusC\ .}

\subsection{Bimodules over a Frobenius Algebra}\label{sec:bimod}

An ${\A_1}\dash{\A_2}$ bimodule \(M\) is defined by a pair of functions \((n^M, P_M)\), where the function \(n^M: L_\Fus \to \NN\) returns the \emph{multiplicity} \(n^M_a\) of \(a \in L_\Fus\) appearing in bimodule \(M\), satisfying \(n^M_a = n^M_{a^*}\), and the basis elements of \(M\) are labeled by \(a_i\), where \(a \in L_\Fus\) satisfies \(n_a^M > 0\), and \(i = 1, 2, \ldots, n^M_a\) labels the multiplicity index. We denote the set of all basis elements in bimodule \(M\) as \(L_M\). 

The left action of $\A_1$ and right action of $\A_2$ on \(M\) are encoded by a function \(P_M: L_{\A_1}\times L_{\A_2} \times L_M \times L_\Fus \times L_M \to \C\), which satisfies:
\eqn[eq:bimod]{&\sum_{uv\in L_\Fus}\ \sum_{y_\upsilon\in L_M}\ [P_M]^{a_\alpha r_\rho}_{x_\chi u y_\upsilon}\ [P_M]^{b_\beta s_\sigma}_{y_\upsilon v z_\zeta}\ G^{v^* by}_{urw}\ G^{w^* bu}_{axc}\ G^{sz^* v}_{wrt^*}\ \sqrt{d_ud_vd_wd_yd_cd_t}\\ 
=\ &\sum_{\gamma = 1}^{n_c}\ \sum_{\tau = 1}^{n_t}\ P^{c_\gamma t_\tau}_{x_\chi w z_\zeta}f^{\A_1}_{a_\alpha c_\gamma^* b_\beta}\ f^{\A_2}_{r_\rho s_\sigma t_\tau},
}
\eqn{\ [P_M]^{00}_{x_\chi y z_\zeta} = \delta_{xy}\delta_{yz}\delta_{\chi\upsilon}\delta_{\upsilon\zeta}.
}
Here, $f^{\A_1}$ and $f^{\A_2}$ are the defining algebra products respectively for Frobeniua algebras ${\A_1}$ and ${\A_2}$.

For convenience, in a lattice model, we use a blue line to indicate that this line is labeled by a simple object in bimodule \(M\) and a wavy blue line to represent summing over all states with labels on this edge in \(L_\Fus\) with coefficient \(P_M\):
\eqn[eq:bimoddef]{
\BimoduleA\ .
}
Here, we use different colors to denote different Frobenius algebras.

The definition \eqref{eq:bimod} of bimodule \(M\) can then be depicted graphically by Pachner moves:
\eq{
\BimoduleB\ .
}

\subsection{The Bimodule Fusion Category over a Frobenius Algebra}\label{sec:bimodcat}

The set of all bimodules over two given Frobenius algebras \({\A_1}\) and \({\A_2}\) in a fusion category \(\Fus\) is denoted as \(_{\A_1}\Bimod_{\A_2}(\Fus)\). In particular, when ${\A_1}={\A_2}$, \({_{\A_1}\Bimod_{\A_1}(\Fus)}\) forms a fusion category. \(_{\A_1}\Bimod_{\A_1}(\Fus)\), \(_{\A_1}\Bimod_{\A_2}(\Fus)\), \(_{\A_2}\Bimod_{\A_1}(\Fus)\) and \(_{\A_2}\Bimod_{\A_2}(\Fus)\) together constitute a multifusion category. In this section, we briefly introduce the categorical data of this multifusion category.

\begin{enumerate}
\item A bimodule \(M\) is \emph{simple} if it cannot be written as a direct sum of two other bimodules. That is, we cannot find two bimodules \(M_1\) and \(M_2\) such that:
\eq{
n^M_a = n^{M_1}_a + n^{M_2}_a,\quad [P_M]^{a_\alpha b_\beta}_{x_\chi y z_\zeta} = \begin{cases}[P_{M_1}]^{a_\alpha b_\beta}_{x_\chi y z_\zeta},\qquad\qquad\qquad (\chi \le n^{M_1}_x, \zeta \le n^{M_1}_z),\\ \\
[P_{M_2}]^{a_\alpha b_\beta}_{x_{(\chi - n^{M_1}_x)} y z_{(\zeta - n^{M_2}_z)}},\quad (\chi > n^{M_1}_x, \zeta > n^{M_1}_z),\\ \\ 0.\qquad\qquad\qquad\qquad\qquad (\rm otherwise).
\end{cases}
}

\item The quantum dimension of a bimodule \(M\) in \(_{\A_1}\Bimod_{\A_2}(\Fus)\) is
\eqn{
d_M = \frac{1}{\sqrt{d_{\A_1} d_{\A_2}}}\sum_{a\in L_\Fus}n^M_ad_a.
}
\item The Frobenius algebra \(\A\) itself is the trivial bimodule \(M_0\) over \(\A\):
\eqn[eq:opposite bimodule]{
L_{M_0} = L_\A,\qquad [P_{M_0}]^{a_\alpha b_\beta}_{x_\chi y z_\zeta} = \sum_{\upsilon = 1}^{n_y}f_{a_\alpha x_\chi y^*_\upsilon} f_{y_\upsilon b_\beta z_\zeta^*}.
}
\item Given a bimodule \(M\), its opposite bimodule \(M^*\) is
\eqn{
L_{M^*} = L_M,\qquad [P_{M^*}]^{a_\alpha b_\beta}_{x_\chi yz_\zeta} = ([P_M]^{a_\alpha b_\beta}_{x_\chi yz_\zeta})^*.
}
Thus, for any $M\in{_{\A_1}\Bimod_{\A_2}(\Fus)}$, $M^*\in{_{\A_2}\Bimod_{\A_1}(\Fus)}$.
\item For any three bimodules \(M_1\), \(M_2\), and \(M_3\), we can represent their fusion rules in terms of their simple objects. Define matrix \(\Delta_{M_1 M_2 M_3}\) to represent how the basis elements in the bimodule spaces are connected when the three bimodules fuse:
\eq{
\BimoduleC\ .
}
From the definition, we can see that if $3$ bimodules can fuse together, we must keep the Frobenius algebra in the same plaquette to be the same. Thus, the general form of a vertex is: $M_1\in{_{\A_1}\Bimod_{\A_2}(\Fus)},\ M_2\in{_{\A_2}\Bimod_{\A_3}(\Fus)},\ M_3\in{_{\A_3}\Bimod_{\A_1}(\Fus)}$\footnote{Here, we allow any two Frobenius algebras in $\{{\A_1},{\A_2},{\A_3}\}$ or three of them to be the same.}. The general form of the fusion matrix $\Delta_{M^1_{{\A_1}{\A_2}}M^2_{{\A_2}{\A_3}}M^3_{{\A_3}{\A_1}}}$ reads:
\eqn{\ [\Delta_{M^1_{{\A_1}{\A_2}}M^2_{{\A_2}{\A_3}}M^3_{{\A_3}{\A_1}}}]_{r_\rho s_\sigma t_\tau}^{x_\chi y_\upsilon z_\zeta} := \frac{1}{d_{\A_1} d_{\A_2} d_{\A_3}}\sum_{b_\beta\in L_{\A_1}}\sum_{c_\gamma \in L_{\A_2} }\sum_{a_\alpha\in L_{\A_3}}\sum_{p\in L_\Fus}\sum_{u_\rho\in L_{M_{{\A_1}{\A_2}}^1}}\sum_{v_\sigma\in L_{M_{{\A_2}{\A_3}}^2}}\\
\sum_{w_\lambda\in L_{M_{{\A_3}{\A_1}}^3}}
[P_1]^{b_\beta c_\gamma^*}_{x_\chi u r_\rho}\ [P_2]^{c_\gamma a_\alpha^*}_{y_\upsilon v s_\sigma}\ 
\ [P_3]^{a_\alpha b_\beta^*}_{z_\zeta w t_\tau}\ G^{bxu^*}_{c^* r^* p}\ G^{swp}_{br^* t^*}\ G^{pvz}_{aw^* s^*}\ G^{xyz}_{vpc}\ \sqrt{d_ud_vd_wd_ad_bd_cd_rd_sd_t}\ d_p\ &,}
where the triple \(x_\chi, y_\upsilon, z_\zeta\) is the row index, and \(r_\rho, s_\sigma, t_\tau\) is the column index.

The fusion rule of three bimodules \(M_1, M_2, M_3\) can be obtained from the trace of the fusion matrix \(\Delta_{M_1 M_2 M_3}\):
\eqn{\delta_{M_1M_2M_3} = {\rm Tr}[\Delta_{M_1M_2M_3}] = \sum_{r_\rho, s_\sigma, t_\tau\in L_M}[\Delta_{M_1M_2M_3}]_{r_\rho s_\sigma t_\tau}^{r_\rho s_\sigma t_\tau}.}
Here, the three indices \(r_\rho, s_\sigma, t_\tau\) of the matrix in the superscripts or subscripts should be understood as a pair, labeling the fusion vertex. In this paper, we focus on the case in which the fusion coefficients \(N_{M_1 M_2}^{M_3}\) in the bimodule category \({_\A\Bimod_\A(\Fus)}\) can only be \(0\) or \(1\), ensuring that \(\delta_{M_1 M_2 M_3}\) is well-defined.
\item The bimodule conditions \eqref{eq:bimod} induce that matrix \(\Delta_{M_1M_2M_3}\) is a projector:
\eq{
\Delta_{M_1M_2M_3}^2 = \Delta_{M_1M_2M_3}.
}
If \(\delta_{M_1M_2M_3}\ne 0\), we can find the normalized eigenvectors \(\V_{M_1M_2M_3}^{r_\rho s_\sigma t_\tau}\in\C\), such that
\eqn{
&\sum_{r_\rho\in L_{M_1}}\sum_{s_\sigma\in L_{M_2}}\sum_{t_\tau\in L_{M_3}}[\Delta_{M_1M_2M_3}]^{x_\chi y_\upsilon z_\zeta}_{r_\rho s_\sigma t_\tau}\ \V_{M_1M_2M_3}^{r_\rho s_\sigma t_\tau} = \V_{M_1M_2M_3}^{x_\chi y_\upsilon z_\zeta},\\
&\sum_{x_\chi\in L_{M_1}}\sum_{y_\upsilon\in L_{M_2}}\sum_{z_\zeta\in L_{M_3}}|\V_{M_1M_2M_3}^{x_\chi y_\upsilon z_\zeta}|^2\ \sqrt{d_xd_yd_z} = N_{M_1M_2M_3}\sqrt{d_{M_1}d_{M_2}d_{M_3}}\ .}
Here, $N_{M_1M_2M_3}$ is a normalization constant which is determined by the Frobenius algebras of the bimodules. In this paper, only $4$ vertex configurations are involved:
\begin{equation}
    \begin{split}
        &N_{M^1_{{\A_1}{\A_1}}M^2_{{\A_1}{\A_1}}M^3_{{\A_1}{\A_1}}}=d_{\A_1}^2,\ N_{M^1_{{\A_2}{\A_2}}M^2_{{\A_2}{\A_2}}M^3_{{\A_2}{\A_2}}}=d_{\A_2}^2,\\
        &N_{M^1_{{\A_1}{\A_1}}M^2_{{\A_1}{\A_2}}M^3_{{\A_2}{\A_1}}}=d_{\A_1}^{\frac{3}{2}}d_{\A_2}^{\frac{1}{2}},\ N_{M^1_{{\A_2}{\A_2}}M^2_{{\A_2}{\A_1}}M^3_{{\A_1}{\A_2}}}=d_{\A_1}^{\frac{1}{2}}d_{\A_2}^{\frac{3}{2}}.\\
    \end{split}
\end{equation}

For convenience, in a lattice model, we use blue lines labeled by a bimodule \(M\) to represent summing over all states with labels in \(L_M\) on this line. A vertex state with a blue dot at the certex is a superposition with coefficients $\mathcal{V}$:
\eqn[eq:vertexdef]{
\BimoduleD\ .
}
Note that two basis states in the RHS of Eq. \eqref{eq:vertexdef} for the same simple object labels $x, y, z$ but different multiplicity indices $\chi, \upsilon, \zeta$ are regarded orthogonal states. Such a state is invariant under \(\Delta_{M_1M_2M_3}\) matrix. 

\item The \(6j\)-symbols can be derived from the eigenvectors \(\V_{M_1M_2M_3}^{r_\rho s_\sigma t_\tau}\). In this paper, we only consider cases there are only two Frobenius algebras -- $\A_1$ and $\A_2$ -- involved, where the \(6j\)-symbols read
\eqn{
\BimoduleE\ .
}
There are four plaquettes in this diagram\footnote{Three of them are marked by {\color{red} $\times$}, while the fourth corresponds to the outside region. In fact, they are the four faces of a tetrahedron.}. We have already established that each plaquette contains exactly one Frobenius algebra. Here, \(N_{\A_1}\) (\(N_{\A_2}\)) denotes the number of plaquettes containing the Frobenius algebra \(\A_1\) (\(\A_2\)).

\end{enumerate}

\section{General Constructions of Dualities and Symmetry Transformations in the Extended String-Net Model}\label{sec:trans}

In this section, we review the general construction of symmetry transformations through Frobenius algebras and bimodules, as proposed in Ref. \cite{zhao2024a}.

Given a fusion category \(\Fus\) and a Frobenius algebra \(\A \in \Fus\), two string-net models with input data \(\Fus\) and \({_{{\A}}\Bimod_{{\A}}(\Fus)}\), respectively, describe the same topological order. Categorically, \({_{{\A}}\Bimod_{{\A}}(\Fus)}\) is defined via an injective functor
\begin{equation}
\D: {_\A\Bimod_\A(\Fus)} \to \Fus, \qquad M \mapsto \bigoplus_{a \in L_\Fus} n^M_a \, a,
\end{equation}
and for a fusion vertex \(\phi_{M_1 M_2}^{M_3}: M_1 \otimes M_2 \to M_3\) and \(\phi_{xy}^z: x \otimes y \to z\), we define
\begin{equation}
\D(\phi_{M_1 M_2}^{M_3}) = \sum_{z_\zeta \in L_{M_3}} \left[ \sum_{x_\chi \in L_{M_1}} \sum_{y_\upsilon \in L_{M_2}} \mathcal{V}_{M_1 M_2 M_3^*}^{x_\chi y_\upsilon z_\zeta^*} \ \phi_{x_\chi y_\upsilon}^{z_\zeta} \right],
\end{equation}
where \(x_\chi\), \(y_\upsilon\), and \(z_\zeta\) denote the \(\chi\)-th copy of \(x\), the \(\upsilon\)-th copy of \(y\), and the \(\zeta\)-th copy of \(z\) appearing in the direct sum decomposition \(\D(M)\), respectively. Note that two fusion vertices \(\phi_{x_\chi y_\upsilon}^{z_\zeta}\) with the same simple objects \(x, y, z\) but different multiplicity indices \(\chi, \upsilon, \zeta\) are regarded as orthogonal.

Such a functor \(\D\) induces a duality between the two models with \(\F\) and \({_\A\Bimod_\A(\Fus)}\) as the input data respectively:
\eqn{
\Edge{M} \qquad \Longrightarrow\quad \sum_{a_\alpha, b_\beta\in L_\A}\sum_{x_\chi, z_\zeta\in L_M}\BimoduleG\ = \sum_{a_\alpha, b_\beta\in L_\A}\sum_{x_\chi, z_\zeta\in L_M}\sum_{y\in L_\Fus}[P_M]^{a_\alpha b_\beta}_{x_\chi yz_\zeta}\BimoduleX.
}
This duality induces a unitary morphism between the Hilbert spaces \(\Hil_{{_\A\Bimod_\A(\Fus)}}\) and \(\Hil_{\Fus}\) of these two models. Such a unitary linear transformation can be understood plaquette by plaquette:
\begin{align}
\BimoduleH\ .\label{eq:pladuality}
\end{align}
Note that the black edges and tails labeled by \(M_i, N_i \in {_\A\Bimod_\A(\Fus)}\) represent basis states in the dual model, where \({_\A\Bimod_\A(\Fus)}\) serves as the input fusion category and \(M_i, N_i\) denote simple objects. In contract, the blue edges and tails labeled by \(M_i, N_i \in {_\A\Bimod_\A(\Fus)}\) represent superposition states in the original model with input fusion category \(\Fus\), where the superpositions are defined in Eqs. \eqref{eq:bimoddef} and \eqref{eq:vertexdef}. In the RHS of Eq. \eqref{eq:pladuality}, each edge and tail of the plaquette is labeled by simple modules' components $x_i, y_i\in L_{M_i}, p_\alpha, q_\beta\in L_M, e_i\in L_{N_i}$, which may carry multiplicity indices. Nevertheless, after performing pachner moves contracting all extra plaquettes marked by ``{\color{red} $\times$}'', the multiplicity indices of degrees of freedoms on edges are reduced. Only $q_\beta$, which appears at the endpoint of the tail, continues to carry a multiplicity index.

After applying the transformation~\eqref{eq:pladuality} to all plaquettes, the resulting basis state \(\ket{\psi}\) satisfies
\begin{equation}
\langle \psi | \psi \rangle = d_\A^{g - 2},
\end{equation}
where \(g\) is the genus of the surface on which the lattice is embedded. By applying the duality map and appropriately normalizing the resulting basis states, we obtain a unitary morphism between the two string-net models.

After the topological moves in Eq.~\eqref{eq:pladuality}, the degrees of freedom on all edges no longer carry multiplicity indices associated with simple objects in bimodules, whereas those on the tails still retain their multiplicity labels. Therefore, to make sense of this duality and ensure unitarity, we are urged to enlarge the Hilbert space of the original Fibonacci string-net model on each tail but not on the edges, such that two simple objects \(a_\alpha, a_\beta \in L_M\) with different multiplicity indices \(\alpha\ne\beta\) are distinguishable when they appear on tails.

\subsection{Enlarging the Hilbert Space}\label{sec:generalenlarge}

In the enlarged Hilbert space, each tail carries a degree of freedom labeled by a pair \(a_\alpha\), where
\eqn{
a \in L_\mathscr{F}, \qquad \alpha = 1, 2, \dots, N^\mathcal{A}_a, \qquad N^\mathcal{A}_a = \max_{M \in {_\A\Bimod_\A(\Fus)}} \{ n^M_a \},
}
and \(L_{_\A\Bimod_\A(\Fus)}\) denotes the set of all simple bimodules over the Frobenius algebra \(\A\). The degrees of freedom on edges, however, remain to take value varying the simple objects of the input fusion category \(\Fus\). The total Hilbert space is spanned by all enlarged degrees of freedom on tails and original degrees of freedom on edges, subject to the fusion constraints at each vertex.

For any bimodule \(M\), its basis element \(x^M_\chi\in L_M\) corresponds to a superposition state \(\ket{x^M_\chi}\) in the local Hilbert space of a tail:
\eqn[eq:localstatemulti]{
\ket{x^M_\chi} := \sum_{i = 1}^{N^\mathcal{A}_x} A^{x,M}_{\chi,i}\ket{x_i}.
}
The coefficients $A_{\chi, i}^{x, M}$ are determined by solving the orthonormality conditions on the local states in Eq. \eqref{eq:localstatemulti}:
\eqn{\BimoduleF\ .}

\subsection{Duality}\label{sec:dual}

The duality map \eqref{eq:pladuality} can be simplified by expressing the unitary transformation vertex by vertex:
\eqn{
\D := \frac{1}{d_\A^{N_P - 1 + \frac{g}{2}}} \prod_{\text{Edge } e} E_e \prod_{\text{Vertex } v} \D_v,
}
where \(N_P\) is the number of plaquettes in the lattice, and \(\D_v\) denotes the local duality transformation acting on vertex \(v\):
\eqn{
\TransB\ .
}
Note that each edge shared by two vertices are acted upon by two \(\D_v\) operators independently. Nevertheless, an edge \(e\) must carry a unique label. We use \(E_e\) to enforce this uniqueness:
\eqn{
\TransF\quad .
}
The \(E_e\) transformations eliminate the multiplicity indices associated with the edge labels, while the multiplicity indices on the tails are retained.

The duality map does not preserve the full Hilbert space:
\eq{
\Hil_{\Fus} \ne \D \Hil_{{_\A\Bimod_\A(\Fus)}},
}
where \(\Hil_{\Fus}\) and \(\Hil_{{_\A\Bimod_\A(\Fus)}}\) denote the Hilbert spaces of the string-net models with input fusion categories \(\Fus\) and \({_\A\Bimod_\A(\Fus)}\), respectively, considered as subspaces of the enlarged Hilbert space. Nevertheless, since the two models describe the same topological order, the ground-state subspace \(\Hil_0\) is preserved under the duality map:
\eqn{
\Hil_{0, \Fus} = \Hil_{0, {_\A\Bimod_\A(\Fus)}},
}
where \(\Hil_{0, \Fus}\) and \(\Hil_{0, {_\A\Bimod_\A(\Fus)}}\) are the ground-state subspaces of the string-net model with input fusion category \(\Fus\) and \({_\A\Bimod_\A(\Fus)}\), respectively.

\subsection{Symmetry Transformation}\label{sec:symm}

In certain cases, the fusion categories \(\Fus\) and \({_\A\Bimod_\A(\Fus)}\) are isomorphic. That is, there exists an isomorphic functor \(\F_\A\) that maps simple objects of \(\Fus\) to simple objects in \({_\A\Bimod_\A(\Fus)}\): 
\eqn{
\varphi_\A(a) = M_a\in{_\A\Bimod_\A(\Fus)}.
}
Such an equivalence induces a linear isomorphism
\(\varphi_\A: \Hil_{\Fus} \to \Hil_{{_\A\Bimod_\A(\Fus)}}\),
which maps the basic degrees of freedom on both edges and tails to corresponding ones:
\eqn{
\varphi_\A: \Hil_{\Fus}\to \Hil_{{_\A\Bimod_\A(\Fus)}},\qquad\qquad\qquad \TransE\ .
}
The composition
\eqn{\G := \D\circ i: \Hil_\Fus\to \Hil_\Fus}
is just a unitary transformation of the same model with \(\Fus\) as the input fusion category, and the symmetry transformation is the composition of the unitary transformation and projection back into the original degrees of freedom. The set of all such symmetry transformations defines the symmetry of the topological order realized by the string-net model with input \(\Fus\).

In particular, consider the trivial Frobenius algebra \(\A_0\):
\eqn{
L_{\A_0} = \{0\},\qquad f_{000} = 1,
}
whose simple bimodules are labeled by the simple objects in \(L_\Fus\): 
\eqn{L_{M_a} = \{a\},\qquad P^{00}_{aaa} = 1.}
In this case, the gauge transformation induced by the Frobenius algebra \(\A_0\) is the identity transformation of the string-net model. 

\section{Half-Braiding \eqs{z}-Tensors in Our Examples}\label{sec:ztensor}

In this section, we provide the anyon excitation $z$-tensors of the models we used in the mainbody.

\subsection{Toric Code Model}\label{subsec:toric ztensor}

The $z$-tensors of the toric code model are:
\begin{equation}
    z^{1;0}_{000}=z^{1;1}_{001}=1;
\end{equation}
\begin{equation}
    z^{m;1}_{110}=z^{m;0}_{111}=1;
\end{equation}
\begin{equation}
    z^{e;0}_{000}=1,\ z^{e;1}_{001}=-1;
\end{equation}
\begin{equation}
    z^{f;1}_{110}=1,\ z^{f;0}_{111}=-1.
\end{equation}

\subsection{\eqs{\Z_3} Quantum-Double Phase}\label{subsec:Z3 ztensor}

The $z$-tensors of the $\Z_3$ quantum-double phase are:
\begin{equation}
    z^{1;0}_{000}=z^{1;1}_{001}=z^{1;2}_{002}=1;
\end{equation}
\begin{equation}
    z^{e;0}_{000}=1,\ z^{e;1}_{001}=e^{i\frac{\pi}{3}},\ z^{e;2}_{002}=e^{-i\frac{\pi}{3}};
\end{equation}
\begin{equation}
    z^{e^2;0}_{000}=1,\ z^{e^2;1}_{001}=e^{-i\frac{\pi}{3}},\ z^{e^2;2}_{002}=e^{i\frac{\pi}{3}};
\end{equation}
\begin{equation}
    z^{m;1}_{110}=z^{m;2}_{111}=z^{m;2}_{112}=1;
\end{equation}
\begin{equation}
    z^{em;1}_{110}=1,\ z^{em;2}_{111}=e^{i\frac{\pi}{3}},\ z^{em;0}_{112}=e^{-i\frac{\pi}{3}};
\end{equation}
\begin{equation}
    z^{e^2m;1}_{110}=1,\ z^{e^2m;2}_{111}=e^{-i\frac{\pi}{3}},\ z^{e^2m;0}_{112}=e^{i\frac{\pi}{3}};
\end{equation}
\begin{equation}
    z^{m^2;2}_{220}=z^{m^2;0}_{221}=z^{m^2;1}_{222}=1;
\end{equation}
\begin{equation}
    z^{em^2;2}_{220}=1,\ z^{em^2;0}_{221}=e^{i\frac{\pi}{3}},\ z^{em^2;1}_{222}=e^{-i\frac{\pi}{3}};
\end{equation}
\begin{equation}
    z^{e^2m^2;2}_{220}=1,\ z^{e^2m^2;0}_{221}=e^{-i\frac{\pi}{3}},\ z^{e^2m^2;1}_{222}=e^{i\frac{\pi}{3}}.
\end{equation}

\subsection{\eqs{\Z_2\times\Z_2} Quantum-Double Phase}\label{subsec:Z2*Z2 ztensor}

The $z$-tensors of the $\Z_2\times\Z_2$ quantum-double phase are:
\begin{equation}
    z^{(1,1);1}_{111}=z^{(1,1);a}_{11a}=z^{(1,1);b}_{11b}=z^{(1,1);c}_{11c}=1;
\end{equation}
\begin{equation}
    z^{(e,1);1}_{111}=z^{(e,1);b}_{11b}=1,\ z^{(e,1);a}_{11a}=z^{(e,1);c}_{11c}=-1;
\end{equation}
\begin{equation}
    z^{(1,e);1}_{111}=z^{(1,e);a}_{11a}=1,\ z^{(1,e);b}_{11b}=z^{(1,e);c}_{11c}=-1;
\end{equation}
\begin{equation}
    z^{(e,e);1}_{111}=z^{(e,e);c}_{11c}=1,\ z^{(e,e);a}_{11a}=z^{(e,e);b}_{11b}=-1;
\end{equation}
\begin{equation}
    z^{(m,1);a}_{aa1}=z^{(m,1);1}_{aaa}=z^{(m,1);c}_{aab}=z^{(m,1);b}_{aac}=1;
\end{equation}
\begin{equation}
    z^{(f,1);a}_{aa1}=z^{(f,1);c}_{aab}=1,\ z^{(f,1);1}_{aaa}=z^{(f,1);b}_{aac}=-1;
\end{equation}
\begin{equation}
    z^{(m,e);a}_{aa1}=z^{(m,e);1}_{aaa}=1,\ z^{(m,e);c}_{aab}=z^{(m,e);b}_{aac}=-1;
\end{equation}
\begin{equation}
    z^{(f,e);1}_{aa1}=z^{(f,e);b}_{aac}=1,\ z^{(f,e);1}_{aaa}=z^{(f,e);c}_{aab}=-1;
\end{equation}
\begin{equation}
    z^{(1,m);b}_{bb1}=z^{(1,m);c}_{bba}=z^{(1,m);1}_{bbb}=z^{(1,m);a}_{bbc}=1;
\end{equation}
\begin{equation}
    z^{(e,m);b}_{bb1}=z^{(e,m);1}_{bbb}=1,\ z^{(e,m);c}_{bba}=z^{(e,m);a}_{bbc}=-1;
\end{equation}
\begin{equation}
    z^{(1,f);b}_{bb1}=z^{(1,f);c}_{bba}=1,\ z^{(1,f);1}_{bbb}=z^{(1,f);a}_{bbc}=-1;
\end{equation}
\begin{equation}
    z^{(e,f);b}_{bb1}=z^{(e,f);a}_{bbc}=1,\ z^{(e,f);c}_{bba}=z^{(e,f);1}_{bbb}=-1;
\end{equation}
\begin{equation}
    z^{(m,m);c}_{cc1}=z^{(m,m);b}_{cca}=z^{(m,m);b}_{11b}=z^{(m,m);c}_{11c}=1;
\end{equation}
\begin{equation}
    z^{(f,m);c}_{cc1}=z^{(f,m);a}_{ccb}=1,\ z^{(f,m);b}_{cca}=z^{(f,m);1}_{ccc}=-1;
\end{equation}
\begin{equation}
    z^{(m,f);c}_{cc1}=z^{(m,f);b}_{cca}=1,\ z^{(m,f);a}_{ccb}=z^{(m,f);1}_{ccc}=-1;
\end{equation}
\begin{equation}
    z^{(f,f);c}_{cc1}=z^{(f,f);1}_{ccc}=1,\ z^{(f,f);b}_{cca}=z^{(f,f);a}_{ccb}=-1.
\end{equation}

\subsection{\eqs{S_3} Quantum-Double Phase}\label{subsec:S3 ztensor}

The $z$-tensors of the $S_3$ quantum-double phase are:
\begin{equation}
    z^{A;1}_{1,1,1}=z^{A;r}_{1,1,r}=z^{A;r^2}_{1,1,r^2}=z^{A;s}_{1,1,s}=z^{A;rs}_{1,1,rs}=z^{A;sr}_{1,1,sr}=1;
\end{equation}
\begin{equation}
    z^{B;1}_{1,1,1}=z^{B;r}_{1,1,r}=z^{B;r^2}_{1,1,r^2}=1,\ z^{B;s}_{1,1,s}=z^{B;rs}_{1,1,rs}=z^{B;sr}_{1,1,sr}=-1;
\end{equation}
\begin{equation}
    \begin{split}
        &z^{C;1}_{1_1,1_1,1}=1,\ z^{C;r}_{1_1,1_1,r}=e^{i\frac{2\pi}{3}},\ z^{C;r^2}_{1_1,1_1,r^2}=e^{-i\frac{2\pi}{3}},\\
        &z^{C;1}_{1_2,1_2,1}=1,\ z^{C;r}_{1_2,1_2,r}=e^{-i\frac{2\pi}{3}},\ z^{C;r^2}_{1_2,1_2,r^2}=e^{i\frac{2\pi}{3}},\\
        &z^{C;s}_{1_1,1_2,s}=1,\ z^{C;rs}_{1_1,1_2,rs}=e^{i\frac{2\pi}{3}},\ z^{C;sr}_{1_1,1_2,sr}=e^{-i\frac{2\pi}{3}},\\
        &z^{C;s}_{1_2,1_1,s}=1,\ z^{C;rs}_{1_2,1_1,rs}=e^{-i\frac{2\pi}{3}},\ z^{C;sr}_{1_2,1_1,sr}=e^{i\frac{2\pi}{3}};\\
    \end{split}
\end{equation}
\begin{equation}
    \begin{split}
        &z^{D;s}_{s,s,1}=z^{D;sr}_{s,rs,r}=z^{D;rs}_{s,sr,r^2}=z^{D;rs}_{rs,rs,1}=z^{D;s}_{rs,sr,r}=z^{D;sr}_{rs,s,r^2}=1,\\
        &z^{D;sr}_{sr,sr,1}=z^{D;rs}_{sr,s,r}=z^{D;s}_{sr,rs,r^2}=z^{D;1}_{s,s,s}=z^{D;r}_{s,rs,sr}=z^{D;r^2}_{s,sr,rs}=1,\\
        &z^{D;1}_{rs,rs,rs}=z^{D;r}_{rs,sr,s}=z^{D;r^2}_{rs,s,sr}=z^{D;1}_{sr,sr,sr}=z^{D;r}_{sr,s,sr}=z^{D;r^2}_{sr,rs,s}=1;\\
    \end{split}
\end{equation}
\begin{equation}
    \begin{split}
        &z^{E;s}_{s,s,1}=z^{E;sr}_{s,rs,r}=z^{E;rs}_{s,sr,r^2}=z^{E;rs}_{rs,rs,1}=z^{E;s}_{rs,sr,r}=z^{E;sr}_{rs,s,r^2}=1,\\
        &z^{E;sr}_{sr,sr,1}=z^{E;rs}_{sr,s,r}=z^{E;s}_{sr,rs,r^2}=1,\ z^{E;1}_{s,s,s}=z^{E;r}_{s,rs,sr}=z^{E;r^2}_{s,sr,rs}=-1,\\
        &z^{E;1}_{rs,rs,rs}=z^{E;r}_{rs,sr,s}=z^{E;r^2}_{rs,s,sr}=z^{E;1}_{sr,sr,sr}=z^{E;r}_{sr,s,sr}=z^{E;r^2}_{sr,rs,s}=-1;\\
    \end{split}
\end{equation}
\begin{equation}
    \begin{split}
        &z^{F;r}_{r,r,1}=z^{F;r^2}_{r,r,r}=z^{F;1}_{r,r,r^2}=z^{F;r^2}_{r^2,r^2,1}=z^{F;1}_{r^2,r^2,r}=z^{F;r}_{r^2,r^2,r^2}=1,\\
        &z^{F;rs}_{r,r^2,s}=z^{F;sr}_{r,r^2,rs}=z^{F;rs}_{r,r^2,sr}=z^{F;sr}_{r^2,r,s}=z^{F;s}_{r^2,r,rs}=z^{F;rs}_{r^2,r,sr}=1;\\
    \end{split}
\end{equation}
\begin{equation}
    \begin{split}
        &z^{G;r}_{r,r,1}=1,\ z^{G;r^2}_{r,r,r}=e^{i\frac{2\pi}{3}},\ z^{G;1}_{r,r,r^2}=e^{-i\frac{2\pi}{3}},\\
        &z^{G;r^2}_{r^2,r^2,1}=1,\ z^{G;1}_{r^2,r^2,r}=e^{-i\frac{2\pi}{3}},\ z^{G;r}_{r^2,r^2,r^2}=e^{i\frac{2\pi}{3}},\\
        &z^{G;rs}_{r,r^2,s}=1,\ z^{G;sr}_{r,r^2,rs}=e^{i\frac{2\pi}{3}},\ z^{G;rs}_{r,r^2,sr}=e^{-i\frac{2\pi}{3}},\\
        &z^{G;sr}_{r^2,r,s}=1,\ z^{G;s}_{r^2,r,rs}=e^{-i\frac{2\pi}{3}},\ z^{G;rs}_{r^2,r,sr}=e^{i\frac{2\pi}{3}};\\
    \end{split}
\end{equation}
\begin{equation}
    \begin{split}
        &z^{H;r}_{r,r,1}=1,\ z^{H;r^2}_{r,r,r}=e^{-i\frac{2\pi}{3}},\ z^{H;1}_{r,r,r^2}=e^{i\frac{2\pi}{3}},\\
        &z^{H;r^2}_{r^2,r^2,1}=1,\ z^{H;1}_{r^2,r^2,r}=e^{i\frac{2\pi}{3}},\ z^{H;r}_{r^2,r^2,r^2}=e^{-i\frac{2\pi}{3}},\\
        &z^{H;rs}_{r,r^2,s}=1,\ z^{H;sr}_{r,r^2,rs}=e^{-i\frac{2\pi}{3}},\ z^{H;rs}_{r,r^2,sr}=e^{i\frac{2\pi}{3}},\\
        &z^{H;sr}_{r^2,r,s}=1,\ z^{H;s}_{r^2,r,rs}=e^{i\frac{2\pi}{3}},\ z^{H;rs}_{r^2,r,sr}=e^{-i\frac{2\pi}{3}}.\\
    \end{split}
\end{equation}

\subsection{EM-Exchange Symmetry Enriched Toric Code Model}\label{subsec:Z2toric ztensor}

The $z$-tensors of the EM-exchange symmetry enriched toric code are:
\begin{equation}
    \begin{split}
        &z^{(1,1);M_0}_{M_0M_0M_0}=z^{(1,1);M_1}_{M_0M_0M_1}=z^{(1,1);M_+}_{M_+M_+M_+}=z^{(1,1);M_-}_{M_+M_+M_-}=1,\\
        &z^{(1,1);M_{\sigma}}_{M_0M_+M_{\sigma}}=z^{(1,1);M^*_{\sigma}}_{M_+M_0M^*_{\sigma}}=1;
    \end{split}
\end{equation}
\begin{equation}
    \begin{split}
        &z^{(m,e);M_1}_{M_1M_1M_0}=z^{(m,e);M_0}_{M_1M_1M_1}=z^{(m,e);M_+}_{M_+M_+M_+}=1,\ z^{(m,e);M_-}_{M_+M_+M_-}=-1,\\
        &z^{(m,e);M_{\sigma}}_{M_1M_+M_{\sigma}}=z^{(m,e);M^*_{\sigma}}_{M_+M_1M^*_{\sigma}}=1;
    \end{split}
\end{equation}
\begin{equation}
    \begin{split}
        &z^{(e,m);M_-}_{M_-M_-M_+}=z^{(e,m);M_+}_{M_-M_-M_-}=z^{(e,m);M_0}_{M_0M_0M_0}=1,\ z^{(e,m);M_1}_{M_0M_0M_1}=-1,\\
        &z^{(e,m);M^*_{\sigma}}_{M_-M_0M^*_{\sigma}}=z^{(e,m);M_{\sigma}}_{M_0M_-M_{\sigma}}=1;
    \end{split}
\end{equation}
\begin{equation}
    \begin{split}
        &z^{(f,f);M_1}_{M_1M_1M_0}=z^{(f,f);M_-}_{M_-M_-M_+}=1,\ z^{(f,f);M_0}_{M_1M_1M_1}=z^{(f,f);M_+}_{M_-M_-M_-}=-1,\\
        &z^{(f,f);M_{\sigma}}_{M_1M_-M_{\sigma}}=z^{(f,f);M^*_{\sigma}}_{M_-M_1M^*_{\sigma}}=1.
    \end{split}
\end{equation}

\subsection{Charge Conjugation Symmetry Enriched \eqs{Z_3} Quantum-Double Phase}\label{subsec:Z2Z3 ztensor}

The $z$-tensors of the charge conjuagtion symmetry enriched $\Z_3$ quantum-double phase are:
\begin{equation}
    \begin{split}
        &z^{(1,1);1}_{1_{++},1_{++},1_{++}}=z^{(1,1);r_{++}}_{1_{++},1_{++},r_{++}}=z^{(1,1);r^2_{++}}_{1_{++},1_{++},r^2_{++}}=1,\\
        &z^{(1,1);s_{+-}}_{1_{++},1_{--},s_{+-}}=z^{(1,1);rs_{+-}}_{1_{++},1_{--},rs_{+-}}=z^{(1,1);sr_{+-}}_{1_{++},1_{--},sr_{+-}}=1;\\
        &z^{(1,1);1}_{1_{--},1_{--},1_{--}}=z^{(1,1);r_{--}}_{1_{--},1_{--},r_{-+}}=z^{(1,1);r^2_{--}}_{1_{--},1_{--},r^2_{--}}=1,\\
        &z^{(1,1);s_{-+}}_{1_{--},1_{++},s_{-+}}=z^{(1,1);rs_{-+}}_{1_{--},1_{++},rs_{-+}}=z^{(1,1);sr_{-+}}_{1_{--},1_{++},sr_{-+}}=1;
    \end{split}
\end{equation}
\begin{equation}
    \begin{split}
        &z^{(e,e^2);1_{++}}_{1_{++},1_{++},1_{++}}=1,\ z^{(e,e^2);r_{++}}_{1_{++},1_{++},r_{++}}=e^{i\frac{2\pi}{3}},\ z^{(e,e^2);r^2_{++}}_{1_{++},1_{++},r^2_{++}}=e^{-i\frac{2\pi}{3}},\\
        &z^{(e,e^2);1_{--}}_{1_{--},1_{--},1_{--}}=1,\ z^{(e,e^2);r_{--}}_{1_{--},1_{--},r_{--}}=e^{-i\frac{2\pi}{3}},\ z^{(e,e^2);r^2_{--}}_{1_{--},1_{--},r^2_{--}}=e^{i\frac{2\pi}{3}},\\
        &z^{(e,e^2);s_{+-}}_{1_{++},1_{--},s_{+-}}=1,\ z^{(e,e^2);rs_{+-}}_{1_{++},1_{--},rs_{+-}}=e^{i\frac{2\pi}{3}},\ z^{(e,e^2);sr_{+-}}_{1_{++},1_{--},sr_{+-}}=e^{-i\frac{2\pi}{3}},\\
        &z^{(e,e^2);s_{-+}}_{1_{--},1_{++},s_{-+}}=1,\ z^{(e,e^2);rs_{-+}}_{1_{--},1_{++},rs_{-+}}=e^{-i\frac{2\pi}{3}},\ z^{(e,e^2);sr_{-+}}_{1_{--},1_{++},sr_{-+}}=e^{i\frac{2\pi}{3}};
    \end{split}
\end{equation}
\begin{equation}
    \begin{split}
        &z^{(e^2,e);1_{++}}_{1_{++},1_{++},1_{++}}=1,\ z^{(e^2,e);r_{++}}_{1_{++},1_{++},r_{++}}=e^{-i\frac{2\pi}{3}},\ z^{(e^2,e);r^2_{++}}_{1_{++},1_{++},r^2_{++}}=e^{i\frac{2\pi}{3}},\\
        &z^{(e^2,e);1_{--}}_{1_{--},1_{--},1_{--}}=1,\ z^{(e^2,e);r_{--}}_{1_{--},1_{--},r_{--}}=e^{i\frac{2\pi}{3}},\ z^{(e^2,e);r^2_{--}}_{1_{--},1_{--},r^2_{--}}=e^{-i\frac{2\pi}{3}},\\
        &z^{(e^2,e);s_{+-}}_{1_{++},1_{--},s_{+-}}=1,\ z^{(e^2,e);rs_{+-}}_{1_{++},1_{--},rs_{+-}}=e^{-i\frac{2\pi}{3}},\ z^{(e^2,e);sr_{+-}}_{1_{++},1_{--},sr_{+-}}=e^{i\frac{2\pi}{3}},\\
        &z^{(e^2,e);s_{-+}}_{1_{--},1_{++},s_{-+}}=1,\ z^{(e^2,e);rs_{-+}}_{1_{--},1_{++},rs_{-+}}=e^{i\frac{2\pi}{3}},\ z^{(e^2,e);sr_{-+}}_{1_{--},1_{++},sr_{-+}}=e^{-i\frac{2\pi}{3}};
    \end{split}
\end{equation}
\begin{equation}
    \begin{split}
        &z^{(m,m^2);r_{++}}_{r_{++},r_{++},1_{++}}=z^{(m,m^2);r^2_{++}}_{r_{++},r_{++},r_{++}}=z^{(m,m^2);1_{++}}_{r_{++},r_{++},r^2_{++}}=1,\\
        &z^{(m,m^2);r^2_{--}}_{r^2_{--},r^2_{--},1_{--}}=z^{(m,m^2);1_{--}}_{r^2_{--},r^2_{--},r_{--}}=z^{(m,m^2);r_{--}}_{r^2_{--},r^2_{--},r^2_{--}}=1,\\
        &z^{(m,m^2);rs_{+-}}_{r_{++},r^2_{--},s_{+-}}=z^{(m,m^2);sr_{+-}}_{r_{++},r^2_{--},rs_{+-}}=z^{(m,m^2);rs_{+-}}_{r_{++},r^2_{--},sr_{+-}}=1,\\
        &z^{(m,m^2);sr_{-+}}_{r^2_{--},r_{++},s_{-+}}=z^{(m,m^2);s_{-+}}_{r^2_{--},r_{++},rs_{-+}}=z^{(m,m^2);rs_{-+}}_{r^2_{--},r_{++},sr_{-+}}=1;
    \end{split}
\end{equation}
\begin{equation}
    \begin{split}
        &z^{(m^2,m);r^2_{++}}_{r^2_{++},r^2_{++},1_{++}}=z^{(m^2,m);1_{++}}_{r^2_{++},r^2_{++},r_{++}}=z^{(m^2,m);r_{++}}_{r^2_{++},r^2_{++},r^2_{++}}=1,\\
        &z^{(m^2,m);r_{--}}_{r_{--},r_{--},1_{--}}=z^{(m^2,m);r^2_{--}}_{r_{--},r_{--},r_{--}}=z^{(m^2,m);1_{--}}_{r_{--},r_{--},r^2_{--}}=1,\\
        &z^{(m^2,m);sr_{+-}}_{r^2_{++},r_{--},s_{+-}}=z^{(m^2,m);s_{+-}}_{r^2_{++},r_{--},rs_{+-}}=z^{(m^2,m);rs_{+-}}_{r^2_{++},r_{--},sr_{+-}}=1,\\
        &z^{(m^2,m);rs_{-+}}_{r_{--},r^2_{++},s_{-+}}=z^{(m^2,m);sr_{-+}}_{r_{--},r^2_{++},rs_{-+}}=z^{(m^2,m);rs_{-+}}_{r_{--},r^2_{++},sr_{-+}}=1;
    \end{split}
\end{equation}
\begin{equation}
    \begin{split}
        &z^{(em,e^2m^2);r_{++}}_{r_{++},r_{++},1_{++}}=1,\ z^{(em,e^2m^2);r^2_{++}}_{r_{++},r_{++},r_{++}}=e^{i\frac{2\pi}{3}},\ z^{(em,e^2m^2);1_{++}}_{r_{++},r_{++},r^2_{++}}=e^{-i\frac{2\pi}{3}},\\
        &z^{(em,e^2m^2);r^2_{--}}_{r^2_{--},r^2_{--},1_{--}}=1,\ z^{(em,e^2m^2);1_{--}}_{r^2_{--},r^2_{--},r_{--}}=e^{-i\frac{2\pi}{3}},\ z^{(em,e^2m^2);r_{--}}_{r^2_{--},r^2_{--},r^2_{--}}=e^{i\frac{2\pi}{3}},\\
        &z^{(em,e^2m^2);rs_{+-}}_{r_{++},r^2_{--},s_{+-}}=1,\ z^{(em,e^2m^2);sr_{+-}}_{r_{++},r^2_{--},rs_{+-}}=e^{i\frac{2\pi}{3}},\ z^{(em,e^2m^2);rs_{+-}}_{r_{++},r^2_{--},sr_{+-}}=e^{-i\frac{2\pi}{3}},\\
        &z^{(em,e^2m^2);sr_{-+}}_{r^2_{--},r_{++},s_{-+}}=1,\ z^{(em,e^2m^2);s_{-+}}_{r^2_{--},r_{++},rs_{-+}}=e^{-i\frac{2\pi}{3}},\ z^{(em,e^2m^2);rs_{-+}}_{r^2_{--},r_{++},sr_{-+}}=e^{i\frac{2\pi}{3}};
    \end{split}
\end{equation}
\begin{equation}
    \begin{split}
        &z^{(e^2m^2,em);r^2_{++}}_{r^2_{++},r^2_{++},1_{++}}=1,\ z^{(e^2m^2,em);1_{++}}_{r^2_{++},r^2_{++},r_{++}}=e^{-i\frac{2\pi}{3}},\ z^{(e^2m^2,em);r_{++}}_{r^2_{++},r^2_{++},r^2_{++}}=e^{i\frac{2\pi}{3}},\\
        &z^{(e^2m^2,em);r_{--}}_{r_{--},r_{--},1_{--}}=1,\ z^{(e^2m^2,em);r^2_{--}}_{r_{--},r_{--},r_{--}}=e^{i\frac{2\pi}{3}},\ z^{(e^2m^2,em);1_{--}}_{r_{--},r_{--},r^2_{--}}=e^{-i\frac{2\pi}{3}},\\
        &z^{(e^2m^2,em);sr_{+-}}_{r^2_{++},r_{--},s_{+-}}=1,\ z^{(e^2m^2,em);s_{+-}}_{r^2_{++},r_{--},rs_{+-}}=e^{-i\frac{2\pi}{3}},\ z^{(e^2m^2,em);rs_{+-}}_{r_{++},r^2_{--},sr_{+-}}=e^{i\frac{2\pi}{3}},\\
        &z^{(e^2m^2,em);rs_{-+}}_{r_{--},r^2_{++},s_{-+}}=1,\ z^{(e^2m^2,em);sr_{-+}}_{r_{--},r^2_{++},rs_{-+}}=e^{i\frac{2\pi}{3}},\ z^{(e^2m^2,em);rs_{-+}}_{r_{--},r^2_{++},sr_{-+}}=e^{-i\frac{2\pi}{3}};
    \end{split}
\end{equation}
\begin{equation}
    \begin{split}
        &z^{(e^2m,em^2);r_{++}}_{r_{++},r_{++},1_{++}}=1,\ z^{(e^2m,em^2);r^2_{++}}_{r_{++},r_{++},r_{++}}=e^{-i\frac{2\pi}{3}},\ z^{(e^2m,em^2);1_{++}}_{r_{++},r_{++},r^2_{++}}=e^{i\frac{2\pi}{3}},\\
        &z^{(e^2m,em^2);r^2_{--}}_{r^2_{--},r^2_{--},1_{--}}=1,\ z^{(e^2m,em^2);1_{--}}_{r^2_{--},r^2_{--},r_{--}}=e^{i\frac{2\pi}{3}},\ z^{(e^2m,em^2);r_{--}}_{r^2_{--},r^2_{--},r^2_{--}}=e^{-i\frac{2\pi}{3}},\\
        &z^{(e^2m,em^2);rs_{+-}}_{r_{++},r^2_{--},s_{+-}}=1,\ z^{(e^2m,em^2);sr_{+-}}_{r_{++},r^2_{--},rs_{+-}}=e^{-i\frac{2\pi}{3}},\ z^{(e^2m,em^2);rs_{+-}}_{r_{++},r^2_{--},sr_{+-}}=e^{i\frac{2\pi}{3}},\\
        &z^{(e^2m,em^2);sr_{-+}}_{r^2_{--},r_{++},s_{-+}}=1,\ z^{(e^2m,em^2);s_{-+}}_{r^2_{--},r_{++},rs_{-+}}=e^{i\frac{2\pi}{3}},\ z^{(e^2m,em^2);rs_{-+}}_{r^2_{--},r_{++},sr_{-+}}=e^{-i\frac{2\pi}{3}};
    \end{split}
\end{equation}
\begin{equation}
    \begin{split}
        &z^{(em^2,e^2m);r^2_{++}}_{r^2_{++},r^2_{++},1_{++}}=1,\ z^{(em^2,e^2m);1_{++}}_{r^2_{++},r^2_{++},r_{++}}=e^{i\frac{2\pi}{3}},\ z^{(em^2,e^2m);r_{++}}_{r^2_{++},r^2_{++},r^2_{++}}=e^{-i\frac{2\pi}{3}},\\
        &z^{(em^2,e^2m);r_{--}}_{r_{--},r_{--},1_{--}}=1,\ z^{(em^2,e^2m);r^2_{--}}_{r_{--},r_{--},r_{--}}=e^{-i\frac{2\pi}{3}},\ z^{(em^2,e^2m);1_{--}}_{r_{--},r_{--},r^2_{--}}=e^{i\frac{2\pi}{3}},\\
        &z^{(em^2,e^2m);sr_{+-}}_{r^2_{++},r_{--},s_{+-}}=1,\ z^{(em^2,e^2m);s_{+-}}_{r^2_{++},r_{--},rs_{+-}}=e^{i\frac{2\pi}{3}},\ z^{(em^2,e^2m);rs_{+-}}_{r_{++},r^2_{--},sr_{+-}}=e^{-i\frac{2\pi}{3}},\\
        &z^{(em^2,e^2m);rs_{-+}}_{r_{--},r^2_{++},s_{-+}}=1,\ z^{(em^2,e^2m);sr_{-+}}_{r_{--},r^2_{++},rs_{-+}}=e^{-i\frac{2\pi}{3}},\ z^{(em^2,e^2m);rs_{-+}}_{r_{--},r^2_{++},sr_{-+}}=e^{i\frac{2\pi}{3}}.
    \end{split}
\end{equation}

\subsection{EM-Exchange Symmetry Enriched \eqs{S_3} Quantum-Double Phase}\label{subsec:Z2S3 ztensor}

The $z$-tensors of the $em$-exchange symmetry enriched $S_3$ quantum-double phase are:
\begin{align*}
        &z^{(A,A);M_1}_{M_1,M_1,M_1}=z^{(A,A);M_r}_{M_1,M_1,M_r}=z^{(A,A);M_{r^2}}_{M_1,M_1,M_{r^2}}=z^{(A,A);M_s}_{M_1,M_1,M_s}=1,\\
        &z^{(A,A);M_{rs}}_{M_1,M_1,M_{rs}}=z^{(A,A);M_{sr}}_{M_1,M_1,M_{sr}}=z^{(A,A);M_{\alpha}}_{M_1,M_I,M_{\alpha}}=z^{(A,A);M_{\beta}}_{M_1,M_I,M_{\beta}}=1,\\
        &z^{(A,A);M_I}_{M_I,M_I,M_I}=z^{(A,A);M_R}_{M_I,M_I,M_R}=z^{(A,A);M_{R^2}}_{M_I,M_I,M_{R^2}}=z^{(A,A);M_S}_{M_I,M_I,M_S}=1,\\
        &z^{(A,A);M_{RS}}_{M_I,M_I,M_{RS}}=z^{(A,A);M_{SR}}_{M_I,M_I,M_{SR}}=z^{(A,A);M^*_{\alpha}}_{M_I,M_1,M^*_{\alpha}}=z^{(A,A);M^*_{\beta}}_{M_I,M_1,M^*_{\beta}}=1;
\end{align*}
\begin{align*}
        &z^{(B,B);M_1}_{M_1,M_1,M_1}=z^{(B,B);M_r}_{M_1,M_1,M_r}=z^{(B,B);M_{r^2}}_{M_1,M_1,M_{r^2}}=1,\\
        &z^{(B,B);M_s}_{M_1,M_1,M_s}=z^{(B,B);M_{rs}}_{M_1,M_1,M_{rs}}=z^{(B,B);M_{sr}}_{M_1,M_1,M_{sr}}=-1,\\
        &z^{(B,B);M_I}_{M_I,M_I,M_I}=z^{(B,B);M_R}_{M_I,M_I,M_R}=z^{(B,B);M_{R^2}}_{M_I,M_I,M_{R^2}}=1,\\
        &z^{(B,B);M_S}_{M_I,M_I,M_S}=z^{(B,B);M_{RS}}_{M_I,M_I,M_{RS}}=z^{(B,B);M_{SR}}_{M_I,M_I,M_{SR}}=-1,\\
        &z^{(B,B);M_{\alpha}}_{M_1,M_I,M_{\alpha}}=z^{(B,B);M_{\beta}}_{M_1,M_I,M_{\beta}}=z^{(B,B);M^*_{\alpha}}_{M_I,M_1,M^*_{\alpha}}=z^{(B,B);M^*_{\beta}}_{M_I,M_1,M^*_{\beta}}=1;
\end{align*}
\begin{align*}
        &z^{(C,F);M_1}_{M_{1_1},M_{1_1},M_1}=1,\ z^{(C,F);M_r}_{M_{1_1},M_{1_1},M_r}=e^{i\frac{2\pi}{3}},\ z^{(C,F);M_{r^2}}_{M_{1_1},M_{1_1},M_{r^2}}=e^{-i\frac{2\pi}{3}},\\
        &z^{(C,F);M_1}_{M_{1_2},M_{1_2},M_1}=1,\ z^{(C,F);M_r}_{M_{1_2},M_{1_2},M_r}=e^{-i\frac{2\pi}{3}},\ z^{(C,F);M_{r^2}}_{M_{1_2},M_{1_2},M_{r^2}}=e^{i\frac{2\pi}{3}},\\
        &z^{(C,F);M_s}_{M_{1_1},M_{1_2},M_s}=1,\ z^{(C,F);M_{rs}}_{M_{1_1},M_{1_2},M_{rs}}=e^{i\frac{2\pi}{3}},\ z^{(C,F);M_{sr}}_{M_{1_1},M_{1_2},M_{sr}}=e^{-i\frac{2\pi}{3}},\\
        &z^{(C,F);M_s}_{M_{1_2},M_{1_1},M_s}=1,\ z^{(C,F);M_{rs}}_{M_{1_2},M_{1_1},M_{rs}}=e^{-i\frac{2\pi}{3}},\ z^{(C,F);M_{sr}}_{M_{1_2},M_{1_1},M_{sr}}=e^{i\frac{2\pi}{3}},\\
        &z^{(C,F);M_R}_{M_R,M_{R},M_I}=z^{(C,F);M_{R^2}}_{M_R,M_{R},M_R}=z^{(C,F);M_I}_{M_R,M_{R},M_{R^2}}=1,\\
        &z^{(C,F);M_{R^2}}_{M_{R^2},M_{R^2},M_I}=z^{(C,F);M_I}_{M_{R^2},M_{R^2},M_R}=z^{(C,F);M_R}_{M_{R^2},M_{R^2},M_{R^2}}=1,\\
        &z^{(C,F);M_{RS}}_{M_R,M_{R^2},M_S}=z^{(C,F);M_{SR}}_{M_R,M_{R^2},M_{RS}}=z^{(C,F);M_{RS}}_{M_R,M_{R^2},M_{SR}}=1,\\
        &z^{(C,F);M_{SR}}_{M_{R^2},M_{R},M_S}=z^{(C,F);M_S}_{M_{R^2},M_{R},M_{RS}}=z^{(C,F);M_{RS}}_{M_{R^2},M_{R},M_{SR}}=1,\\
        &z^{(C,F);M_{\alpha}}_{M_{1_1},M_{R},M_{\alpha}}=z^{(C,F);M_{\alpha}}_{M_{1_2},M_{R^2},M_{\alpha}}=z^{(C,F);M^*_{\alpha}}_{M_R,M_{1_1},M^*_{\alpha}}=z^{(C,F);M^*_{\alpha}}_{M_{R^2},M_{1_2},M^*_{\alpha}}=1,\\
        &z^{(C,F);M_{\beta}}_{M_{1_1},M_{R^2},M_{\beta}}=z^{(C,F);M_{\beta}}_{M_{1_2},M_{R},M_{\beta}}=z^{(C,F);M^*_{\beta}}_{M_{R^2},M_{1_1},M^*_{\beta}}=z^{(C,F);M^*_{\beta}}_{M_R,M_{1_2},M^*_{\beta}}=1;
\end{align*}
\begin{align*}
        &z^{(F,C);M_I}_{M_{I_1},M_{I_1},M_I}=1,\ z^{(F,C);M_R}_{M_{I_1},M_{I_1},M_R}=e^{i\frac{2\pi}{3}},\ z^{(F,C);M_{R^2}}_{M_{I_1},M_{I_1},M_{R^2}}=e^{-i\frac{2\pi}{3}},\\
        &z^{(F,C);M_I}_{M_{I_2},M_{I_2},M_I}=1,\ z^{(F,C);M_R}_{M_{I_2},M_{I_2},M_R}=e^{-i\frac{2\pi}{3}},\ z^{(F,C);M_{R^2}}_{M_{I_2},M_{I_2},M_{R^2}}=e^{i\frac{2\pi}{3}},\\
        &z^{(F,C);M_S}_{M_{I_1},M_{I_2},M_S}=1,\ z^{(F,C);M_{RS}}_{M_{I_1},M_{I_2},M_{RS}}=e^{i\frac{2\pi}{3}},\ z^{(F,C);M_{SR}}_{M_{I_1},M_{I_2},M_{SR}}=e^{-i\frac{2\pi}{3}},\\
        &z^{(F,C);M_S}_{M_{I_2},M_{I_1},M_S}=1,\ z^{(F,C);M_{RS}}_{M_{I_2},M_{I_1},M_{RS}}=e^{-i\frac{2\pi}{3}},\ z^{(F,C);M_{SR}}_{M_{I_2},M_{I_1},M_{SR}}=e^{i\frac{2\pi}{3}},\\
        &z^{(F,C);M_r}_{M_r,M_{r},M_1}=z^{(F,C);M_{r^2}}_{M_r,M_{r},M_r}=z^{(F,C);M_1}_{M_r,M_{r},M_{r^2}}=1,\\
        &z^{(F,C);M_{r^2}}_{M_{r^2},M_{r^2},M_1}=z^{(F,C);M_1}_{M_{r^2},M_{r^2},M_r}=z^{(F,C);M_r}_{M_{r^2},M_{r^2},M_{r^2}}=1,\\
        &z^{(F,C);M_{rs}}_{M_r,M_{r^2},M_s}=z^{(F,C);M_{sr}}_{M_r,M_{r^2},M_{rs}}=z^{(F,C);M_{rs}}_{M_r,M_{r^2},M_{sr}}=1,\\
        &z^{(F,C);M_{sr}}_{M_{r^2},M_{r},M_s}=z^{(F,C);M_s}_{M_{r^2},M_{r},M_{rs}}=z^{(F,C);M_{rs}}_{M_{r^2},M_{r},M_{sr}}=1,\\
        &z^{(F,C);M^*_{\alpha}}_{M_{I_1},M_{r},M^*_{\alpha}}=z^{(F,C);M^*_{\alpha}}_{M_{I_2},M_{r^2},M^*_{\alpha}}=z^{(F,C);M_{\alpha}}_{M_r,M_{I_1},M_{\alpha}}=z^{(F,C);M_{\alpha}}_{M_{r^2},M_{I_2},M_{\alpha}}=1,\\
        &z^{(F,C);M^*_{\beta}}_{M_{I_1},M_{r^2},M^*_{\beta}}=z^{(F,C);M^*_{\beta}}_{M_{I_2},M_{r},M^*_{\beta}}=z^{(F,C);M_{\beta}}_{M_{r^2},M_{I_1},M_{\beta}}=z^{(F,C);M_{\beta}}_{M_r,M_{I_2},M_{\beta}}=1;
\end{align*}
\begin{align*}
        &z^{(G,G);M_r}_{M_r,M_{r},M_1}=1,\ z^{(G,G);M_{r^2}}_{M_r,M_{r},M_r}=e^{i\frac{2\pi}{3}},\ z^{(G,G);M_1}_{M_r,M_{r},M_{r^2}}=e^{-i\frac{2\pi}{3}},\\
        &z^{(G,G);M_{r^2}}_{M_{r^2},M_{r^2},M_1}=1,\ z^{(G,G);M_1}_{M_{r^2},M_{r^2},M_r}=e^{-i\frac{2\pi}{3}},\ z^{(G,G);M_r}_{M_{r^2},M_{r^2},M_{r^2}}=e^{i\frac{2\pi}{3}},\\
        &z^{(G,G);M_{rs}}_{M_r,M_{r^2},M_s}=1,\ z^{(G,G);M_{sr}}_{M_r,M_{r^2},M_{rs}}=e^{i\frac{2\pi}{3}},\ z^{(G,G);M_{rs}}_{M_r,M_{r^2},M_{sr}}=e^{-i\frac{2\pi}{3}},\\
        &z^{(G,G);M_{sr}}_{M_{r^2},M_{r},M_s}=1,\ z^{(G,G);M_s}_{M_{r^2},M_{r},M_{rs}}=e^{-i\frac{2\pi}{3}},\ z^{(G,G);M_{rs}}_{M_{r^2},M_{r},M_{sr}}=e^{i\frac{2\pi}{3}},\\
        &z^{(G,G);M_R}_{M_R,M_{R},M_I}=1,\ z^{(G,G);M_{R^2}}_{M_R,M_{R},M_R}=e^{i\frac{2\pi}{3}},\ z^{(G,G);M_I}_{M_R,M_{R},M_{R^2}}=e^{-i\frac{2\pi}{3}},\\
        &z^{(G,G);M_{R^2}}_{M_{R^2},M_{R^2},M_I}=1,\ z^{(G,G);M_I}_{M_{R^2},M_{R^2},M_R}=e^{-i\frac{2\pi}{3}},\ z^{(G,G);M_R}_{M_{R^2},M_{R^2},M_{R^2}}=e^{i\frac{2\pi}{3}},\\
        &z^{(G,G);M_{RS}}_{M_R,M_{R^2},M_S}=1,\ z^{(G,G);M_{SR}}_{M_R,M_{R^2},M_{RS}}=e^{i\frac{2\pi}{3}},\ z^{(G,G);M_{RS}}_{M_R,M_{R^2},M_{SR}}=e^{-i\frac{2\pi}{3}},\\
        &z^{(G,G);M_{SR}}_{M_{R^2},M_{R},M_S}=1,\ z^{(G,G);M_S}_{M_{R^2},M_{R},M_{RS}}=e^{-i\frac{2\pi}{3}},\ z^{(G,G);M_{RS}}_{M_{R^2},M_{R},M_{SR}}=e^{i\frac{2\pi}{3}}, \\
        &z^{(G,G);M_{\alpha}}_{M_r,M_{R},M_{\alpha}}=z^{(G,G);M_{\alpha}}_{M_{r^2},M_{R^2},M_{\alpha}}=z^{(G,G);M_{\beta}}_{M_r,M_{R^2},M_{\beta}}=z^{(G,G);M_{\beta}}_{M_{r^2},M_{R},M_{\beta}}=e^{i\frac{2\pi}{3}},\\
        &z^{(G,G);M^*_{\alpha}}_{M_R,M_{r},M^*_{\alpha}}=z^{(G,G);M^*_{\alpha}}_{M_{R^2},M_{r^2}M^*_{\alpha}}=z^{(G,G);M^*_{\beta}}_{M_R,M_{r^2},M^*_{\beta}}=z^{(G,G);M^*_{\beta}}_{M_{R^2},M_{r},M^*_{\beta}}=e^{i\frac{2\pi}{3}};
\end{align*}
\begin{align*}
        &z^{(H,H);M_r}_{M_r,M_{r},M_1}=1,\ z^{(H,H);M_{r^2}}_{M_r,M_{r},M_r}=e^{-i\frac{2\pi}{3}},\ z^{(H,H);M_1}_{M_r,M_{r},M_{r^2}}=e^{i\frac{2\pi}{3}},\\
        &z^{(H,H);M_{r^2}}_{M_{r^2},M_{r^2},M_1}=1,\ z^{(H,H);M_1}_{M_{r^2},M_{r^2},M_r}=e^{i\frac{2\pi}{3}},\ z^{(H,H);M_r}_{M_{r^2},M_{r^2},M_{r^2}}=e^{-i\frac{2\pi}{3}},\\
        &z^{(H,H);M_{rs}}_{M_r,M_{r^2},M_s}=1,\ z^{(H,H);M_{sr}}_{M_r,M_{r^2},M_{rs}}=e^{-i\frac{2\pi}{3}},\ z^{(H,H);M_{rs}}_{M_r,M_{r^2},M_{sr}}=e^{i\frac{2\pi}{3}},\\
        &z^{(H,H);M_{sr}}_{M_{r^2},M_{r},M_s}=1,\ z^{(H,H);M_s}_{M_{r^2},M_{r},M_{rs}}=e^{i\frac{2\pi}{3}},\ z^{(H,H);M_{rs}}_{M_{r^2},M_{r},M_{sr}}=e^{-i\frac{2\pi}{3}},\\
        &z^{(H,H);M_R}_{M_R,M_{R},M_I}=1,\ z^{(H,H);M_{R^2}}_{M_R,M_{R},M_R}=e^{-i\frac{2\pi}{3}},\ z^{(H,H);M_I}_{M_R,M_{R},M_{R^2}}=e^{i\frac{2\pi}{3}},\\
        &z^{(H,H);M_{R^2}}_{M_{R^2},M_{R^2},M_I}=1,\ z^{(H,H);M_I}_{M_{R^2},M_{R^2},M_R}=e^{i\frac{2\pi}{3}},\ z^{(H,H);M_R}_{M_{R^2},M_{R^2},M_{R^2}}=e^{-i\frac{2\pi}{3}},\\
        &z^{(H,H);M_{RS}}_{M_R,M_{R^2},M_S}=1,\ z^{(H,H);M_{SR}}_{M_R,M_{R^2},M_{RS}}=e^{-i\frac{2\pi}{3}},\ z^{(H,H);M_{RS}}_{M_R,M_{R^2},M_{SR}}=e^{i\frac{2\pi}{3}},\\
        &z^{(H,H);M_{SR}}_{M_{R^2},M_{R},M_S}=1,\ z^{(H,H);M_S}_{M_{R^2},M_{R},M_{RS}}=e^{i\frac{2\pi}{3}},\ z^{(H,H);M_{RS}}_{M_{R^2},M_{R},M_{SR}}=e^{-i\frac{2\pi}{3}},\\
        &z^{(H,H);M_{\alpha}}_{M_r,M_{R^2},M_{\alpha}}=z^{(H,H);M_{\alpha}}_{M_{r^2},M_{R},M_{\alpha}}=z^{(H,H);M_{\beta}}_{M_r,M_{R},M_{\beta}}=z^{(H,H);M_{\beta}}_{M_{r^2},M_{R},M_{\beta}}=e^{-i\frac{2\pi}{3}},\\
        &z^{(H,H);M^*_{\alpha}}_{M_R,M_{r^2},M^*_{\alpha}}=z^{(H,H);M^*_{\alpha}}_{M_{R^2},M_{r},M^*_{\alpha}}=z^{(H,H);M^*_{\beta}}_{M_R,M_{r},M^*_{\beta}}=z^{(H,H);M^*_{\beta}}_{M_{R^2},M_{r^2},M^*_{\beta}}=e^{-i\frac{2\pi}{3}};
\end{align*}

\begin{align*}
        &z^{(D,D);M_s}_{M_s,M_s,M_1}=z^{(D,D);M_{sr}}_{M_s,M_{rs},M_r}=z^{(D,D);M_{rs}}_{M_s,M_{sr},M_{r^2}}=1,\\
        &z^{(D,D);M_{rs}}_{M_{rs},M_{rs},M_1}=z^{(D,D);M_s}_{M_{rs},M_{sr},M_r}=z^{(D,D);M_{sr}}_{M_{rs},M_s,M_{r^2}}=1,\\
        &z^{(D,D);M_{sr}}_{M_{sr},M_{sr},M_1}=z^{(D,D);M_{rs}}_{M_{sr},M_s,M_r}=z^{(D,D);M_s}_{M_{sr},M_{rs},M_{r^2}}=1,\\
        &z^{(D,D);M_1}_{M_s,M_s,M_s}=z^{(D,D);M_r}_{M_s,M_{rs},M_{sr}}=z^{(D,D);M_{r^2}}_{M_s,M_{sr},M_{rs}}=1,\\
        &z^{(D,D);M_1}_{M_{rs},M_{rs},M_{rs}}=z^{(D,D);M_r}_{M_{rs},M_{sr},M_s}=z^{(D,D);M_{r^2}}_{M_{rs},M_s,M_{sr}}=1,\\
        &z^{(D,D);M_1}_{M_{sr},M_{sr},M_{sr}}=z^{(D,D);r}_{M_{sr},M_s,M_{sr}}=z^{(D,D);M_{r^2}}_{M_{sr},M_{rs},M_s}=1,\\
        &z^{(D,D);M_S}_{M_S,M_S,M_I}=z^{(D,D);M_{SR}}_{M_S,M_{RS},M_R}=z^{(D,D);M_{RS}}_{M_S,M_{SR},M_{R^2}}=1,\\
        &z^{(D,D);M_{RS}}_{M_{RS},M_{RS},M_I}=z^{(D,D);M_S}_{M_{RS},M_{SR},M_R}=z^{(D,D);M_{SR}}_{M_{RS},M_S,M_{R^2}}=1,\\
        &z^{(D,D);M_{SR}}_{M_{SR},M_{SR},M_I}=z^{(D,D);M_{RS}}_{M_{SR},M_S,M_R}=z^{(D,D);M_S}_{M_{SR},M_{RS},M_{R^2}}=1,\\
        &z^{(D,D);M_I}_{M_S,M_S,M_S}=z^{(D,D);M_R}_{M_S,M_{RS},M_{SR}}=z^{(D,D);M_{R^2}}_{M_S,M_{SR},M_{RS}}=1,\\
        &z^{(D,D);M_I}_{M_{RS},M_{RS},M_{RS}}=z^{(D,D);M_R}_{M_{RS},M_{SR},M_S}=z^{(D,D);M_{R^2}}_{M_{RS},M_S,M_{SR}}=1,\\
        &z^{(D,D);M_I}_{M_{SR},M_{SR},M_{SR}}=z^{(D,D);R}_{M_{SR},M_S,M_{SR}}=z^{(D,D);M_{R^2}}_{M_{SR},M_{RS},M_S}=1,\\
        &z^{(D,D);M_{\beta}}_{M_s,M_S,M_{\alpha}}=z^{(D,D);M_{\beta}}_{M_s,M_{RS},M_{\alpha}}=z^{(D,D);M_{\beta}}_{M_{rs},M_S,M_{\alpha}}=z^{(D,D);M_{\beta}}_{M_s,M_{SR},M_{\alpha}}=z^{(D,D);M_{\beta}}_{M_{sr},M_S,M_{\alpha}}=\frac{1}{\sqrt{3}},\\
        &z^{(D,D);M_{\beta}}_{M_{rs},M_{RS},M_{\alpha}}=z^{(D,D);M_{\beta}}_{M_{sr},M_{SR},M_{\alpha}}=\frac{e^{i\frac{2\pi}{3}}}{\sqrt{3}},\ z^{(D,D);M_{\beta}}_{M_{rs},M_{SR},M_{\alpha}}=z^{(D,D);M_{\beta}}_{M_{sr},M_{RS},M_{\alpha}}=\frac{e^{-i\frac{2\pi}{3}}}{\sqrt{3}},\\
        &z^{(D,D);M_{\alpha}}_{M_s,M_S,M_{\beta}}=z^{(D,D);M_{\alpha}}_{M_s,M_{RS},M_{\beta}}=z^{(D,D);M_{\alpha}}_{M_{rs},M_S,M_{\beta}}=z^{(D,D);M_{\alpha}}_{M_s,M_{SR},M_{\beta}}=z^{(D,D);M_{\alpha}}_{M_{sr},M_S,M_{\beta}}=\frac{1}{\sqrt{3}},\\
        &z^{(D,D);M_{\alpha}}_{M_{rs},M_{RS},M_{\beta}}=z^{(D,D);M_{\alpha}}_{M_{sr},M_{SR},M_{\beta}}=\frac{e^{-i\frac{2\pi}{3}}}{\sqrt{3}},\ z^{(D,D);M_{\alpha}}_{M_{rs},M_{SR},M_{\beta}}=z^{(D,D);M_{\alpha}}_{M_{sr},M_{RS},M_{\beta}}=\frac{e^{i\frac{2\pi}{3}}}{\sqrt{3}},\\
        &z^{(D,D);M^*_{\beta}}_{M_S,M_s,M^*_{\alpha}}=z^{(D,D);M^*_{\beta}}_{M_S,M_{rs},M^*_{\alpha}}=z^{(D,D);M^*_{\beta}}_{M_{RS},M_s,M^*_{\alpha}}=z^{(D,D);M^*_{\beta}}_{M_S,M_{sr},M^*_{\alpha}}=z^{(D,D);M^*_{\beta}}_{M_{SR},M_s,M^*_{\alpha}}=\frac{1}{\sqrt{3}},\\
        &z^{(D,D);M^*_{\beta}}_{M_{RS},M_{rs},M^*_{\alpha}}=z^{(D,D);M^*_{\beta}}_{M_{SR},M_{sr},M^*_{\alpha}}=\frac{e^{i\frac{2\pi}{3}}}{\sqrt{3}},\ z^{(D,D);M^*_{\beta}}_{M_{RS},M_{sr},M^*_{\alpha}}=z^{(D,D);M^*_{\beta}}_{M_{SR},M_{rs},M^*_{\alpha}}=\frac{e^{-i\frac{2\pi}{3}}}{\sqrt{3}},\\
        &z^{(D,D);M^*_{\alpha}}_{M_S,M_s,M^*_{\beta}}=z^{(D,D);M^*_{\alpha}}_{M_S,M_{rs},M^*_{\beta}}=z^{(D,D);M^*_{\alpha}}_{M_{RS},M_s,M^*_{\beta}}=z^{(D,D);M^*_{\alpha}}_{M_S,M_{sr},M^*_{\beta}}=z^{(D,D);M^*_{\alpha}}_{M_{SR},M_s,M^*_{\beta}}=\frac{1}{\sqrt{3}},\\
        &z^{(D,D);M^*_{\alpha}}_{M_{RS},M_{rs},M^*_{\beta}}=z^{(D,D);M^*_{\alpha}}_{M_{SR},M_{sr},M^*_{\beta}}=\frac{e^{-i\frac{2\pi}{3}}}{\sqrt{3}},\ z^{(D,D);M^*_{\alpha}}_{M_{RS},M_{sr},M^*_{\beta}}=z^{(D,D);M^*_{\alpha}}_{M_{SR},M_{rs},M^*_{\beta}}=\frac{e^{i\frac{2\pi}{3}}}{\sqrt{3}};
\end{align*}

\begin{align*}
        &z^{(E,E);M_s}_{M_s,M_s,M_1}=z^{(E,E);M_{sr}}_{M_s,M_{rs},M_r}=z^{(E,E);M_{rs}}_{M_s,M_{sr},M_{r^2}}=1,\\
        &z^{(E,E);M_{rs}}_{M_{rs},M_{rs},M_1}=z^{(E,E);M_s}_{M_{rs},M_{sr},M_r}=z^{(E,E);M_{sr}}_{M_{rs},M_s,M_{r^2}}=1,\\
        &z^{(E,E);M_{sr}}_{M_{sr},M_{sr},M_1}=z^{(E,E);M_{rs}}_{M_{sr},M_s,M_r}=z^{(E,E);M_s}_{M_{sr},M_{rs},M_{r^2}}=1,\\
        &z^{(E,E);M_1}_{M_s,M_s,M_s}=z^{(E,E);M_r}_{M_s,M_{rs},M_{sr}}=z^{(E,E);M_{r^2}}_{M_s,M_{sr},M_{rs}}=-1,\\
        &z^{(E,E);M_1}_{M_{rs},M_{rs},M_{rs}}=z^{(E,E);M_r}_{M_{rs},M_{sr},M_s}=z^{(E,E);M_{r^2}}_{M_{rs},M_s,M_{sr}}=-1,\\
        &z^{(E,E);M_1}_{M_{sr},M_{sr},M_{sr}}=z^{(E,E);r}_{M_{sr},M_s,M_{sr}}=z^{(E,E);M_{r^2}}_{M_{sr},M_{rs},M_s}=-1,\\
        &z^{(E,E);M_S}_{M_S,M_S,M_I}=z^{(E,E);M_{SR}}_{M_S,M_{RS},M_R}=z^{(E,E);M_{RS}}_{M_S,M_{SR},M_{R^2}}=1,\\
        &z^{(E,E);M_{RS}}_{M_{RS},M_{RS},M_I}=z^{(E,E);M_S}_{M_{RS},M_{SR},M_R}=z^{(E,E);M_{SR}}_{M_{RS},M_S,M_{R^2}}=1,\\
        &z^{(E,E);M_{SR}}_{M_{SR},M_{SR},M_I}=z^{(E,E);M_{RS}}_{M_{SR},M_S,M_R}=z^{(E,E);M_S}_{M_{SR},M_{RS},M_{R^2}}=1,\\
        &z^{(E,E);M_I}_{M_S,M_S,M_S}=z^{(E,E);M_R}_{M_S,M_{RS},M_{SR}}=z^{(E,E);M_{R^2}}_{M_S,M_{SR},M_{RS}}=-1,\\
        &z^{(E,E);M_I}_{M_{RS},M_{RS},M_{RS}}=z^{(E,E);M_R}_{M_{RS},M_{SR},M_S}=z^{(E,E);M_{R^2}}_{M_{RS},M_S,M_{SR}}=-1,\\
        &z^{(E,E);M_I}_{M_{SR},M_{SR},M_{SR}}=z^{(E,E);R}_{M_{SR},M_S,M_{SR}}=z^{(E,E);M_{R^2}}_{M_{SR},M_{RS},M_S}=-1,\\
        &z^{(E,E);M_{\beta}}_{M_s,M_S,M_{\alpha}}=z^{(E,E);M_{\beta}}_{M_s,M_{RS},M_{\alpha}}=z^{(E,E);M_{\beta}}_{M_{rs},M_S,M_{\alpha}}=z^{(E,E);M_{\beta}}_{M_s,M_{SR},M_{\alpha}}=z^{(E,E);M_{\beta}}_{M_{sr},M_S,M_{\alpha}}=\frac{1}{\sqrt{3}},\\
        &z^{(E,E);M_{\beta}}_{M_{rs},M_{RS},M_{\alpha}}=z^{(E,E);M_{\beta}}_{M_{sr},M_{SR},M_{\alpha}}=\frac{e^{i\frac{2\pi}{3}}}{\sqrt{3}},\ z^{(E,E);M_{\beta}}_{M_{rs},M_{SR},M_{\alpha}}=z^{(E,E);M_{\beta}}_{M_{sr},M_{RS},M_{\alpha}}=\frac{e^{-i\frac{2\pi}{3}}}{\sqrt{3}},\\
        &z^{(E,E);M_{\alpha}}_{M_s,M_S,M_{\beta}}=z^{(E,E);M_{\alpha}}_{M_s,M_{RS},M_{\beta}}=z^{(E,E);M_{\alpha}}_{M_{rs},M_S,M_{\beta}}=z^{(E,E);M_{\alpha}}_{M_s,M_{SR},M_{\beta}}=z^{(E,E);M_{\alpha}}_{M_{sr},M_S,M_{\beta}}=\frac{1}{\sqrt{3}},\\
        &z^{(E,E);M_{\alpha}}_{M_{rs},M_{RS},M_{\beta}}=z^{(E,E);M_{\alpha}}_{M_{sr},M_{SR},M_{\beta}}=\frac{e^{-i\frac{2\pi}{3}}}{\sqrt{3}},\ z^{(E,E);M_{\alpha}}_{M_{rs},M_{SR},M_{\beta}}=z^{(E,E);M_{\alpha}}_{M_{sr},M_{RS},M_{\beta}}=\frac{e^{i\frac{2\pi}{3}}}{\sqrt{3}},\\
        &z^{(E,E);M^*_{\beta}}_{M_S,M_s,M^*_{\alpha}}=z^{(E,E);M^*_{\beta}}_{M_S,M_{rs},M^*_{\alpha}}=z^{(E,E);M^*_{\beta}}_{M_{RS},M_s,M^*_{\alpha}}=z^{(E,E);M^*_{\beta}}_{M_S,M_{sr},M^*_{\alpha}}=z^{(E,E);M^*_{\beta}}_{M_{SR},M_s,M^*_{\alpha}}=\frac{1}{\sqrt{3}},\\
        &z^{(E,E);M^*_{\beta}}_{M_{RS},M_{rs},M^*_{\alpha}}=z^{(E,E);M^*_{\beta}}_{M_{SR},M_{sr},M^*_{\alpha}}=\frac{e^{i\frac{2\pi}{3}}}{\sqrt{3}},\ z^{(E,E);M^*_{\beta}}_{M_{RS},M_{sr},M^*_{\alpha}}=z^{(E,E);M^*_{\beta}}_{M_{SR},M_{rs},M^*_{\alpha}}=\frac{e^{-i\frac{2\pi}{3}}}{\sqrt{3}},\\
        &z^{(E,E);M^*_{\alpha}}_{M_S,M_s,M^*_{\beta}}=z^{(E,E);M^*_{\alpha}}_{M_S,M_{rs},M^*_{\beta}}=z^{(E,E);M^*_{\alpha}}_{M_{RS},M_s,M^*_{\beta}}=z^{(E,E);M^*_{\alpha}}_{M_S,M_{sr},M^*_{\beta}}=z^{(E,E);M^*_{\alpha}}_{M_{SR},M_s,M^*_{\beta}}=\frac{1}{\sqrt{3}},\\
        &z^{(E,E);M^*_{\alpha}}_{M_{RS},M_{rs},M^*_{\beta}}=z^{(E,E);M^*_{\alpha}}_{M_{SR},M_{sr},M^*_{\beta}}=\frac{e^{-i\frac{2\pi}{3}}}{\sqrt{3}},\ z^{(E,E);M^*_{\alpha}}_{M_{RS},M_{sr},M^*_{\beta}}=z^{(E,E);M^*_{\alpha}}_{M_{SR},M_{rs},M^*_{\beta}}=\frac{e^{i\frac{2\pi}{3}}}{\sqrt{3}}.
\end{align*}

\section{Data of Frobenius Algebras and Their Bimodules in Our Examples}

This section records the data of Frobenius algebras and bimodules of the examples in Section \ref{sec:bimodule}.

\subsection{Frobenius Algebras and Bimodules over the \eqs{\Z_2} Toric Code String-Net Model}\label{Z2}

\subsubsection{Frobenius algebra}

There are $2$ Frobenius algebras in $\Vec(Z_2)$:
\begin{enumerate}
    \item The trivial one $\A_+$: $L_{\A_+}=\{0\},\ f_{000}=1$;
    \item The nontrivial one $\A_-$: $L_{\A_-}=\{0,1\},\ f_{000}=f_{011}=1$.
\end{enumerate}
Their quantum dimesions:
\eqn{d_{\A_+}=1,\ d_{\A_-}=2.}

\subsubsection{Bimodules}

This section lists all kinds of bimodules over $\A_+$ and $\A_-$:
\begin{enumerate}
    \item There are $2$ simple ${\A_+}\dash{\A_+}$ bimodules $M_0,M_1$, which are respectively identified with the $\Vec(\Z_2)$ simple objects $0,1$ according to \eqref{eq:trivial bimodule};
    \item There are $2$ simple ${\A_-}\dash{\A_-}$ bimodules: $M_+,M_-$:
    \eqn{
    &\qquad\qquad\quad L_{M_+} = L_{M_-} = \{0, 1\},\\
        &[P_+]^{00}_{000} = [P_+]^{00}_{111} = [P_+]^{11}_{010} = [P_+]^{11}_{101} = 1,\\
        &[P_+]^{01}_{001} = [P_+]^{10}_{100} = [P_+]^{01}_{001} = [P_+]^{10}_{011} = 1,\\
        [P_-&]^{00}_{000} = [P_-]^{00}_{111} = 1, \qquad [P_-]^{11}_{010} = [P_-]^{11}_{101} = -1,\\
        [P_-&]^{01}_{001} = [P_-]^{10}_{100} = i, \qquad [P_-]^{01}_{110} = [P_-]^{10}_{011} = -i.
    }
    Their quantum dimensions are $d_{M_+}=d_{M_-}=1$.
    \item There is only $1$ simple ${\A_+}\dash{\A_+}$ bimodule: $M_\sigma$:
    \eqn{
    &\qquad\qquad\qquad\qquad L_{M_{\sigma}}=\{0, 1\},\\
        &[P_{M_{\sigma}}]^{00}_{000}=[P_{M_{\sigma}}]^{00}_{111}=[P_{M_{\sigma}}]^{01}_{001}=[P_{M_{\sigma}}]^{01}_{110}=1.
    }
    The quantum dimension is $d_{M_\sigma}=\sqrt{2}$.
    \item There is only $1$ simple ${\A_+}\dash{\A_-}$ bimodule: $M^*_\sigma$, which is the opposite bimodule of $M_\sigma$. $d_{M^*_{\sigma}}=\sqrt{2}$.
\end{enumerate}

\subsubsection{The Fusion Properties Between Bimodules}

The fusion rules between these bimodules are
\eqn{\label{SET_toric}
\delta_{M_+M_+M_+}=\delta_{M_+M_-M_-}=\delta_{M^*_{\sigma}M_0M_{\sigma}}=\delta_{M^*_{\sigma}M_1M_{\sigma}}=\delta_{M^*_{\sigma}M_{\sigma}M_+}=\delta_{M^*_{\sigma}M_{\sigma}M_-}=1.
}

The corresponding vertex coefficients are:
\eqn{
&\mathcal{V}^{000}_{M_+M_+M_+}=\mathcal{V}^{011}_{M_+M_+M_+}=\mathcal{V}^{101}_{M_+M_+M_+}=\mathcal{V}^{110}_{M_+M_+M_+}=1,\\
&\mathcal{V}^{000}_{M_+M_-M_-}=\mathcal{V}^{011}_{M_+M_-M_-}=1,\ \mathcal{V}^{101}_{M_+M_-M_-}=-i,\ \mathcal{V}^{110}_{M_+M_-M_-}=i,\\
&\mathcal{V}^{000}_{M^*_{\sigma}M_0M_{\sigma}}=\mathcal{V}^{101}_{M^*_{\sigma}M_0M_{\sigma}}=\mathcal{V}^{011}_{M^*_{\sigma}M_1M_{\sigma}}=\mathcal{V}^{110}_{M^*_{\sigma}M_1M_{\sigma}}=1,\\
&\mathcal{V}^{000}_{M^*_{\sigma}M_{\sigma}M_+}=\mathcal{V}^{011}_{M^*_{\sigma}M_{\sigma}M_+}=\mathcal{V}^{101}_{M^*_{\sigma}M_{\sigma}M_+}=\mathcal{V}^{110}_{M^*_{\sigma}M_{\sigma}M_+}=1,\\
&\mathcal{V}^{000}_{M^*_{\sigma}M_{\sigma}M_-}=1,\ \mathcal{V}^{011}_{M^*_{\sigma}M_{\sigma}M_-}=i,\ \mathcal{V}^{101}_{M^*_{\sigma}M_{\sigma}M_-}=-i,\ \mathcal{V}^{110}_{M^*_{\sigma}M_{\sigma}M_-}=-1.
}

From the vertex coefficients, we can obtain the nonzero $6j$-symbols:
\eqn{\label{SET_toric_6j}
&G^{M_+M_+M_+}_{M_+M_+M_+}=G^{M_+M_+M_+}_{M_-M_-M_-}=G^{M_+M_-M_-}_{M_+M_-M_-}=1,\\
&G^{M_{\sigma}M^*_{\sigma}M_0}_{M_{\sigma}M^*_{\sigma}M_+}=G^{M_{\sigma}M^*_{\sigma}M_1}_{M_{\sigma}M^*_{\sigma}M_+}=G^{M_{\sigma}M^*_{\sigma}M_0}_{M_{\sigma}M^*_{\sigma}M_-}=\frac{1}{\sqrt{2}},\ G^{M_{\sigma}M^*_{\sigma}M_1}_{M_{\sigma}M^*_{\sigma}M_-}=-\frac{1}{\sqrt{2}},\\
&G^{M_0M_0M_0}_{M_{\sigma}M^*_{\sigma}M_{\sigma}}=G^{M_0M_1M_1}_{M_{\sigma}M^*_{\sigma}M_{\sigma}}=G^{M_+M_+M_+}_{M^*_{\sigma}M_{\sigma}M^*_{\sigma}}=G^{M_+M_-M_-}_{M^*_{\sigma}M_{\sigma}M^*_{\sigma}}=\frac{1}{\sqrt[4]{2}}.
}

\subsection{Frobenius Algebras and Bimodules over the \eqs{\Z_2\times\Z_2} Quantum-Double Phase}\label{Z2timesZ2}

\subsubsection{Frobenius algebra}

We consider $2$ Frobenius algebras here:
\begin{enumerate}
    \item The trivial one ${\A_+}$: $L_{\A}=\{0\},\ f_{000}=1$;
    \item The nontrivial one ${\A_-}$: $L_{\A_-}=\{1,a,b,c\}$, $f_{xyz}=\delta_{xyz}$, corresponding to the trivial $2$-cocycle of $\Z_2\times\Z_2$.
\end{enumerate}
Their quantum dimensions are
\eqn{
d_{\A_+}=1,\ d_{\A_-}=4.
}

\subsubsection{Bimodules}

This section lists all kinds of bimodules over $\A_+$ and $\A_-$:
\begin{enumerate}
    \item There are $4$ simple ${\A_+}\dash{\A_+}$ bimodules $M_1,M_a,M_b,M_c$, which are respectively identified with the $\Vec(\Z_2\times\Z_2)$ simple objects $1,a,b,c$ according to \eqref{eq:trivial bimodule}.
    \item There are $4$ simple ${\A_-}\dash{\A_-}$ bimodules: $M_I,M_A,M_B,M_C$:
    \begin{enumerate}
        \item $L_{M_I}=\{1,a,b,c\},\ [P_{M_I}]^{\mu\nu}_{xyz}=\delta_{\mu xy}\delta_{\nu zy}$;
        \item $L_{M_A}=\{1,a,b,c\},\ [P_{M_A}]^{\mu\nu}_{xyz}=\delta_{\mu xy}\delta_{\nu zy}\frac{\omega_a(x)\omega_a(z)}{\omega_a^2(y)}$;
        \item $L_{M_B}=\{1,a,b,c\},\ [P_{M_B}]^{\mu\nu}_{xyz}=\delta_{\mu xy}\delta_{\nu zy}\frac{\omega_b(x)\omega_b(z)}{\omega_b^2(y)}$;
        \item $L_{M_C}=\{1,a,b,c\},\ [P_{M_C}]^{\mu\nu}_{xyz}=\delta_{\mu xy}\delta_{\nu zy}\frac{\omega_c(x)\omega_c(z)}{\omega_c^2(y)}$.
    \end{enumerate}
    The function $\omega$ here is defined by
    \begin{equation}
        \omega_{m}(x)=\begin{cases}
            1, & x=1\ \text{or}\ m=1\ \text{or}\ x=m,\\
            i, & \text{otherwise}.
        \end{cases}\qquad m\in\{1,a,b,c\}
    \end{equation}
    Their dimensions
    \begin{equation}
        d_{M_I}=d_{M_A}=d_{M_B}=d_{M_C}=1.
    \end{equation}
    \item There is only $1$ simple ${\A_+}\dash{\A_-}$ bimodule $M_{\chi}$:
    \begin{equation}
        L_{M_{\chi}}=\{1,a,b,c\},\ [P_{M_{\chi}}]^{1\mu}_{xxy}=\delta_{\mu yx}.
    \end{equation}
    The dimension
    \begin{equation}
        d_{M_{\chi}}=2.
    \end{equation}
    \item There is only $1$ simple ${\A_-}\dash{\A_+}$ bimodule $M^*_{\chi}$, which is just the opposite bimodule of $M_{\chi}$. $d_{M^*_{\chi}}=2$.
\end{enumerate}

\subsubsection{The Fusion Properties Between Bimodules}

The fusion rules for these bimodules are
\eqn{
&\delta_{M_IM_IM_I}=\delta_{M_IM_AM_A}=\delta_{M_IM_BM_B}=1,\\
&\delta_{M_IM_CM_C}=\delta_{M_AM_BM_C}=\delta_{M_AM_CM_B}=1,\\
&\delta_{M^*_{\chi'}M_1M_{\chi'}}=\delta_{M^*_{\chi'}M_aM_{\chi'}}=\delta_{M^*_{\chi'}M_bM_{\chi'}}=\delta_{M^*_{\chi'}M_cM_{\chi'}}=1,\\
&\delta_{M^*_{\chi'}M_{\chi'}M_I}=\delta_{M^*_{\chi'}M_{\chi'}M_A}=\delta_{M^*_{\chi'}M_{\chi'}M_B}=\delta_{M^*_{\chi'}M_{\chi'}M_C}=1.
}
The corresponding fusion coefficients are given by
\eqn{
\mathcal{V}^{xyz}_{M_KM_LM_N}=\delta_{xyz}\frac{\omega_{k}(x)\omega_{l}(y)\omega_n(z)}{\omega_{k}^2(z)\omega^2_{l}(x)\omega_n^2(y)}.
}
From the fusion coefficients, we can obtain the nonzero $6j$-symbols:
\eqn{
&G^{M_KM_LM_N}_{M_XM_YM_Z}=\delta_{kln}\delta_{xyn}\delta_{kyz}\delta_{xlz},\\
&G^{M_{\chi}M^*_{\chi}M_x}_{M_{\chi}M^*_{\chi}M_Y}=\frac{\omega^2_y(x)}{2},\ G^{M_{\chi}M^*_{\chi}M_y}_{M_xM_zM^*_{\chi}}=G^{M_XM_YM_Z}_{M^*_{\chi}M_{\chi}M^*_{\chi}}=\frac{1}{\sqrt{2}}.
}
Here, $M_K,M_L,M_N,M_X,M_Y,M_Z\in{_{\A_-}\Bimod_{\A_-}(\Fus)}$ and $M_x,M_y,M_z\in{_{\A_+}\Bimod_{\A_+}(\Fus)}$.

\subsection{Frobenius Algebras and Bimodules over the \eqs{\Vec(S_3)} String-Net Model}\label{S3}

\subsubsection{Frobenius algebra}

We consider $2$ Frobenius algebras in our SET constructing procrdure:
\begin{enumerate}
    \item The trivial one ${\A_+}$: $L_{\A}=\{0\},\ f_{000}=1$;
    \item The nontrivial one ${\A_-}$: $L_\B=\{1,r,r^2\},\ f_{abc}=\delta_{abc},\ \forall a,b,c\in L_{\mathcal{B}}$.
\end{enumerate}
Their quantum dimensions are
\eqn{
d_{\A_+}=1,\ d_{\A_-}=3.
}

\subsubsection{Bimodules}

This section lists all kinds of bimodules over $\A_+$ and $\A_-$:
\begin{enumerate}
    \item There are $6$ simple ${\A_+}\dash{\A_+}$ bimodules $M_1,M_r,M_{r^2},M_{s},M_{rs},M_{sr}$, which are respectively identified with the $\Vec(S_3)$ simple objects $1,r,r^2,s,rs,sr$ according to \eqref{eq:trivial bimodule}.
    \item There are $6$ ${\A_-}\dash{\A_-}$ bimodules $M_I,M_R,M_{R^2},M_S,M_{RS},M_{SR}$:
    \begin{enumerate}
        \item $M_{I},\ L_{M_I}=\{1,r,r^2\},\ [P_{M_I}]^{ab}_{xyz}=\delta_{axy^*}\delta_{ybz^*}$;
        \item $M_{R},\ L_{M_R}=\{1,r,r^2\},\ [P_{M_I}]^{ab}_{xyz}=\omega(ab^*)\delta_{axy^*}\delta_{ybz^*}$;
        \item $M_{R^2},\ L_{M_{R^2}}=\{1,r,r^2\},\ [P_{M_I}]^{ab}_{xyz}=\omega^*(ab^*)\delta_{axy^*}\delta_{ybz^*}$;
        \item $M_{S},\ L_{M_S}=\{s,rs,sr\},\ [P_{M_I}]^{ab}_{xyz}=\delta_{axy^*}\delta_{ybz^*}$;
        \item $M_{RS},\ L_{M_{RS}}=\{s,rs,sr\},\ [P_{M_I}]^{ab}_{xyz}=\omega(ab)\delta_{axy^*}\delta_{ybz^*}$;
        \item $M_{SR},\ L_{M_{SR}}=\{s,rs,sr\},\ [P_{M_I}]^{ab}_{xyz}=\omega^*(ab)\delta_{axy^*}\delta_{ybz^*}$.
    \end{enumerate}
    Here, the function $\omega$ is defined by
    \begin{equation}
        \omega(x)=\begin{cases}
            1, &x=1,\\
            e^{i\frac{2\pi}{3}},&x=r,\\
            e^{-i\frac{2\pi}{3}},&x=r^2.
        \end{cases}
    \end{equation}
    Their quantum dimensions are
    \eqn{
    d_{M_I}=d_{M_R}=d_{M_{R^2}}=d_{M_S}=d_{M_{RS}}=d_{M_{SR}}=1.
    }
    \item There are $2$ simple ${\A_+}\dash{\A_-}$ bimodules, $M_\alpha$ and $M_\beta$:
    \eqn{
    &L_{M_{\alpha}}=\{1,r,r^2\},\ L_{M_{\beta}}=\{s,rs,sr\},\\
    &[P_{M_{\alpha}}]^{1,1}_{1,1,1}=[P_{M_{\alpha}}]^{1,1}_{r,r,r}=[P_{M_{\alpha}}]^{1,1}_{r^2,r^2,r^2}=1,\\
    &[P_{M_{\alpha}}]^{1,r}_{1,1,r}=[P_{M_{\alpha}}]^{1,r}_{r,r,r^2}=[P_{M_{\alpha}}]^{1,r}_{r^2,r^2,1}=1,\\
    &[P_{M_{\alpha}}]^{1,r^2}_{1,1,r^2}=[P_{M_{\alpha}}]^{1,r^2}_{r,r,1}=[P_{M_{\alpha}}]^{1,r^2}_{r^2,r^2,r}=1,\\
    &[P_{M_{\beta}}]^{1,1}_{s,s,s}=[P_{M_{\beta}}]^{1,1}_{rs,rs,rs}=[P_{M_{\beta}}]^{1,1}_{sr,sr,sr}=1,\\
    &[P_{M_{\beta}}]^{1,r}_{s,s,sr}=[P_{M_{\beta}}]^{1,r}_{rs,rs,s}=[P_{M_{\beta}}]^{1,r}_{sr,sr,rs}=1,\\
    &[P_{M_{\beta}}]^{1,r^2}_{s,s,rs}=[P_{M_{\beta}}]^{1,r^2}_{rs,rs,sr}=[P_{M_{\beta}}]^{1,r^2}_{sr,sr,s}=1.\\
    }
    Their quantum dimensions are
    \eqn{
    d_{M_\alpha}=d_{M_\beta}=\sqrt{3}.
    }
    \item There are $2$ simple ${\A_-}\dash{\A_+}$ bimodules: $M^*_\alpha$ and $M^*_\beta$. They are the opposite bimodules of $M_\alpha$ and $M_\beta$ respectively. $d_{M^*_\alpha}=d_{M^*_\beta}\sqrt{3}$.
\end{enumerate}

\subsubsection{The Fusion Properties Between Bimodules}

The fusion rules between these bimodules are
\eqn{
&\delta_{M_KM_LM_N}=\delta_{kln},\ \forall M_K,M_L,M_N\in\{M_I,M_R,M_{R^2},M_S,M_{RS},M_{SR}\},\\
&\delta_{M^*_{\alpha}M_1M_{\alpha}}=\delta_{M^*_{\alpha}M_rM_{\alpha}}=\delta_{M^*_{\alpha}M_{r^2}M_{\alpha}}=\delta_{M^*_{\beta}M_1M_{\beta}}=\delta_{M^*_{\beta}M_rM_{\beta}}=\delta_{M^*_{\beta}M_{r^2}M_{\beta}}=1,\\
&\delta_{M^*_{\alpha}M_sM_{\beta}}=\delta_{M^*_{\alpha}M_{rs}M_{\beta}}=\delta_{M^*_{\alpha}M_{sr}M_{\beta}}=\delta_{M^*_{\beta}M_sM_{\alpha}}=\delta_{M^*_{\beta}M_{rs}M_{\alpha}}=\delta_{M^*_{\beta}M_{sr}M_{\alpha}}=1,\\
&\delta_{M^*_{\alpha}M_{\alpha}M_I}=\delta_{M^*_{\alpha}M_{\alpha}M_{R}}=\delta_{M^*_{\alpha}M_{\alpha}M_{R^2}}=\delta_{M^*_{\beta}M_{\beta}M_I}=\delta_{M^*_{\beta}M_{\beta}M_R}=\delta_{M^*_{\beta}M_{\beta}M_{R^2}}=1,\\
&\delta_{M^*_{\alpha}M_{\beta}M_S}=\delta_{M^*_{\alpha}M_{\beta}M_{RS}}=\delta_{M^*_{\alpha}M_{\beta}M_{SR}}=\delta_{M^*_{\beta}M_{\alpha}M_S}=\delta_{M^*_{\beta}M_{\alpha}M_{RS}}=\delta_{M^*_{\beta}M_{\alpha}M_{SR}}=1.\\
}

The corresponding vertex coefficients are
\begin{equation*}
    \mathcal{V}^{abc}_{M^*_{\alpha}M_xM_{\alpha}}=\mathcal{V}^{abc}_{M^*_{\beta}M_xM_{\beta}}=\delta_{abc},\ \forall M_x\in\{M_1,M_r,M_{r^2}\},
\end{equation*}
\begin{equation*}
    \mathcal{V}^{abc}_{M^*_{\alpha}M_yM_{\beta}}=\mathcal{V}^{abc}_{M^*_{\beta}M_yM_{\alpha}}=\delta_{abc},\ \forall M_y\in\{M_s,M_{rs},M_{sr}\},
    \end{equation*}
\begin{equation*}
        \mathcal{V}^{abc}_{M^*_{\alpha}M_{\alpha}M_I}=\delta_{abc},\ \mathcal{V}^{abc}_{M^*_{\alpha}M_{\alpha}M_R}=\omega(ba^*)\delta_{abc},\ \mathcal{V}^{abc}_{M^*_{\alpha}M_{\alpha}M_{R^2}}=\omega^*(ba^*)\delta_{abc},
    \end{equation*}
\begin{equation*}
    \mathcal{V}^{abc}_{M^*_{\beta}M_{\beta}M_I}=\delta_{abc},\ \mathcal{V}^{abc}_{M^*_{\beta}M_{\beta}M_R}=\omega(sbsa)\delta_{abc},\ \mathcal{V}^{abc}_{M^*_{\beta}M_{\beta}M_{R^2}}=\omega^*(sbsa)\delta_{abc},
\end{equation*}
\begin{equation*}
    \mathcal{V}^{abc}_{M^*_{\alpha}M_{\beta}M_S}=\delta_{abc},\ \mathcal{V}^{abc}_{M^*_{\alpha}M_{\beta}M_{RS}}=\omega(sba)\delta_{abc},\ \mathcal{V}^{abc}_{M^*_{\alpha}M_{\beta}M_{SR}}=\omega^*(sba)\delta_{abc},
\end{equation*}
\begin{equation*}
    \mathcal{V}^{abc}_{M^*_{\beta}M_{\alpha}M_S}=\delta_{abc},\ \mathcal{V}^{abc}_{M^*_{\beta}M_{\alpha}M_{RS}}=\omega(bas)\delta_{abc},\ \mathcal{V}^{abc}_{M^*_{\beta}M_{\alpha}M_{SR}}=\omega^*(bas)\delta_{abc},
\end{equation*}
\begin{equation*}
    \mathcal{V}^{abc}_{M_1M_1M_1}=\delta_{abc},\ \mathcal{V}^{abc}_{M_1M_rM_{r^2}}=\omega(a)\delta_{abc},
\end{equation*}
\begin{equation*}
    \mathcal{V}^{abc}_{M_rM_rM_r}=\omega(bc^*)\delta_{abc},\ \mathcal{V}^{abc}_{M_{r^2}M_{r^2}M_{r^2}}=\omega^*(bc^*)\delta_{abc},
\end{equation*}
\begin{equation*}
    \mathcal{V}^{abc}_{M_1M_sM_s}=\delta_{abc},\ \mathcal{V}^{abc}_{M_1M_{rs}M_{rs}}=\omega(a)\delta_{abc},\ \mathcal{V}^{abc}_{M_1M_{sr}M_{sr}}=\omega^*(a)\delta_{abc},
\end{equation*}
\begin{equation*}
    \mathcal{V}^{abc}_{M_rM_sM_{rs}}=\omega(sb)\delta_{abc},\ \mathcal{V}^{abc}_{M_rM_{rs}M_{sr}}=\omega(bscs)\delta_{abc},\ \mathcal{V}^{abc}_{M_rM_{sr}M_{s}}=\omega(sc)\delta_{abc},
\end{equation*}
\begin{equation*}
    \mathcal{V}^{abc}_{M_{r^2}M_sM_{sr}}=\omega^*(sb)\delta_{abc},\ \mathcal{V}^{abc}_{M_{r^2}M_{rs}M_{s}}=\omega^*(sc)\delta_{abc},\ \mathcal{V}^{abc}_{M_{r^2}M_{sr}M_{rs}}=\omega^*(bscs)\delta_{abc}.
\end{equation*}

From these vertex coefficients, we can obtain the nonzero $6j$-symbols:
\begin{equation*}
    G^{M_{\alpha}M^*_{\alpha}M_1}_{M_{\alpha}M^*_{\alpha}M_{I}}=G^{M_{\alpha}M^*_{\alpha}M_1}_{M_{\alpha}M^*_{\alpha}M_{R}}=G^{M_{\alpha}M^*_{\alpha}M_1}_{M_{\alpha}M^*_{\alpha}M_{R^2}}=\frac{1}{\sqrt{3}},
\end{equation*}
\begin{equation*}
    G^{M_{\alpha}M^*_{\alpha}M_r}_{M_{\alpha}M^*_{\alpha}M_{I}}=\frac{1}{\sqrt{3}},\ G^{M_{\alpha}M^*_{\alpha}M_r}_{M_{\alpha}M^*_{\alpha}M_{R}}=\frac{e^{-i\frac{2\pi}{3}}}{\sqrt{3}},\ G^{M_{\alpha}M^*_{\alpha}M_r}_{M_{\alpha}M^*_{\alpha}M_{R^2}}=\frac{e^{i\frac{2\pi}{3}}}{\sqrt{3}},
\end{equation*}
\begin{equation*}
    G^{M_{\alpha}M^*_{\alpha}M_{r^2}}_{M_{\alpha}M^*_{\alpha}M_{I}}=\frac{1}{\sqrt{3}},\ G^{M_{\alpha}M^*_{\alpha}M_{r^2}}_{M_{\alpha}M^*_{\alpha}M_{R}}=\frac{e^{i\frac{2\pi}{3}}}{\sqrt{3}},\ G^{M_{\alpha}M^*_{\alpha}M_{r^2}}_{M_{\alpha}M^*_{\alpha}M_{R^2}}=\frac{e^{-i\frac{2\pi}{3}}}{\sqrt{3}},
\end{equation*}
\begin{equation*}
    G^{M_{\beta}M^*_{\beta}M_1}_{M_{\beta}M^*_{\beta}M_{I}}=G^{M_{\beta}M^*_{\beta}M_1}_{M_{\beta}M^*_{\beta}M_{R}}=G^{M_{\beta}M^*_{\beta}M_1}_{M_{\beta}M^*_{\beta}M_{R^2}}=\frac{1}{\sqrt{3}},
\end{equation*}
\begin{equation*}
    G^{M_{\beta}M^*_{\beta}M_r}_{M_{\beta}M^*_{\beta}M_{I}}=\frac{1}{\sqrt{3}},\ G^{M_{\beta}M^*_{\beta}M_r}_{M_{\beta}M^*_{\beta}M_{R}}=\frac{e^{i\frac{2\pi}{3}}}{\sqrt{3}},\ G^{M_{\beta}M^*_{\beta}M_r}_{M_{\beta}M^*_{\beta}M_{R^2}}=\frac{e^{-i\frac{2\pi}{3}}}{\sqrt{3}},
\end{equation*}
\begin{equation*}
    G^{M_{\beta}M^*_{\beta}M_{r^2}}_{M_{\beta}M^*_{\beta}M_{I}}=\frac{1}{\sqrt{3}},\ G^{M_{\beta}M^*_{\beta}M_{r^2}}_{M_{\beta}M^*_{\beta}M_{R}}=\frac{e^{-i\frac{2\pi}{3}}}{\sqrt{3}},\ G^{M_{\beta}M^*_{\beta}M_{r^2}}_{M_{\beta}M^*_{\beta}M_{R^2}}=\frac{e^{i\frac{2\pi}{3}}}{\sqrt{3}},
\end{equation*}
\begin{equation*}
    G^{M_{\alpha}M^*_{\beta}M_s}_{M_{\beta}M^*_{\alpha}M_{I}}=G^{M_{\alpha}M^*_{\beta}M_s}_{M_{\beta}M^*_{\alpha}M_{R}}=G^{M_{\alpha}M^*_{\beta}M_s}_{M_{\beta}M^*_{\alpha}M_{R^2}}=\frac{1}{\sqrt{3}},
\end{equation*}
\begin{equation*}
    G^{M_{\alpha}M^*_{\beta}M_{rs}}_{M_{\beta}M^*_{\alpha}M_{I}}=\frac{1}{\sqrt{3}},\ G^{M_{\alpha}M^*_{\beta}M_{rs}}_{M_{\beta}M^*_{\alpha}M_{R}}=\frac{e^{i\frac{2\pi}{3}}}{\sqrt{3}},\ G^{M_{\alpha}M^*_{\beta}M_{rs}}_{M_{\beta}M^*_{\alpha}M_{R^2}}=\frac{e^{-i\frac{2\pi}{3}}}{\sqrt{3}},
\end{equation*}
\begin{equation*}
    G^{M_{\alpha}M^*_{\beta}M_{sr}}_{M_{\beta}M^*_{\alpha}M_{I}}=\frac{1}{\sqrt{3}},\ G^{M_{\alpha}M^*_{\beta}M_{sr}}_{M_{\beta}M^*_{\alpha}M_{R}}=\frac{e^{-i\frac{2\pi}{3}}}{\sqrt{3}},\ G^{M_{\alpha}M^*_{\beta}M_{sr}}_{M_{\beta}M^*_{\alpha}M_{R^2}}=\frac{e^{i\frac{2\pi}{3}}}{\sqrt{3}},
\end{equation*}
\begin{equation*}
    G^{M_{\alpha}M^*_{\alpha}M_1}_{M_{\beta}M^*_{\beta}M_{S}}=G^{M_{\alpha}M^*_{\alpha}M_1}_{M_{\beta}M^*_{\beta}M_{RS}}=G^{M_{\alpha}M^*_{\alpha}M_1}_{M_{\beta}M^*_{\beta}M_{SR}}=\frac{1}{\sqrt{3}},
\end{equation*}
\begin{equation*}
    G^{M_{\alpha}M^*_{\alpha}M_r}_{M_{\beta}M^*_{\beta}M_{S}}=\frac{1}{\sqrt{3}},\ G^{M_{\alpha}M^*_{\alpha}M_r}_{M_{\beta}M^*_{\beta}M_{RS}}=\frac{e^{-i\frac{2\pi}{3}}}{\sqrt{3}},\ G^{M_{\alpha}M^*_{\alpha}M_r}_{M_{\beta}M^*_{\beta}M_{SR}}=\frac{e^{i\frac{2\pi}{3}}}{\sqrt{3}},
\end{equation*}
\begin{equation*}
    G^{M_{\alpha}M^*_{\alpha}M_{r^2}}_{M_{\beta}M^*_{\beta}M_{S}}=\frac{1}{\sqrt{3}},\ G^{M_{\alpha}M^*_{\alpha}M_{r^2}}_{M_{\beta}M^*_{\beta}M_{RS}}=\frac{e^{i\frac{2\pi}{3}}}{\sqrt{3}},\ G^{M_{\alpha}M^*_{\alpha}M_{r^2}}_{M_{\beta}M^*_{\beta}M_{SR}}=\frac{e^{-i\frac{2\pi}{3}}}{\sqrt{3}},
\end{equation*}
\begin{equation*}
    G^{M_{\alpha}M^*_{\beta}M_s}_{M_{\alpha}M^*_{\beta}M_{S}}=G^{M_{\alpha}M^*_{\beta}M_s}_{M_{\alpha}M^*_{\beta}M_{RS}}=G^{M_{\alpha}M^*_{\beta}M_s}_{M_{\alpha}M^*_{\beta}M_{SR}}=\frac{1}{\sqrt{3}},
\end{equation*}
\begin{equation*}
    G^{M_{\alpha}M^*_{\beta}M_{rs}}_{M_{\alpha}M^*_{\beta}M_{S}}=\frac{1}{\sqrt{3}},\ G^{M_{\alpha}M^*_{\beta}M_{rs}}_{M_{\alpha}M^*_{\beta}M_{RS}}=\frac{e^{i\frac{2\pi}{3}}}{\sqrt{3}},\ G^{M_{\alpha}M^*_{\beta}M_{rs}}_{M_{\alpha}M^*_{\beta}M_{SR}}=\frac{e^{-i\frac{2\pi}{3}}}{\sqrt{3}},
\end{equation*}
\begin{equation*}
    G^{M_{\alpha}M^*_{\beta}M_{sr}}_{M_{\alpha}M^*_{\beta}M_{S}}=\frac{1}{\sqrt{3}},\ G^{M_{\alpha}M^*_{\beta}M_{sr}}_{M_{\alpha}M^*_{\beta}M_{RS}}=\frac{e^{-i\frac{2\pi}{3}}}{\sqrt{3}},\ G^{M_{\alpha}M^*_{\beta}M_{sr}}_{M_{\alpha}M^*_{\beta}M_{SR}}=\frac{e^{i\frac{2\pi}{3}}}{\sqrt{3}},
\end{equation*}
\begin{equation*}
    G^{M_aM_bM_c}_{M_{\mu}M^*_{\nu}M^*_{\gamma}}=\frac{1}{\sqrt[4]{3}}\delta_{abc}\delta_{M^*_{\nu}M^*_{c}M_{\mu}}\delta_{M^*_{\gamma}M^*_{a}M_{\nu}}\delta_{M^*_{\mu}M^*_{b}M_{\gamma}},
\end{equation*}
\begin{equation*}
    \forall M_a,M_b,M_c\in\{M_1,M_r,M_{r^2},M_s,M_{rs},M_{sr}\},\ M_{\mu},M_{\nu},M_{\gamma}\in \{M_{\alpha},M_{\beta}\},
\end{equation*}
\begin{equation*}
    G^{M_AM_BM_C}_{M^*_{\mu}M_{\nu}M_{\gamma}}=\frac{1}{\sqrt[4]{3}}\delta_{abc}\delta_{M_{\nu}M^*_{c}M^*_{\mu}}\delta_{M_{\gamma}M^*_{a}M^*_{\nu}}\delta_{M_{\mu}M^*_{b}M^*_{\gamma}},
\end{equation*}
\begin{equation*}
    \forall M_A,M_B,M_C\in\{M_I,M_R,M_{R^2},M_S,M_{RS},M_{SR}\},\ M_{\mu},M_{\nu},M_{\gamma}\in \{M_{\alpha},M_{\beta}\}.
\end{equation*}

\section{The Morita Equivalence Between \eqs{\MA} and \eqs{A_5}}\label{Morita_equivalence}

In this section proves that \(\MA\) is Morita equivalent to \(A_5\)\footnote{The \(A_5\) fusion category is a modification of the \(SU(2)_4\) fusion category. While \(A_5\) shares the same fusion rules as \(SU(2)_4\), it differs in its \(6j\)-symbols, which possess tetrahedral symmetry. See Ref.~\cite{aasen2020topological} for details.}. As shown in Ref.~\cite{etingof2016}, the category of bimodules over a Frobenius algebra in a fusion category $\Fus$ is Morita equivalent to $\Fus$. Therefore, it suffices to show that the \(A_5\) fusion category is isomorphic to the category of bimodules over some Frobenius algebra in the \(\mathcal{M}_A\) category.

Consider Frobenius algebra $\A$ in $\MA$ fusion category:
\begin{equation}
    L_{\A}=\{1,s\},\qquad f_{111}=f_{1ss}=1,\ f_{11s}=f_{111}=0.
\end{equation}
The quantum dimension is
\begin{equation}
    d_{\A}=2.
\end{equation}

There are five simple bimodules over $\A$: $M_0,M_1,M_2,M_3,M_4$, where
\begin{equation}
    L_{M_0}=L_{M_4}=\{1,s\},\ L_{M_1}=L_{M_3}=\{\alpha,\beta\},\ L_{M_2}=\{r,r^2,rs,sr\},
\end{equation}
and their nonzero functions are
\begin{equation}
    \begin{split}
        &[P_{M_0}]^{1,1}_{1,1,1}=[P_{M_0}]^{1,1}_{s,s,s}=[P_{M_0}]^{s,s}_{1,s,1}=[P_{M_0}]^{s,s}_{s,1,s}=1,\\
        &[P_{M_0}]^{1,s}_{1,1,s}=[P_{M_0}]^{1,s}_{s,s,1}=[P_{M_0}]^{s,1}_{1,s,s}=[P_{M_0}]^{s,1}_{s,1,1}=1;\\
        &[P_{M_1}]^{1,1}_{\alpha,\alpha,\alpha}=[P_{M_1}]^{1,1}_{\beta,\beta,\beta}=[P_{M_1}]^{s,s}_{\alpha,\beta,\alpha}=[P_{M_1}]^{s,s}_{\beta,\alpha,\beta}=1,\\
        &[P_{M_1}]^{1,s}_{\alpha,\alpha,\beta}=[P_{M_1}]^{1,s}_{\beta,\beta,\alpha}=[P_{M_1}]^{s,1}_{\alpha,\beta,\beta}=[P_{M_1}]^{s,1}_{\beta,\alpha,\alpha}=1;\\
        &[P_{M_2}]^{x,y}_{a,b,c}=\delta_{xab^*}\delta_{byc^*},\ \forall x,y\in L_{\mathcal{C}},\ a,b,c\in L_{M_2};\\
        &[P_{M_3}]^{1,1}_{\alpha,\alpha,\alpha}=[P_{M_3}]^{1,1}_{\beta,\beta,\beta}=1,\ [P_{M_3}]^{s,s}_{\alpha,\beta,\alpha}=[P_{M_3}]^{s,s}_{\beta,\alpha,\beta}=-1,\\
        &[P_{M_3}]^{1,s}_{\alpha,\alpha,\beta}=[P_{M_3}]^{s,1}_{\beta,\alpha,\alpha}=i,\ [P_{M_3}]^{1,s}_{\beta,\beta,\alpha}=[P_{M_3}]^{s,1}_{\alpha,\beta,\beta}=-i;\\
        &[P_{M_4}]^{1,1}_{1,1,1}=[P_{M_4}]^{1,1}_{s,s,s}=1,\ [P_{M_4}]^{s,s}_{1,s,1}=[P_{M_4}]^{s,s}_{s,1,s}=-1,\\
        &[P_{M_4}]^{1,s}_{1,1,s}=[P_{M_4}]^{s,1}_{s,1,1}=i,\ [P_{M_4}]^{1,s}_{s,s,1}=[P_{M_4}]^{s,1}_{1,s,s}=-i.\\
    \end{split}
\end{equation}
Their quantum dimensions are:
\begin{equation}
    d_{M_0}=d_{M_4}=1,\ d_{M_1}=d_{M_3}=\sqrt{3},\ d_{M_2}=2.
\end{equation}

The fusion rules between them are
\begin{equation}
    \begin{split}
        &\delta_{M_0M_0M_0}=\delta_{M_0M_4M_4}=\delta_{M_0M_2M_2}=\delta_{M_4M_2M_2}=\delta_{M_2M_2M_2}=1,\\
        &\delta_{M_2M_1M_1}=\delta_{M_2M_3M_3}=\delta_{M_2M_1M_3}=\delta_{M_2M_3M_1}=1,\\
        &\delta_{M_0M_1M_1}=\delta_{M_0M_3M_3}=\delta_{M_4M_1M_3}=\delta_{M_4M_3M_1}=1.\\
    \end{split}
\end{equation}
They can be summarized in the Table \ref{tab:morita}.
\begin{table}[ht]
    \centering
    \begin{tabular}{|c|c|c|c|c|c|}
    \hline
        $\times$ & $M_0$ & $M_1$ & $M_2$ & $M_3$ & $M_4$ \\
        \hline
        $M_0$ & $M_0$ & $M_1$ & $M_2$ & $M_3$ & $M_4$ \\
        \hline
        $M_1$ & $M_1$ & $M_0+M_2$ & $M_1+M_3$ & $M_2+M_4$ & $M_3$ \\
        \hline
        $M_2$ & $M_2$ & $M_1+M_3$ & $M_0+M_2+M_4$ & $M_1+M_3$ & $M_2$ \\
        \hline
        $M_3$ & $M_3$ & $M_2+M_4$ & $M_1+M_3$ & $M_0+M_2$ & $M_1$ \\
        \hline
        $M_4$ & $M_4$ & $M_3$ & $M_2$ & $M_1$ & $M_0$ \\
        \hline
    \end{tabular}
    \caption{The fusion table of input fusion category of parent phase}
    \label{tab:morita}
\end{table}
It is the fusion rules of $SU(2)_4$, or equivalently, the fusion rules of $A_5$.

The corresponding vertex coefficients are
\begin{equation*}
    \mathcal{V}^{111}_{M_0M_0M_0}=\mathcal{V}^{1ss}_{M_0M_0M_0}=\mathcal{V}^{s1s}_{M_0M_0M_0}=\mathcal{V}^{ss1}_{M_0M_0M_0}=1;
\end{equation*}
\begin{equation*}
    \mathcal{V}^{111}_{M_0M_4M_4}=\mathcal{V}^{1ss}_{M_0M_4M_4}=1,\ \mathcal{V}^{s1s}_{M_0M_4M_4}=-i,\ \mathcal{V}^{ss1}_{M_0M_4M_4}=i;
\end{equation*}
\begin{equation*}
    \mathcal{V}^{1\alpha\alpha}_{M_0M_1M_1}=\mathcal{V}^{1\beta\beta}_{M_0M_1M_1}=\mathcal{V}^{s\alpha\beta}_{M_0M_1M_1}=\mathcal{V}^{s\beta\alpha}_{M_0M_1M_1}=1;
\end{equation*}
\begin{equation*}
    \mathcal{V}^{1\alpha\alpha}_{M_0M_3M_3}=\mathcal{V}^{1\beta\beta}_{M_0M_3M_3}=1,\ \mathcal{V}^{s\alpha\beta}_{M_0M_3M_3}=-i,\ \mathcal{V}^{s\beta\alpha}_{M_0M_3M_3}=i;
\end{equation*}
\begin{equation*}
    \mathcal{V}^{1\alpha\alpha}_{M_4M_3M_1}=\mathcal{V}^{s\beta\alpha}_{M_4M_3M_1}=e^{-i\frac{\pi}{4}},\ \mathcal{V}^{1\beta\beta}_{M_4M_3M_1}=-e^{i\frac{\pi}{4}},\ \mathcal{V}^{s\alpha\beta}_{M_4M_3M_1}=e^{i\frac{\pi}{4}};
\end{equation*}
\begin{equation*}
    \mathcal{V}^{1\alpha\alpha}_{M_4M_1M_3}=\mathcal{V}^{s\alpha\beta}_{M_4M_1M_3}=e^{i\frac{\pi}{4}},\ \mathcal{V}^{s\beta\alpha}_{M_4M_1M_3}=e^{-i\frac{\pi}{4}},\ \mathcal{V}^{1\beta\beta}_{M_4M_1M_3}=-e^{-i\frac{\pi}{4}};
\end{equation*}
\begin{equation*}
    \mathcal{V}^{1,r,r^2}_{M_0M_2M_2}=\mathcal{V}^{1,r^2,r}_{M_0M_2M_2}=\mathcal{V}^{1,rs,rs}_{M_0M_2M_2}=\mathcal{V}^{1,sr,sr}_{M_0M_2M_2}=1,
\end{equation*}
\begin{equation*}
    \mathcal{V}^{s,r,sr}_{M_0M_2M_2}=\mathcal{V}^{s,r^2,rs}_{M_0M_2M_2}=\mathcal{V}^{s,rs,r}_{M_0M_2M_2}=\mathcal{V}^{s,sr,r^2}_{M_0M_2M_2}=1,
\end{equation*}
\begin{equation*}
    \mathcal{V}^{1,r,r^2}_{M_4M_2M_2}=\mathcal{V}^{1,rs,rs}_{M_4M_2M_2}=1,\ \mathcal{V}^{1,r^2,r}_{M_4M_2M_2}=\mathcal{V}^{1,sr,sr}_{M_4M_2M_2}=-1,
\end{equation*}
\begin{equation*}
    \mathcal{V}^{s,r,sr}_{M_4M_2M_2}=\mathcal{V}^{s,rs,r}_{M_4M_2M_2}=i,\ \mathcal{V}^{s,r^2,rs}_{M_4M_2M_2}=\mathcal{V}^{s,sr,r^2}_{M_4M_2M_2}=-i,
\end{equation*}
\begin{equation*}
    \mathcal{V}^{r,r,r}_{M_2M_2M_2}=\mathcal{V}^{r,rs,sr}_{M_2M_2M_2}=\mathcal{V}^{r^2,r^2,r^2}_{M_2M_2M_2}=\mathcal{V}^{r^2,sr,rs}_{M_2M_2M_2}=\sqrt[4]{2},
\end{equation*}
\begin{equation*}
    \mathcal{V}^{rs,r^2,sr}_{M_2M_2M_2}=\mathcal{V}^{rs,sr,r}_{M_2M_2M_2}=\mathcal{V}^{sr,r,rs}_{M_2M_2M_2}=\mathcal{V}^{sr,rs,r^2}_{M_2M_2M_2}=\sqrt[4]{2},
\end{equation*}
\begin{equation*}
    \mathcal{V}^{r,\alpha,\alpha}_{M_2M_1M_1}=\mathcal{V}^{r,\beta,\beta}_{M_2M_1M_1}=\mathcal{V}^{r^2,\alpha,\alpha}_{M_2M_1M_1}=\mathcal{V}^{r^2,\beta,\beta}_{M_2M_1M_1}=\frac{1}{\sqrt[4]{2}},
\end{equation*}
\begin{equation*}
    \mathcal{V}^{rs,\alpha,\beta}_{M_2M_1M_1}=\mathcal{V}^{rs,\beta,\alpha}_{M_2M_1M_1}=\mathcal{V}^{sr,\alpha,\beta}_{M_2M_1M_1}=\mathcal{V}^{sr,\beta,\alpha}_{M_2M_1M_1}=\frac{1}{\sqrt[4]{2}},
\end{equation*}
\begin{equation*}
    \mathcal{V}^{r,\alpha,\alpha}_{M_2M_3M_3}=\mathcal{V}^{r,\beta,\beta}_{M_2M_3M_3}=\mathcal{V}^{r^2,\alpha,\alpha}_{M_2M_3M_3}=\mathcal{V}^{r^2,\beta,\beta}_{M_2M_3M_3}=-\frac{1}{\sqrt[4]{2}},
\end{equation*}
\begin{equation*}
    \mathcal{V}^{rs,\alpha,\beta}_{M_2M_3M_3}=\frac{i}{\sqrt[4]{2}},\ \mathcal{V}^{rs,\beta,\alpha}_{M_2M_3M_3}=-\frac{i}{\sqrt[4]{2}},\ \mathcal{V}^{sr,\alpha,\beta}_{M_2M_3M_3}=\frac{i}{\sqrt[4]{2}},\ \mathcal{V}^{sr,\beta,\alpha}_{M_2M_3M_3}=-\frac{i}{\sqrt[4]{2}},
\end{equation*}
\begin{equation*}
    \mathcal{V}^{r,\alpha,\alpha}_{M_2M_1M_3}=\frac{e^{i\frac{\pi}{4}}}{\sqrt[4]{2}},\ \mathcal{V}^{r,\beta,\beta}_{M_2M_1M_3}=-\frac{e^{-i\frac{\pi}{4}}}{\sqrt[4]{2}},\ \mathcal{V}^{r^2,\alpha,\alpha}_{M_2M_1M_3}=-\frac{e^{i\frac{\pi}{4}}}{\sqrt[4]{2}},\ \mathcal{V}^{r^2,\beta,\beta}_{M_2M_1M_3}=\frac{e^{-i\frac{\pi}{4}}}{\sqrt[4]{2}},
\end{equation*}
\begin{equation*}
    \mathcal{V}^{rs,\alpha,\beta}_{M_2M_1M_3}=-\frac{e^{-i\frac{\pi}{4}}}{\sqrt[4]{2}},\ \mathcal{V}^{rs,\beta,\alpha}_{M_2M_1M_3}=\frac{e^{i\frac{\pi}{4}}}{\sqrt[4]{2}},\ \mathcal{V}^{sr,\alpha,\beta}_{M_2M_1M_3}=\frac{e^{-i\frac{\pi}{4}}}{\sqrt[4]{2}},\ \mathcal{V}^{sr,\beta,\alpha}_{M_2M_1M_3}=-\frac{e^{i\frac{\pi}{4}}}{\sqrt[4]{2}},
\end{equation*}
\begin{equation*}
    \mathcal{V}^{r,\alpha,\alpha}_{M_2M_3M_1}=-\frac{e^{-i\frac{\pi}{4}}}{\sqrt[4]{2}},\ \mathcal{V}^{r,\beta,\beta}_{M_2M_3M_1}=\frac{e^{i\frac{\pi}{4}}}{\sqrt[4]{2}},\ \mathcal{V}^{r^2,\alpha,\alpha}_{M_2M_3M_1}=\frac{e^{-i\frac{\pi}{4}}}{\sqrt[4]{2}},\ \mathcal{V}^{r^2,\beta,\beta}_{M_2M_1M_3}=-\frac{e^{i\frac{\pi}{4}}}{\sqrt[4]{2}},
\end{equation*}
\begin{equation*}
    \mathcal{V}^{rs,\alpha,\beta}_{M_2M_3M_1}=\frac{e^{-i\frac{\pi}{4}}}{\sqrt[4]{2}},\ \mathcal{V}^{rs,\beta,\alpha}_{M_2M_3M_1}=-\frac{e^{i\frac{\pi}{4}}}{\sqrt[4]{2}},\ \mathcal{V}^{sr,\alpha,\beta}_{M_2M_3M_1}=-\frac{e^{-i\frac{\pi}{4}}}{\sqrt[4]{2}},\ \mathcal{V}^{sr,\beta,\alpha}_{M_2M_3M_1}=\frac{e^{i\frac{\pi}{4}}}{\sqrt[4]{2}}.
\end{equation*}

From the vertex coefficients, we can obtain the nonzero 6$j$-symbols:
\begin{equation}
    \begin{split}
        &G^{M_0M_0M_0}_{M_0M_0M_0}=G^{M_0M_0M_0}_{M_4M_4M_4}=G^{M_0M_4M_4}_{M_0M_4M_4}=1,\ G^{M_0M_0M_0}_{M_2M_2M_2}=G^{M_0M_4M_4}_{M_2M_2M_2}=\frac{1}{\sqrt{2}}\\
        &G^{M_0M_2M_2}_{M_0M_2M_2}=G^{M_0M_2M_2}_{M_4M_2M_2}=G^{M_4M_2M_2}_{M_4M_2M_2}=G^{M_0M_2M_2}_{M_2M_2M_2}=\frac{1}{2},\ G^{M_4M_2M_2}_{M_2M_2M_2}=-\frac{1}{2},\\
        &G^{M_0M_0M_0}_{M_1M_1M_1}=G^{M_0M_0M_0}_{M_3M_3M_3}=G^{M_0M_4M_4}_{M_3M_1M_1}=G^{M_0M_4M_4}_{M_1M_3M_3}=\frac{1}{\sqrt[4]{3}},\\
        &G^{M_0M_1M_1}_{M_0M_1M_1}=G^{M_0M_3M_3}_{M_0M_3M_3}=G^{M_0M_1M_1}_{M_4M_3M_3}=\frac{1}{\sqrt{3}},\ G^{M_4M_1M_3}_{M_4M_1M_3}=-\frac{1}{\sqrt{3}},\\
        &G^{M_0M_1M_1}_{M_2M_1M_1}=G^{M_0M_3M_3}_{M_2M_3M_3}=G^{M_0M_1M_1}_{M_2M_3M_3}=G^{M_4M_3M_1}_{M_2M_3M_1}=\frac{1}{\sqrt{3}},\ G^{M_4M_1M_3}_{M_2M_3M_1}=-\frac{1}{\sqrt{3}},\\
        &G^{M_2M_1M_3}_{M_2M_3M_3}=-\frac{1}{2},\ G^{M_2M_1M_3}_{M_2M_3M_3}=\frac{1}{2},\\
        &G^{M_1M_1M_2}_{M_1M_1M_2}=G^{M_3M_3M_2}_{M_3M_3M_2}=-\frac{1}{2\sqrt{3}},\ G^{M_1M_3M_2}_{M_1M_3M_2}=G^{M_1M_3M_2}_{M_3M_1M_2}=\frac{1}{2\sqrt{3}},\\
        &G^{M_0M_2M_2}_{M_1M_1M_1}=G^{M_0M_2M_2}_{M_3M_3M_3}=G^{M_0M_2M_2}_{M_1M_3M_3}=G^{M_0M_2M_2}_{M_3M_1M_1}=\frac{1}{\sqrt{2\sqrt{3}}},\\
        &G^{M_4M_2M_2}_{M_1M_3M_1}=G^{M_4M_2M_2}_{M_3M_1M_3}=G^{M_4M_2M_2}_{M_3M_3M_1}=G^{M_4M_2M_2}_{M_1M_1M_3}=\frac{1}{\sqrt{2\sqrt{3}}},\\
        &G^{M_2M_2M_2}_{M_1M_1M_1}=G^{M_2M_2M_2}_{M_3M_3M_1}=\frac{1}{2\sqrt[4]{3}},\ G^{M_2M_2M_2}_{M_3M_3M_3}=G^{M_2M_2M_2}_{M_1M_1M_3}=-\frac{1}{2\sqrt[4]{3}},
    \end{split}
\end{equation}
which agree with the \(6j\)-symbols of \(A_5\), computed using the formula proposed in Ref.~\cite{aasen2020topological}. Hence, there is
\eqn{
A_5\cong{_\A\Bimod_\A(\MA)}.
}
Thus, $\MA$ model is Morita equivalent to $A_5$ model.

\section{Derivation of Local Excitation Creation Operators}\label{sec:local excitation derivation}

In this section introduces how to derive the local excitation creation operators in detail.

For an SET model with a generic input multifusion category $\M$ and a set of isomorphisms that yield a symmetry group $G$, the unconstrained $z$-tensors for trivial anyons contain in general more than one parameters, collectively denoted by $\Theta = (\theta_1, \theta_2, \cdots)$. Then, the trivial anyon creation operators $W^{L,\Theta}_P$ are defined by
\begin{equation}\label{eq:general local excitation}
    [W^{L,\Theta}_P]_{a_0a_n}=\sum_{a_1,a_2,\cdots ,a_{n-1}=1}^{N_\Theta}\prod_{k=0}^{n-1}[W^{L,\Theta}_{l_k}]_{a_ka_{k+1}},
\end{equation}
\begin{equation}
    [W^{L,\Theta}_l]_{ab} \ExcitedC\ :=\  \Tilde{\rho}^{\Theta}_{ab}(g_i,g_j)\ExcitedD.
\end{equation}
where $N_\Theta$ is the dimension of the matrix coefficients $\tilde\rho^\Theta(g_i,g_j)$, and path $P$ intersects $n$ lattice edges, $l_0$ through $l_{n-1}$. Here, \((\idm_{g_i g_i})_a\) denotes the internal space degree of freedom of the trivial anyon. $\idm_{g_ig_i}$ is a simple object of $\M$, as defined in Appendix \ref{sec:multifusion}. The index \(a\) arises from the noncommutative fusion rules, where the \(z\)-tensors take matrix values, as discussed in Appendix~\ref{appendix:noab}. Here The coefficients $\tilde\rho_{ab}^\Theta(g_i, g_j)$ are independent of the degree of freedom $x_{g_ig_j}$ on edge $l$ but only on the $G$-graded indices $g_i, g_j$:
$$\Tilde{\rho}^{\Theta}_{ab}(g_i, g_j) = z^{\mathcal{I};x_{g_ig_j}}_{(\idm_{g_ig_i})_a(\idm_{g_jg_j})_bx_{g_ig_j}}(\Theta),\ \forall x_{g_ig_j}\in\M.$$
The matrix $\Tilde{\rho}^{\Theta}(g_i, g_j)$ satisfies 
\begin{equation}\label{eq:matrix constraint}
    \begin{split}
        &\Tilde{\rho}^{\Theta}(g_i,g_i) = \idm,\ \forall g_i\in G,\\
        &\Tilde{\rho}^{\Theta}(g_i,g_j)\Tilde{\rho}^{\Theta}(g_j,g_k)=\Tilde{\rho}^{\Theta}(g_i,g_k),\ \forall g_i,g_j,g_k\in G.
    \end{split}
\end{equation}
The first equation comes from $z^{\mathcal{I};\idm_{g_ig_i}}_{(\idm_{g_ig_i})_a(\idm_{g_ig_i})_b\idm_{g_ig_i}}(\Theta)=\idm,\ \forall g_i\in G$.
The second equation directly comes from $z$-tensors' definition \eqref{eq:halfA}.

Now we can check how the commutativity \eqref{eq:WcommuteG} constrains $\Theta$ in general. For any $g\in G$, there is a global symmetry transformation that maps $x_{g_ig_j}$ to $x_{(gg_i)(gg_j)}$, so by \eqref{eq:WcommuteG}, we should have:
\begin{equation}\label{eq:constraint1}
    \Tilde{\rho}^{\Theta}(g_i,g_j)=\Tilde{\rho}^{\Theta}(gg_i,gg_j),\ \forall g\in G,
\end{equation}
For each parameters solution $\Theta_i$ to constraint \eqref{eq:constraint1}, there must be a matrix $\rho_i(g)$, such that:
\begin{equation}\label{eq:rep}
    \rho_i(g):=\Tilde{\rho}^{\Theta_i}(e,g)=\Tilde{\rho}^{\Theta_i}(h,hg),\ \forall h\in G,
\end{equation}
According to \eqref{eq:matrix constraint} and $\eqref{eq:rep}$, $\rho_i$ is a representation of symmetry group $G$:
\begin{align*}
    & \rho_i(e)=\Tilde{\rho}^{\Theta_i}(e,e)=\idm,\\
    & \rho_i(g)\rho_i(h)=\Tilde{\rho}^{\Theta_i}(e,g)\Tilde{\rho}^{\Theta_i}(g,gh)=\Tilde{\rho}^{\Theta_i}(e,gh)=\rho_i(gh),\\
    & \rho_i(g^{-1})=\Tilde{\rho}^{\Theta_i}(e,g^{-1})=[\Tilde{\rho}^{\Theta_i}(g^{-1},e)]^{-1}=\rho^{-1}_i(g).
\end{align*}

In conclusion, the local-excitation types in the SET model are characterized by irreducible representations $\rho$ of group $G$. The creation operators for local excitation $\rho$ are
\begin{equation}
    [W^{L,\rho}_P]_{a_0a_n}=\sum_{a_1,a_2,\cdots ,a_{n-1}=1}^{N_\rho}\prod_{k=0}^{n-1}[W^{L,\rho}_{l_{k}}]_{a_ka_{k+1}},
\end{equation}
\begin{equation}
    [W^{L,\rho}_l]_{ab}\ExcitedC\ :=\rho_{ab}(g_i^{-1}g_j)\ExcitedD,
\end{equation}
where $N_{\rho}$  --  the dimension of internal space of the local excitation  --  is the dimension of representation $\rho$.

\bibliographystyle{apsrev4-1}
\bibliography{StringNet}
\end{document}